\documentclass{article}
\usepackage{a4}
\usepackage[utf8]{inputenc}
\usepackage{graphicx,caption,subcaption}
\usepackage{verbatim}
\usepackage[total={6in, 9in}, top=1in,left=1.1in]{geometry} 
\usepackage{moreverb}
\usepackage{psfrag}
\usepackage{amsmath,amssymb,amsfonts,mathrsfs,mathcomp}
\usepackage{upquote}
\usepackage{fancyhdr}
\usepackage[numbers]{natbib}

\usepackage{float}
\usepackage{tikz, graphicx}

\usepackage{colortbl}
\usepackage{setspace}
\usepackage{import}
\usepackage{multirow}
\usepackage[pagewise]{lineno}

\makeatletter
\newcounter{reaction}
\renewcommand\thereaction{R\,\arabic{reaction}}

\newcommand\reactiontag%
{\refstepcounter{reaction}\tag{\thereaction}}
\newcommand\reaction@[2][]%
{\begin{equation}\ce{#2}%
		\ifx\@empty#1\@empty\else\label{#1}\fi%
		\reactiontag\end{equation}}
\newcommand\reaction@nonumber[1]%
{\begin{equation*}\ce{#1}\end{equation*}}
\newcommand\reaction%
{\@ifstar{\reaction@nonumber}{\reaction@}}
\makeatother

\makeatletter
\renewcommand*{\@fnsymbol}[1]{\ensuremath{\ifcase#1\or 1 \or* \or 3  \or  4 \else\@ctrerr\fi}}
\makeatother

\usepackage{imakeidx}
\makeindex

\usepackage{upquote} 
\usepackage[pagebackref=false,colorlinks,linkcolor=blue,citecolor=blue,urlcolor=blue]{hyperref}

\usepackage{xcolor}

\usepackage{color,soul}

\usepackage[version=4]{mhchem}

\usepackage{comment}

\usepackage{listings}

\usepackage{accsupp}
\newcommand{\noncopynumber}[1]{%
	\BeginAccSupp{method=escape,ActualText={}}%
	#1%
	\EndAccSupp{}%
}

\colorlet{mygray}{black!4}
\definecolor{mygreen}{rgb}{0,0.4,0}
\colorlet{myred}{red!80!blue}

\lstdefinestyle{cpp}{
	language=C++,
	basicstyle=\ttfamily\footnotesize,
	keepspaces=true,
	columns=fixed,
	fontadjust=true,
	basewidth=0.5em,
	backgroundcolor=\color{mygray},
	tabsize=4,
	captionpos=b,
	frame=single,
	numbers=left,
	numberstyle=\tiny\noncopynumber,
	numbersep=5pt,
	breaklines=true,
	showstringspaces=false,
	keywordstyle=\color{blue},
	commentstyle=\color{mygreen},
	stringstyle=\color{myred}
}

\lstdefinestyle{OpenFOAMDict}{
	language=C++,
	basicstyle=\ttfamily\footnotesize,
	keepspaces=true,
	columns=fixed,
	fontadjust=true,
	basewidth=0.5em,
	backgroundcolor=\color{mygray},
	tabsize=4,
	captionpos=b,
	frame=single,
	numbers=left,
	numberstyle=\tiny\noncopynumber,
	numbersep=5pt,
	breaklines=true,
	showstringspaces=false,
	commentstyle=\color{mygreen},
}

\makeatletter
\newcommand*{\textalltt}{}
\DeclareRobustCommand*{\textalltt}{%
	\begingroup
	\let\do\@makeother
	\dospecials
	\catcode`\\=\z@
	\catcode`\{=\@ne
	\catcode`\}=\tw@
	\verbatim@font\@noligs
	\@vobeyspaces
	\frenchspacing
	\@textalltt
}
\newcommand*{\@textalltt}[1]{%
	#1%
	\endgroup
}
\makeatother

\usepackage{siunitx}
\usepackage{nomencl}
\usepackage{ifthen}
\renewcommand\nomgroup[1]{
	\ifthenelse{\equal{#1}{C}}{%
		\vspace{0.5cm}%
		\item[\textbf{Constants}]}{%
		\ifthenelse{\equal{#1}{A}}{%
			\vspace{0.5cm}%
			\item[\textbf{Acronyms}]}{
			\ifthenelse{\equal{#1}{E}}{%
				\vspace{0.5cm}%
				\item[{\textbf{English symbols}}]}{
				\ifthenelse{\equal{#1}{G}}{%
					\vspace{0.5cm}
					\item[{\textbf{Greek symbols}}]}{
					\ifthenelse{\equal{#1}{S}}{%
						\vspace{0.5cm}%
						\item[\textbf{Superscripts}]}{
						\ifthenelse{\equal{#1}{U}}{%
							\vspace{0.5cm}%
							\item[\textbf{Subscripts}]}{
							\ifthenelse{\equal{#1}{X}}{%
								\vspace{0.5cm}%
								\item[\textbf{Other symbols}]}{
								{}}}}}}}}}

\newcommand{\nomunit}[1]{\renewcommand{\nomentryend}{\dotfill#1}}
\makenomenclature

\usepackage{titlesec}

\titleformat{\paragraph}
{\normalfont\normalsize\bfseries}{\theparagraph}{1em}{}
\titlespacing*{\paragraph}
{0pt}{3.25ex plus 1ex minus .2ex}{1.5ex plus .2ex}

\newcommand{\beginsupplement}{%
	\setcounter{figure}{0}%
	\renewcommand{\thefigure}{S\arabic{figure}}%
	\setcounter{table}{0}%
	\renewcommand{\thetable}{S\arabic{table}}%
	\setcounter{equation}{0}%
	\renewcommand{\theequation}{S\arabic{equation}}%
	\setcounter{section}{0}%
	\renewcommand{\thesection}{S\arabic{section}}%
	\setcounter{page}{1}%
	\renewcommand{\thepage}{S.\arabic{page}}%
	\renewcommand{\theHfigure}{S\arabic{figure}}%
	\renewcommand{\theHtable}{S\arabic{table}}%
	\renewcommand{\theHequation}{S\arabic{equation}}%
	\renewcommand{\theHsection}{S\arabic{section}}%
}

\newcommand{\citeColored}[2]{%
	{\hypersetup{citecolor=#1}%
		\cite{#2}}%
}

\begin{document}

\pagenumbering{arabic}
\setcounter{page}{1} 

\title{\renewcommand\baselinestretch{1}\bf
	Omnisoot: an object-oriented process design package for gas-phase synthesis of carbonaceous nanoparticles
}
\renewcommand\baselinestretch{0.8}
\author{
	Mohammad Adib{\vspace{0.4em}}\footnote{\scriptsize{Department of Mechanical and Aerospace Engineering, Carleton University, 1125 Colonel By Dr, Ottawa, ON K1S 5B6, Canada}}{\hspace{0.3em}}\textsuperscript{,}{\hspace{0.01em}}\footnote{\scriptsize{Correspondeing author}},
	Sina Kazemi\textsuperscript{1}{\vspace{0.4em}},  
	M. Reza Kholghy\textsuperscript{1,*} 
}
\date{}
\maketitle
\renewcommand\baselinestretch{1.3}

\begin{abstract}
	A computational tool, Omnisoot, was developed utilizing the chemical kinetics capabilities of Cantera to model the formation of carbonaceous nanoparticles, such as soot and Carbon Black (CB), from the reactions of gaseous hydrocarbons. Omnisoot integrates constant volume, constant pressure, perfectly stirred, and plug flow reactor models with four inception models from the literature, as well as two population balance models: a monodisperse model and a sectional model. This package serves as an integrated process design tool to predict soot mass, morphology, and composition under varying process conditions. The modeling approach accounts for soot inception, surface growth, and oxidation, coupled with detailed gas-phase chemistry, to close the mass and energy balances of the gas-particle system; subsequently, soot and gas-phase chemistry are linked to the particle dynamics models that consider the evolving fractal-like structure of soot agglomerates. The developed tool was employed to highlight the similarities and differences among the implemented inception models in predicting soot mass, morphology, and size distribution for three use-cases: methane pyrolysis in a shock tube, ethylene pyrolysis in a flow reactor, and ethylene combustion in a perfectly stirred reactor. The simulations of 5\% $\mathrm{CH_4}$ pyrolysis in shock-tube with short residence times ($\approx1.5$ ms) demonstrated that multiple combinations of inception and surface growth rates minimized the prediction error for carbon yield but led to markedly different morphologies, emphasizing the need for measured data on soot morphology to constrain inception and surface growth rates. The comparison of simulation results in a pyrolysis flow reactor at three different flow rates suggested that only irreversible models can predict bimodality in particle size distribution. 
	
\end{abstract}
	
\maketitle
\nomenclature[a]{MPBM}{Monodisperse Population Balance Model}
\nomenclature[a]{CFD}{Computational Fluid Dynamics}
\nomenclature[a]{SPBM}{Sectional Population Balance Model}
\nomenclature[a]{DEM}{Discrete Element Modelling}
\nomenclature[a]{HACA}{H-abstraction-$\mathrm{C_2H_2}$/Carbon-addition}

\nomenclature[a]{PSD}{Particle Size Distribution}
\nomenclature[a]{SPSD}{Self-Preserving Size Distribution}
\nomenclature[a]{MOM}{Method of Moment}
\nomenclature[a]{TEM}{Transmission Electron Microscopy}

\nomenclature[a]{CMF}{Carbon Mass Fraction}

\nomenclature[c]{$k_B$}{Boltzmann constant \nomunit{$1.3806488 \times 10^{-23}$ \si{J \per K}}}
\nomenclature[c]{$Av$}{Avogadro's number\nomunit{6.02214076 $\times 10^{23}$ \si{1 / mol}}}
\nomenclature[c]{$R$}{Global gas constant\nomunit{8.314 \si{\joule\per\mol\per\kelvin}}}

\nomenclature[e]{$h$}{Enthalpy\nomunit{\si{J\per\kg}}}
\nomenclature[e]{$e$}{Internal energy\nomunit{\si{J\per\kg}}}

\nomenclature[e]{$c_p$}{Heat capacity at constant pressure\nomunit{\si{\joule\per\kg\per\kelvin}}}
\nomenclature[e]{$c_v$}{Heat capacity at constant volume\nomunit{\si{\joule\per\kg\per\kelvin}}}
\nomenclature[e]{$c_{soot}$}{Heat capacity of soot\nomunit{\si{\joule\per\kg\per\kelvin}}}

\nomenclature[e]{$u$}{Velocity\nomunit{\si{\meter\per\second}}}

\nomenclature[e]{$V$}{Volume\nomunit{\si{\meter\cubed}}}

\nomenclature[e]{$p$}{Pressure\nomunit{\si{\pascal}}}

\nomenclature[e]{$P_c$}{Wetted perimeter of the flow reactor\nomunit{\si{\meter}}}

\nomenclature[e]{$A$}{Surface area\nomunit{\si{\meter\squared}}}

\nomenclature[e]{$k_{dep}$}{Deposition velocity\nomunit{\si{\meter\per\second}}}

\nomenclature[e]{$N_{agg}$}{Number density of agglomerates\nomunit{\si{mol\per\kg}}}
\nomenclature[e]{$N_{pri}$}{Number density of primary particles\nomunit{\si{mol\per\kg}}}
\nomenclature[e]{$C_{tot}$}{Total carbon content of soot particles (per section)\nomunit{\si{mol\per\kg}}}
\nomenclature[e]{$H_{tot}$}{Total hydrogen content of soot particles (per section)\nomunit{\si{mol\per\kg}}}

\nomenclature[e]{$E(m)$}{Absorption function of soot\nomunit{\si{-}}}

\nomenclature[e]{$T$}{Temperature\nomunit{K}}
\nomenclature[e]{$I$}{Source terms for soot variables\nomunit{\si{\mol\per\kilogram\per\second}}}
\nomenclature[e]{$d$}{Diameter\nomunit{m}}
\nomenclature[e]{$m$}{Mass\nomunit{\si{\kilogram}}}
\nomenclature[e]{$W$}{Molecular weight\nomunit{kg/mol}}
\nomenclature[e]{$n_p$}{Number of primary particles of per agglomerate\nomunit{\si{-}}}
\nomenclature[e]{$d_p$}{Primary particle diameter \nomunit{\si{\metre}}}
\nomenclature[e]{$d_m$}{Mobility diameter \nomunit{\si{\metre}}}
\nomenclature[e]{$d_g$}{Gyration diameter \nomunit{\si{\metre}}}
\nomenclature[e]{$d_c$}{Collision diameter \nomunit{\si{\metre}}}
\nomenclature[e]{$k$}{Reaction rate constant \nomunit{\si{\meter\cubed\second\per\mol}}}
\nomenclature[e]{$D$}{Diffusion coefficient of particles \nomunit{\si{\meter\squared\per\second}}}
\nomenclature[e]{$SF$}{Sectional spacing factor \nomunit{\si{-}}}
\nomenclature[e]{$f$}{Friction factor \nomunit{\si{-}}}
\nomenclature[e]{$R_H$}{Duct hydraulic radius \nomunit{\si{\meter}}}
\nomenclature[e]{$D_H$}{Duct hydraulic diameter \nomunit{\si{\meter}}}
\nomenclature[e]{$Kn$}{Knudsen number \nomunit{\si{-}}}
\nomenclature[e]{$f_v$}{Soot volume fraction based on gas volume\nomunit{\si{\metre\cubed\per\metre\cubed}}}
\nomenclature[e]{$Y$}{Mass fraction of gaseous species\nomunit{\si{-}}}
\nomenclature[e]{$Re$}{Reynolds number\nomunit{\si{-}}}
\nomenclature[e]{$Sc$}{Schmidt number\nomunit{\si{-}}}
\nomenclature[e]{$Sh$}{Sherwood number\nomunit{\si{-}}}

\nomenclature[g]{$\rho$}{Density\nomunit{\si{\kg\per\metre^3}}}
\nomenclature[g]{$\varphi$}{Soot volume fraction based on reactor volume\nomunit{\si{\metre\cubed\per\metre\cubed
}}}

\nomenclature[g]{$\nu$}{Gas kinematic viscosity\nomunit{\si{\metre^2\per\second}}}
\nomenclature[g]{$\beta$}{Collision frequency\nomunit{\si{\metre^3\per\second}}}
\nomenclature[g]{$\lambda$}{Mean free path\nomunit{\si{\metre}}}
\nomenclature[g]{$\lambda_a$}{Mean stopping distance of particles\nomunit{\si{\metre}}}
\nomenclature[g]{$\delta_a$}{Mean distance of particles\nomunit{\si{\metre}}}
\nomenclature[g]{$\omega$}{The rate of production of gaseous species\nomunit{\si{\mol\per\metre\cubed\second}}}
\nomenclature[g]{$\tau_w$}{Wall shear stress\nomunit{\si{\pascal}}}
\nomenclature[g]{$\tau_{psr}$}{Nominal residence time of perfectly stirred reactor\nomunit{\si{\second}}}
\nomenclature[g]{$\epsilon$}{Wall roughness\nomunit{\si{\meter}}}

\nomenclature[u]{agg}{Agglomerate}
\nomenclature[u]{pri}{Primary particle}
\nomenclature[u]{fm}{Free-molecular}
\nomenclature[u]{cont}{Continuum}
\nomenclature[u]{inc}{Inception}
\nomenclature[u]{grow}{Surface growth}
\nomenclature[u]{ox}{Oxidation}
\nomenclature[u]{coag}{Coagulation}
\nomenclature[u]{ads}{Adsorption}
\nomenclature[u]{f}{Forward}
\nomenclature[u]{r}{Reverse}
\nomenclature[u]{reac}{Reactive}


\setlength{\nomitemsep}{-\parsep}
\printnomenclature

\section{Introduction}
\begin{figure}[H]
	\centering
	\begin{tikzpicture}
		\draw (0, 0) node[inner sep=0] 	{\includegraphics[width=0.7\textwidth]{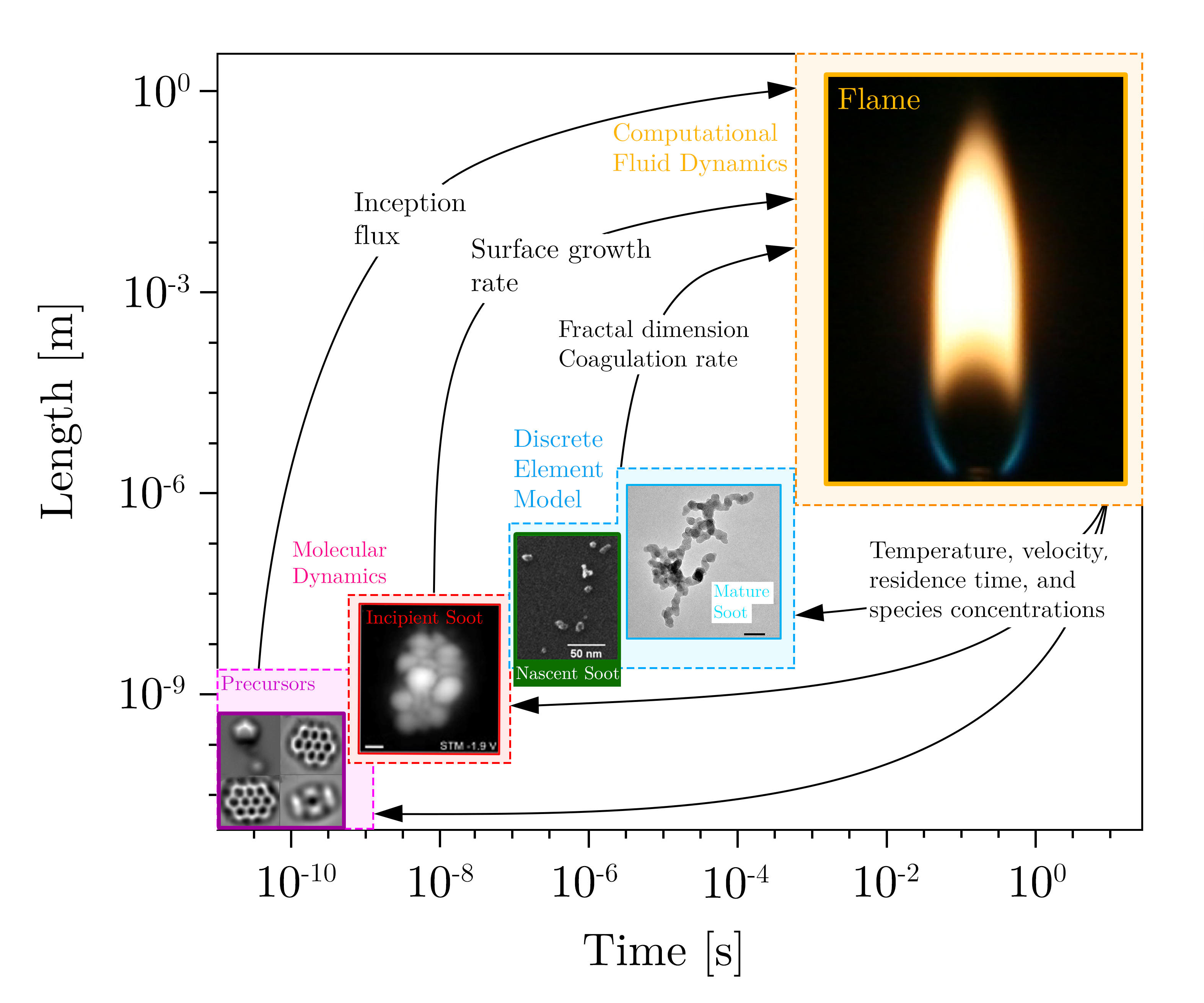}};
		\draw (-2.45, -1.63) node[text=magenta] {\tiny{
				\citeColored{magenta}{commodo2019early}
		}};
		\draw (-1.99, -1.25) node[text=red] {\tiny{
				\citeColored{red}{schulz2019insights}
		}};
		\draw (-0.65, -0.7) node[text=green] {\tiny{
				\citeColored{green}{schenk2013imaging}
		}};
		\draw (0.39, -0.03) node[text=cyan] {\tiny{
				\citeColored{cyan}{kholghy2021morphology}
		}};					
		\draw (2.3, 3.2) node[text=orange] {\footnotesize{
				\citeColored{orange}{johansson2017evolution}
		}};		
	\end{tikzpicture}
	\caption{The processes involved in soot formation span a wide range of time and length scales, including molecular reactions between precursors, surface growth, coagulation of incipient and nascent soot particles, and their maturation, which is governed by the temperature-time history determined by flame velocity and temperature.}
	\label{fig:sootscales}
\end{figure}
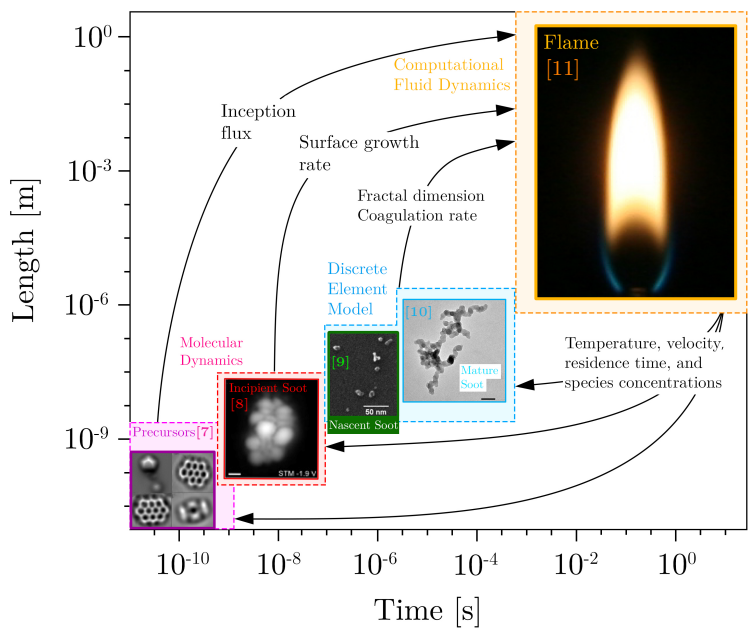

The formation of carbonaceous nanoparticles such as soot and Carbon Black (CB) is a complex, multiscale process involving chemical reactions, heat transfer, fluid dynamics, and particle dynamics, and spans wide spatial ($\sim 10^{-9}$ to 1~m) and temporal ($\sim 10^{-10}$ to 1~s) scales. Figure~\ref{fig:sootscales} illustrates the length and time scales relevant to different stages of soot formation, from PAH precursors to incipient, nascent, and mature soot in flames. Understanding the effect of process parameters on the morphology and composition of carbonaceous nanoparticles is crucial to assess the health and environmental impacts of soot and the functional properties of CB.

Soot is a broadband light absorber~\cite{d2009combustion} and is emitted on a large scale ($\sim 9.5$~megatons per year~\citep{myhre2014anthropogenic}), making it the third strongest contributor to climate change after methane and carbon dioxide. Its fine particulate nature ($\mathrm{PM_{2.5}}$) also poses serious health risks~\citep{borm2004inhaled}. In contrast, CB is the largest flame-made nanomaterial by production volume ($\sim15$~megatons annually~\citep{rosner2024techno}) and plays a vital role in industrial applications, including rubber reinforcement~\citep{international2016carbon} and lithium-ion batteries~\citep{Palomares2010}.

Controlling CB yield, structure, morphology, and composition is essential for producing specific grades of CB tailored to target applications in both conventional (e.g., furnace process~\citep{dames2023plasma}) and emerging production methods~\citep{li2017experimental, fulcheri2023energy, patlolla2023review}. However, the influence of process parameters such as feedstock composition, pressure, and temperature-time history on CB properties remains incompletely understood due to the complexity of gas-phase chemistry and the intricate nature of CB formation (as discussed in the following sections). This underscores the need for robust computational models to predict yield, particle structure, and composition under varying process conditions, and to support the mitigation of soot emissions from combustion systems.

\textit{"Soot"} and \textit{"Carbon Black"} are distinct materials in terms of chemical properties and synthesis processes~\cite{watson2001carbon}. While, soot usually refers to unwanted particulate matter formed during incomplete combustion with variable organic content and a large variation in carbon to hydrogen ratio (C/H)~\citep{watson2001carbon}, CB is commercially produced under highly controlled partial combustion or thermal decomposition of hydrocarbons. However, soot particles generated under controlled laboratory conditions can have similar structure and composition to CB. Mature soot formed in methane and ethylene premixed flames can reach elemental $\mathrm{C/H}\approx20$~\cite{russo2015dehydrogenation}, which is close to C/H of CB. Additionally, the comparison of Transmission Electron Microscopy (TEM) images of industrially produced CB~\citep{singh2018nanostructure} with soot sampled from diesel fuel~\citep{vander2007hrtem, lapuerta2017morphological} indicates similarities in their morphology and structure. Hereafter, soot will be used to collectively refer to carbonaceous nanoparticles produced during combustion/pyrolysis processes.

Soot inception is believed to begin with the formation of Polycyclic Aromatic Hydrocarbons (PAHs) in the gas phase and followed by their transition into incipient particles. Soot inception remains poorly understood at the level of pathways and elementary reactions~\citep{Wang2011}. This lack of understanding stems primarily from uncertainties in PAH chemistry and the kinetics of PAH growth into soot particles, a process that is highly reversible and thus sensitive to local conditions such as temperature, pressure, and the concentrations of intermediate species~\citep{Wang2011}.

The classic description of soot inception relies on PAH dimerization where collision of two PAH molecules (monomers in this context) forms a dimer held together by van der Waals forces (vdW)~\citep{frenklach1991detailed}. The dimerization is an irreversible process with an efficiency that accounts for the reversibility or dissociation of dimers. The theory postulates that PAH growth continues by sequential addition of monomers forming stacks of dimers, trimers, tetramers, and so on to reach a certain mass threshold that marks the emergence of incipient soot~\citep{frenklach1991detailed}, but for practical purposes, a dimer is usually considered incipient soot. Here, we call this model \textit{Irreversible Dimerization}. 
Irreversible Dimerization has been used to predict soot formation in burner-stabilized premixed~\citep{salenbauch2015modeling, desgroux2017comparative}, counterflow diffusion flames~\citep{wang2015soot, xu2021experimental}, and coflow diffusion flames~\citep{kholghy2016core, veshkini2016understanding}. A collision efficiency factor ranging between $10^{-6}$ to 1 is also employed to adjust the inception flux and PAH adsorption rates to achieve desired soot mass and size distribution. PAHs of moderate sizes, such as pyrene to coronene, are typically considered as the starting point of inception due to their thermodynamic stability, which justifies the irreversibility at high temperatures~\citep{frenklach1991detailed}. However, the theoretical calculations~\citep{miller1985calculations} and experiments~\citep{sabbah2010exploring} indicated that PAH dimerization is highly reversible in flame conditions. The inception flux of irreversible dimerization is mainly controlled by PAH concentration due to its weak temperature dependence, so it produces new particles at low temperatures (even below 500 K)~\citep{naseri2022simulating} despite experimental evidence for termination of inception below 1200 K~\citep{sanchez2012polycyclic, cho2016synthesis}.

\citet{kholghy2019role} emphasized the necessity of chemical bond formation following physical PAH clustering for accurate prediction of volume fraction, primary particle diameter, and Particle Size Distribution (PSD) in ethylene coflow diffusion flames. \citet{kholghy2018reactive} proposed the \textit{Reactive Dimerization} model, which begins with the reversible collision of PAHs to form physical dimers stabilized by van der Waals forces. These dimers are subsequently graphitized and form chemically bonded dimers that act as incipient soot, which then grow via coagulation and surface reactions.

However, \citet{frenklach2020mechanism} argued that an inception model initiated by a highly reversible step, such as that in Reactive Dimerization~\citep{kholghy2018reactive}, cannot generate a sufficient particle flux to match measurements in the benchmark burner-stabilized stagnation flame~\citep{abid2009quantitative}. Instead, they proposed the H-abstraction-$\mathrm{C_2H_2}$/Carbon-addition~\citep{frenklach1991detailed, appel2000kinetic} (HACA)-driven mechanism where addition of a monomer molecule to its radical activated by hydrogen abstraction forms a stable dimer via an E-Bridge bond formation, and this sequential process continues to form trimers, tetramers, and larger PAH clusters. 

The gas-phase chemistry of aromatics can be extended to account for chemical growth of incipient soot via surface reactions~\citep{frenklach2002reaction}. This hypothesis, known as “chemical similarity”, postulates that the reactions occurring on the soot surface are similar to those involving large molecules of PAHs in the gas phase. It also provides a means to describe the rates of surface growth and particle oxidation in
terms of elementary chemical reactions. In other words, it is assumed that the surface of soot particles consists of lateral faces of large PAHs covered with C-H bonds that can be transformed into radical sites through H-abstraction~\citep{frenklach1991detailed, appel2000kinetic}. These sites can then be attacked by acetylene molecules, leading to an increase in particle mass. The reactivity of these sites changes with time and temperature~\citep{woods1991soot, dasch1985decay}, described as soot aging. For modeling purposes, a temperature-dependent surface reactivity, usually represented by $\alpha$, was introduced to account for the depletion of reactive sites with soot maturity. \citet{appel2000kinetic} showed $\alpha$ changes with temperature and particle size in laminar premixed flames.

Adsorption of PAHs on the surface of soot particles is also a viable growth mechanism~\citep{frenklach1991detailed}, more specifically called physisorption or chemisorption depending on the mechanisms driving the adsorption process~\citep{michelsen2020review}. There is still debate over the stability of adsorbed PAH molecules on soot surface~\citep{obolensky2007interplay}. Following the hypothesis that PAHs are building blocks of soot particles, a mechanism similar to inception is often used to describe PAH-soot growth.

In typical soot formation processes, such as those occurring in flames and reactors, inception and surface growth are active for only a relatively short duration compared to the total residence time of soot particles. Subsequently, due to high particle concentrations, coagulation becomes the dominant mechanism, rapidly leading to the development of both a Self-Preserving Size Distribution (SPSD)~\citep{lai1972self} and an asymptotic fractal-like structure~\citep{mountain1986simulation, Goudeli2016}.
The collision frequency of agglomerates depends on their evolving fractal-like morphology, with polydisperse agglomerates colliding more frequently than monodisperse ones. This enhancement reaches an asymptotic value of 35$\%$~\citep{Goudeli2016} in the free molecular regime or 82$\%$~\citep{kelesidis2021self} in the transition regime at SPSD, irrespective of primary particle polydispersity.

Particle morphology, governed by inception, surface growth, and agglomeration, can be precisely tracked using mesoscale simulations such as Discrete Element Modeling (DEM)~\citep{Kelesidis2017Flame} provided that the gas-to-particle mass flux through inception and surface growth is known a priori. However, these methods are computationally expensive, and integrating them with chemical kinetics in Computational Fluid Dynamics (CFD) frameworks is challenging~\citep{kelesidis2021perspective}. Consequently, their application is often limited to canonical scenarios where particle dynamics are solved independently of chemistry and flow dynamics, typically by populating the simulation domain with incipient particles and neglecting the removal or addition of gaseous species due to soot formation~\citep{Kelesidis2017} as it is difficult to have a particle dynamics modeled fully coupled with gas chemistry. 

As a computationally efficient alternative, particle dynamics can be tracked by Eulerian approaches such as the Method of Moments (MOM)~\citep{kazakov1998dynamic} or Monodisperse Population Balance Models (MPBM)~\citep{kruis1993simple}. Such models only track average particle properties and their accuracy can be limited if unrealistic assumptions such as approximating agglomerates as spheres are used. However, when inception and surface growth are short~\citep{Spicer2002} and high particle number concentrations are formed~\cite{Kelesidis2017}, they lead to rapid attainment of SPSD and agglomerates having asymptotic structure~\citep{Goudeli2016}. In this case, a MPBM or MOM can be assembled on a firm scientific basis with accuracy on par with DEM~\citep{Kelesidis2017Flame} and experimental data ~\citep{abid2008evolution, ma2013soot, camacho2015mobility}. Such models can be readily interfaced with CFD simulations~\citep{grohn2012fluid} without significant computational cost, making them ideal for three-dimensional and even turbulent flame simulations.

The main difficulty of MOM is the closure of transport equations terms rising from expressing inception, surface growth and coagulation source terms based on the tracked moments~\citep{frenklach2002method}. In contrast, MPBMs do not have the closure problem and calculate average particle properties by tracking their total concentration, mass \citep{kruis1993simple} and area~\citep{tsantilis2004soft, lindstedt1994simplified}. 

Sectional Population Balance Models (SPBMs) are similar to MPBMs but capable of tracking PSDs~\citep{Xiong1993}. SPBMs can solve one, two, or three equations per section, with their capabilities depending on the number of equations solved: single-equation models track basic properties like total mass or number concentration~\citep{gelbard1980sectional}, two-equation models capture additional details such as surface area or primary particle number~\citep{park2004novel}, and three-equation models enable detailed tracking of complex properties like composition or morphology~\citep{kholghy2016core}. Coupled with relations for fractal-like structure~\citep{matsoukas1991dynamics} and collision frequency~\citep{fuchs1964mechanics}, SPBMs can accurately model particle size distribution, morphology, and composition of soot particles. However, their computational cost rises significantly with the number of sections~\citep{xiong1993formation} and tracked properties~\citep{kholghy2016core}.

SPBMs are gaining attention, due to the recent increase in computational resources, for simulating laminar and turbulent benchmark flames in two-dimensional domains, even with moderately large chemical mechanisms. For example, the CoFlame solver~\citep{eaves2016coflame}, designed to simulate laminar diffusion flames in axisymmetric domains using a SPBM, has been employed in numerous studies, demonstrating its ability to accurately capture soot morphological properties~\citep{dworkin2011application, liu2015numerical}. The CRECK group developed a sectional approach in which discrete size bins are embedded in the kinetic mechanism as lumped pseudo-species (BINs), covering the mass range from $\mathrm{C_{20}}$ PAHs to the largest agglomerates ($\mathrm{C_{10^7}}$)~\citep{cuoci2015opensmoke++, cuoci2013computational}. This hybrid soot-gas mechanism, available in standard Chemkin file format, has been validated against a wide range of flame and reactor configurations~\cite{ramalli2023automatic}. However, portability and flexibility remain major challenges of this approach. The base gas-phase chemistry (excluding BINs) cannot be readily replaced with other mechanisms, which is important given the large uncertainty in the prediction of even small hydrocarbons (as discussed in Section~\ref{sec:uncergaschem}). In addition, the large size of the mechanism comprising 709 species and 109500 reactions~\citep{nobili2022modeling} poses a significant computational burden, particularly in parametric studies.

Here, we develop a computational package, called Omnisoot, which integrates Cantera~\citep{cantera} as a chemistry solver to simulate soot formation in reduced-order dimensions. The package provides a versatile soot model that can be coupled with all reaction mechanisms. Both particle dynamics models (MPBM and SPBM) are coupled with various soot inception and surface growth models,  offering flexibility to study soot formation alongside gas phase chemistry. This package facilitates fundamental investigations of soot formation, including pathway analysis, reaction mechanism evaluation, and inception flux estimation, while also supporting process design and optimization of CB production in industrial reactors under diverse fuel compositions, temperatures, pressures, and residence times. The theoretical background and governing equations for sub-models of Omnisoot are detailed in subsequent sections, followed by validation against benchmark DEM simulations and verification of mass and energy conservation across all models. Finally, three use cases of Omnisoot were demonstrated by predicting gas chemistry, soot yield, morphology, and size distribution in shock tubes, flow reactors, and perfectly stirred reactors.

\section{Theoretical foundation and governing equations}
\begin{figure}[!htbp]
	\centering
	\includegraphics[height=90mm, ]{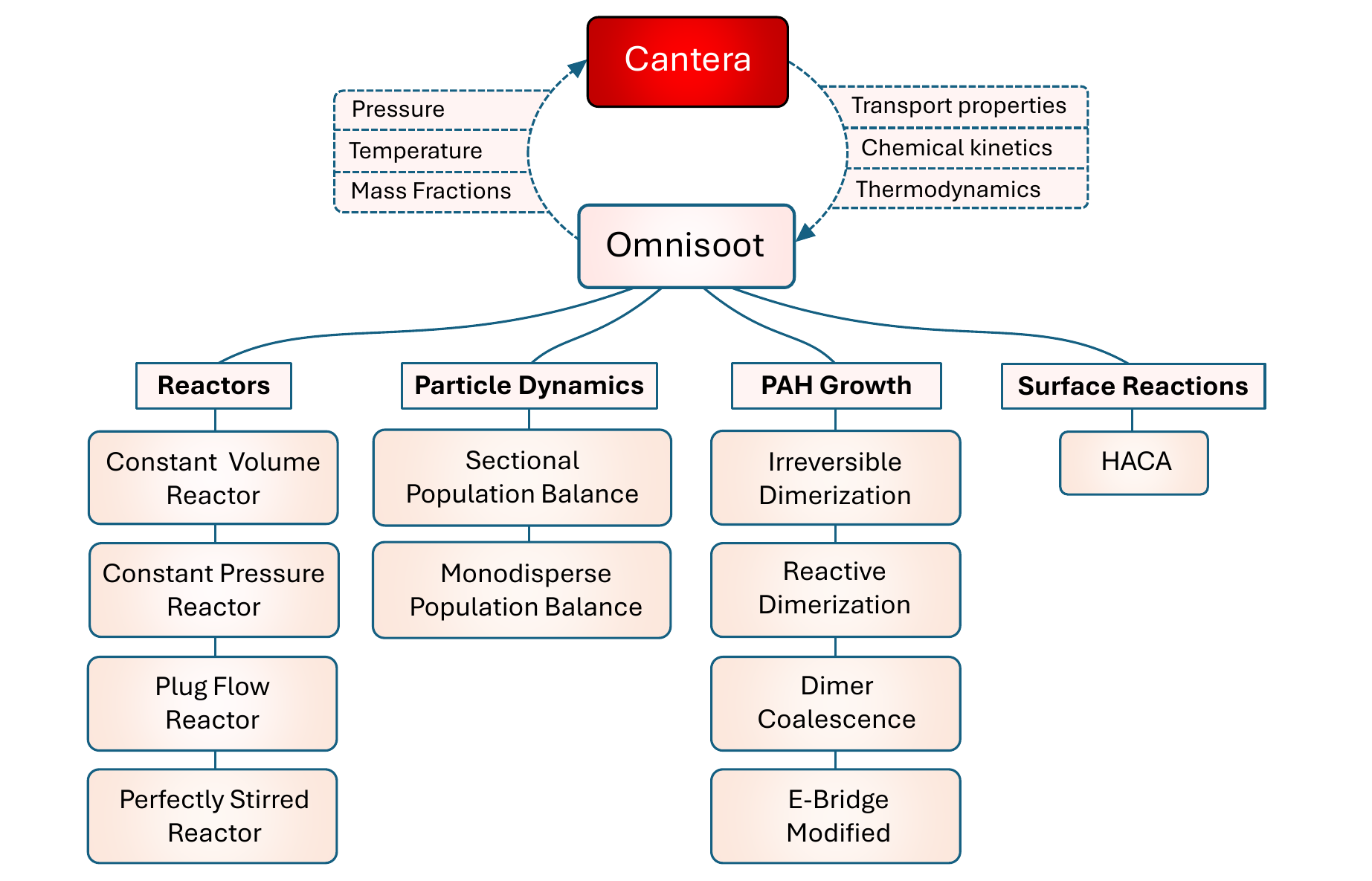}
	\caption{The structure of Omnisoot that illustrates the coupling with Cantera and sub-models including the reactors, the particle dynamic models, the PAH growth models and the surface reactions model.}
	\label{fig:structure}
\end{figure}

The mathematical basis for Omnisoot is explained in the top-to-bottom hierarchical order. First, the governing equations for constant volume, constant pressure, perfectly stirred and plug flow reactor models are reviewed in Sections~\ref{sec:cvr}-\ref{sec:pfr}. The transport equations of ``soot variables" include the source terms that account for change in each soot variable due to inception, surface growth, oxidation and coagulation. Then, particle dynamics models are explained in Sections~\ref{sec:particledynamics} that entail the description of size distribution and morphology of soot particles and their collision rate, which is used to determine the coagulation source terms. Section~\ref{sec:pahgrowmodel} focuses on the \textit{``PAH growth"} models that take care of inception and adsorption from designated precursors and calculate the corresponding source terms. Similarly, the \textit{"surface reactions"} model is detailed in Section~\ref{sec:surfreacmodel}, which elaborates on the surface growth and oxidation rates based on HACA mechanism. Figure~\ref{fig:structure} illustrates the general structure of Omnisoot and its sub-models.

\subsection{Assumptions and conventions}
Here, the main conventions and assumptions used in the derivation of the mathematical models are listed below.

\begin{enumerate}

\item $f_v$ and $\varphi$ denote the volume of soot particles normalized by the gas volume and reactor volume, respectively. Their relationship can be expressed as:

\begin{equation}
	\begin{split}
		f_v = \frac{V_{soot}}{V_{gas}}, \\
		\varphi = \frac{V_{soot}}{V_{gas} + V_{soot}}, \\
		\varphi = \frac{f_v}{1 + f_v}
		\label{eqn:fvdef}.
	\end{split}
\end{equation}

\item ${\dot{s}_k}$ denotes the rate production/consumption of $k_{th}$ gaseous species (i.e., gas scrubbing) due to soot inception, surface growth and oxidation. It is positive when the species is released to the gas mixture.

\item Each soot agglomerate consists of monodisperse spherical primary particles, which are in point contact. So, the model does not consider formation of necking (or sintering) in soot agglomerates by surface growth.


\item The word \textit{``particle"} refers to soot both in spherical and agglomerate shape. 

\item The density of soot is assumed constant at the value of 1800 $\mathrm{kg/m^3}$. Soot density changes with its maturity level, which is often linked to the elemental C/H ratio of soot particles~\citep{michelsen2021effects}. Here, the considered value represents an average between density of mature soot with high C/H ratios ($\mathrm{\rho=2000\;kg/m^3}$) and that of nascent soot with low C/H ratios ($\mathrm{\rho=1600\;kg/m^3}$)~\citep{jensen2007measurement, michelsen2021effects}.

\item The incipient soot particles are 2~nm in diameter, so the model does not allow particles with a primary particle diameter smaller than 2~nm. The number of carbon atoms in the incipient soot particle, $n_{c,min}$, is calculated from the mass of a sphere with a diameter of $d_{p,min}=2$~nm assuming pure carbon content, which results in:
\begin{equation}
	\begin{split}
	n_{c,min}& =\frac{\pi}{6}\rho_{soot}d^3_{p,min}\frac{1}{W_{carbon}}\approx378.
	\label{eqn:nc_min}
	\end{split}
\end{equation}

\item The specific heat, internal energy and enthalpy of soot are approximated by those of pure graphite, and are employed to close the energy balance in the system~\cite{mcbride1993coefficients}.

\item Soot particles and gas are in thermal equilibrium during soot formation processes, and there is no temperature gradient within each agglomerate.

\item $\psi$ denotes a \textit{soot variable} that represents a mean property of soot particles in each section tracked in Omnisoot by solving transport equations for the total concentration of agglomerates, $N_{agg}$ and primary particles, $N_{pri}$, and the total carbon, $C_{tot}$ and hydrogen content, $H_{tot}$ of soot. $S_{\psi}$ is the source term of the soot variable, $\psi$, which appears in soot equations.  

\item \textit{PAH growth} is a sub-model of Omnisoot with a set of pathways that determine the rate of inception and adsorption from PAHs (designated as soot precursors) in the gas mixture.

\item \textit{Surface reactions} is a sub-model of Omnisoot that describes the addition of acetylene to soot surface, and removal of carbon via oxidation by OH and $\mathrm{O_2}$ following the HACA scheme. The model does not consider soot oxidation with $\mathrm{CO_2}$, $\mathrm{H_2O}$ and $\mathrm{NO_x}$.

\item The single superscript \textit{i} denotes the section number of a soot variable or a derived property. For example, $d^i_p$ represents the primary particle diameter of section \textit{i}. The double superscript \textit{ij} indicates a property related to two sections; for example, $\beta^{ij}$ denotes the collision frequency between sections \textit{i} and \textit{j}. In the case of the monodisperse model, the section number can be omitted because it is equivalent to the sectional model with only one section.

\item The computation of morphological parameters is done similarly in both particle dynamics models, but they are explained separately in Section~\ref{sec:sootmorphology}.

\item \textit{precursors} refer to the PAHs larger than naphthalene used for inception and PAH adsorption. The list of precursors with their chemical formula and molecular mass is provided in Table~\ref{tab:precursors_list}. It should be noted that the precursors can be dynamically changed by Omnisoot's user interface.

\renewcommand{\arraystretch}{1.5}
\begin{table}
	\caption{The names, symbols, chemical formula and molecular weight of the soot precursors used by Omnisoot.}
	\label{tab:precursors_list}
	\centering
	\begin{tabular}{l l l l}
		\hline
		Species name & Symbol & Chemical formula & W~[kg/mol] \\
		\hline
		Naphthalene       & A2   &  $\mathrm{C_{10}H_{8}}$   & 0.128 \\
		Phenanthrene      & A3   &  $\mathrm{C_{14}H_{10}}$  & 0.178 \\
		Pyrene            & A4   &  $\mathrm{C_{16}H_{10}}$  & 0.202 \\
		Acenaphthylene    & A2R5 &  $\mathrm{C_{12}H_{8}}$   & 0.152 \\
		Acephenanthrylene & A3R5 &  $\mathrm{C_{16}H_{10}}$  & 0.202 \\
		Cyclopentapyrene  & A4R5 &  $\mathrm{C_{18}H_{10}}$  & 0.226 \\
		\hline
	\end{tabular}
\end{table}

\end{enumerate}

\subsection{Reactors}

\begin{figure}[H]
	\centering
	\includegraphics[width=1\textwidth]{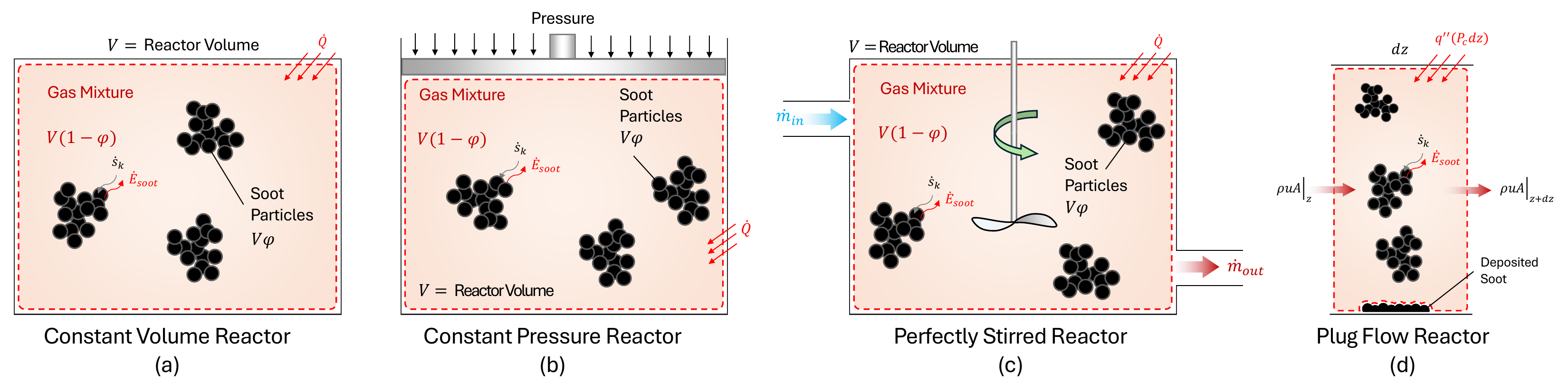}
	\caption{The schematics of constant volume reactor (a), constant pressure reactor (b) perfectly stirred reactor (c), and plug flow reactor (d).}
	\label{fig:reactors} 
\end{figure}

The governing equations of reactor models implemented in Omnisoot are briefly presented in the following sections. The control volume encompasses the gas mixture and soot particles, as illustrated in Figure~\ref{fig:reactors}. The equations ensure conservation of total mass and energy of the gas and particle system, which can also receive or lose heat through the reactor walls.

\subsubsection{Constant Volume Reactor (CVR)}
\label{sec:cvr}
As shown in Figure~\ref{fig:reactors}a, CVR represents a closed system constant volume where chemical reactions converts part of the gas mixture to soot particles. The mass balance equation is written as:

\begin{equation}
	\frac{\mathrm{d}}{\mathrm{d}t}(m) = (1-\varphi)V \sum_i \dot s_i W_i,
	\label{eqn:contconstuv}
\end{equation} 

\noindent where $m$ is the mass of the gas mixture. The rate of change of $m$ is equal to the rate of production of soot mass.
Similarly, the species equation for species $k$ is expressed as:

\begin{equation}
	\frac{\mathrm{d}Y_k}{\mathrm{d}t}
	=
	\frac{1}{\rho}
	\left(
		{\dot{\omega}}_k
		+
		{\dot{s}}_k
	\right)W_k
	-\frac{1}{\rho}Y_k\sum_{i}{{\dot{s}}_i W_i}
	\label{eqn:speciesconstuv}.
\end{equation}

The transport equation for a generic soot variable, $\psi$, can be written as:
\begin{equation}
	\frac{\mathrm{d} \psi}{\mathrm{d} t}= S_{\psi} - \frac{\psi}{\rho} \sum_i \dot{s}_i W_i
	\label{eqn:sootconstuv}.
\end{equation}

The second term on RHS of Equations~\eqref{eqn:speciesconstuv} and~\eqref{eqn:sootconstuv} denotes the change in $Y_k$ and $\psi$, respectively, due to the removal or addition of gas mass by soot-related processes.
The energy conservation for the gas mixture is written in terms of the rate of change of temperature. An external heat source of $\mathrm{\dot{Q}}$ is considered to account for possible heat loss/gain of the reactor.
\begin{equation}
	\begin{split}
		\frac{\mathrm{d} T}{\mathrm{d} t}=
		\frac{1}{\rho c_v+\rho_{soot}f_v c_{soot}}
		\left[
			-\sum_k e_k
				\left(
					\dot{\omega}_k+\dot{s}_k
				\right) W_k
		\right. \\
		\left.
			+e_{soot}\sum_k \dot{s}_k W_k
			+\frac{\dot{Q}}{V(1-\varphi)}
		\right],
	\end{split}
	\label{eqn:energyconstuv}
\end{equation}

\noindent where $\rho_{soot}f_v c_{soot}$ and $e_{soot}\sum_k \dot{s}_k W_k$ represent the formation and sensible energy of soot, respectively. We investigated the effect of considering soot formation and sensible energy on gas and soot properties by simulating the pyrolysis of 30\%~$\mathrm{CH_4}$-Ar with and without considering the above mentioned term. As shown in Figure~\ref{fig:sseeffect}a, neglecting soot sensible energy results in the overprediction of temperature by nearly 150 K and mobility diameter by a factor of 3 during the 80 ms of the simulation. The overprediction of temperature changes gas chemistry, leading to a noticeable decrease in the residual methane and benzene (Figure~\ref{fig:sseeffect}b).

\subsubsection{Constant Pressure Reactor (CPR)}

As shown in Figure~\ref{fig:reactors}b, CPR is a closed system similar to CVR, but the pressure stays constant throughout the process, which means the boundaries of the system can move, changing its volume. The equations of mass, species, and soot variables in CPR are the same as those in CVR, but the energy equation is written in terms of enthalpy, $h$, instead of internal energy $e$, as:

\begin{equation}
	\begin{split}
		\frac{\mathrm{d} T}{\mathrm{d} t}=
		\frac{1}{\rho c_p+\rho_{soot}f_v c_{soot}}
		\left[
		-\sum_k h_k
		\left(
		\dot{\omega}_k+\dot{s}_k
		\right) W_k \right. \\
		\left.
		+h_{soot}\sum_k \dot{s}_k W_k
		+\frac{\dot{Q}}{V(1-\varphi)}
		\right]
		\label{eqn:energypressure}.
	\end{split}
\end{equation}

\subsubsection{Perfectly Stirred Reactor (PSR)}
In this reactor, gas enters with a mass flow rate, composition, and temperature of $\dot{m}_{in}$, $Y_{in}$, and $T_{in}$, respectively, and homogeneously reacts with the mixture inside the reactor. The reacting gas reaches a spatially uniform temperature and composition described by $T$ and $Y$. It is assumed that temperature, composition and soot properties of the outflow are the same as the mixture inside reactor. Without soot formation, the inlet and outlet mass flow rates are equal (i.e. ${\dot{m}_{in}}={\dot{m}_{out}}$), but when soot is formed, ${\dot{m}_{out}}$ is slightly less than ${\dot{m}_{in}}$. The nominal residence time of PSR is calculated as:

\begin{equation}
	\tau_{psr} = \frac{\rho V}{\dot{m}_{in}}
	\label{eqn:taupsr}.
\end{equation} 

The conservation of mass of PSR can be written by considering the mass flux of in- and outflow, and the removal of mass due to soot formation as:

\begin{equation}
	\frac{\mathrm{d} m}{\mathrm{d} t}
	=
	\dot{m}_{in} - \dot{m}_{out} 
	+ V(1 - \varphi)\sum_i \dot{s}_i W_i 
	\label{eqn:contpsr}.
\end{equation}

Gas composition is obtained by solving the species transport equations as:

\begin{equation}
	\frac{\mathrm{d} Y_k}{\mathrm{d} t}
	=
	\frac{{\dot{m}}_{in}}{\rho V
	\left(1-\varphi\right)}
	\left(Y_{k,in}-Y_k \right)+
	\frac{1}{\rho}\left[\left(\dot{\omega}_k+\dot{s}_k\right) W_k-Y_k \sum_i \dot{s}_i W_i\right]
	\label{eqn:speciespsr}.
\end{equation}

The soot transport equations can also be expressed as:
\begin{equation}
	\frac{\mathrm{d}\psi}{\mathrm{d}t}
	=
	\frac{{\dot{m}}_{in}}{\rho V
		\left(1-\varphi\right)}
	\left(\psi_{in}-\psi\right)
	+
	S_{\psi}
	-\frac{1}{\rho}\psi\sum_{i}{{\dot{s}}_i W_i}
	\label{eqn:sootpsr}.
\end{equation}

The energy equation for this reactor is written as:
\begin{equation}
	\begin{split}
		\frac{\mathrm{d}T}{\mathrm{d}t}
		=
		\frac{1}
		{
			\rho c_p+\rho_{soot}c_{p,soot}f_v
		}
		\left[
		\frac{{\dot{m}}_{in}}{V(1 - \varphi)}
		\left(h_{in}-h\right)
		-
		\frac{{\dot{m}}_{in}}{V (1 - \varphi)}\sum_{k}\left(Y_{k,in}-Y_k\right)h_k
		\right.\\
		\left.	
		-
		\sum_{k}{
			\left(
			{\dot{\omega}}_k
			+
			{\dot{s}}_k
			\right) W_k h_k}
		+\sum_{i}{{\dot{s}}_i W_i} h_{soot}+\frac{\dot{Q}}{V(1 - \varphi)}
		\right].
	\end{split}
		\label{eqn:energypsr}
\end{equation}

\subsubsection{Plug Flow Reactor (PFR)}
\label{sec:pfr}
PFR is an ideal representation of a channel or duct where the temperature, composition, and soot properties of a steady-state one-dimensional flow evolve along the reactor. There is no spatial gradient over cross-section due to strong mixing. The diffusion along reactor is negligible.

The continuity equation for PFR is written as:
\begin{equation}
	\frac{\mathrm{d}\dot{m}}{\mathrm{d}z} =(1-\varphi)A \sum_i \dot s_i W_i
	\label{eqn:contpfr}.
\end{equation}

The momentum equation can also be established as:
\begin{equation}
	u (1-f_v) \sum_i \dot s_i W_i + \rho u (1-\varphi) \frac{\mathrm{d}u}{\mathrm{d}z}
	=-\frac{\mathrm{d}}{\mathrm{d}z}(p(1-\varphi))-\frac{\tau_{w}}{R_H} 
	\label{eqn:momenpfr},
\end{equation}
 \noindent where $\tau_w$, and $R_H$ are the wall shear stress and hydraulic radius of the reactor. $\tau_w$ can be determined from the friction factor, $f$, as:
\begin{equation}
	\tau_w = \frac{1}{2}\rho u^2 f, 
	\label{eqn:wallshearpfr}
\end{equation}

\noindent where $f$ can be accurately calculated over the entire range of Reynolds numbers, from laminar to turbulent flow, using the explicit formula provided by \citet{haaland1983simple}:

\begin{equation}
	\frac{1}{f^{1/2}} = -1.8 \mathrm{log}
	\left(
		\frac{6.9}{Re}+
		\left[ \frac{\epsilon/D_H}{3.7} \right]^{1.11}
	\right)
	\label{eqn:fpfr},
\end{equation}
\noindent where $\epsilon$ and ${D_H}$ are the wall roughness and the hydraulic radius of the reactor. The species equation can be expressed as:
\begin{equation}
	\frac{\mathrm{d} Y_k}{\mathrm{d} z}=\frac{1}{\rho u}\left[\left(\dot{\omega}_k+\dot{s}_k\right) W_k-Y_k \sum_i \dot{s}_i W_i\right]
	\label{eqn:speciespfr}.
\end{equation}

 The soot transport equations can also be written as:
\begin{equation}
	\frac{\mathrm{d} \psi}{\mathrm{d} z}=
	\frac{S_{\psi}}{u}
	-\frac{\psi}{\rho u}\sum_i \dot{s}_i W_i
	-\frac{4}{D_H}\frac{k^i_{dep}\psi}{u},
	\label{eqn:sootpfr}
\end{equation}
\noindent where $k^i_{dep}$ is the deposition velocity of soot particles of section $i$, which is calculated as:

\begin{equation}
	k_{dep}=
	\frac{Sh\cdot D^i}{D_H},
	\label{eqn:kdep}
\end{equation}

\noindent where $Sh=3.66$ for laminar flows, and it calculated using the Berger and Hau correlation~\citep{berger1977mass} for the turbulent flow as:

\begin{equation}
	Sh=
	0.0165Re^{0.86} Sc^{1/3}
	\label{eqn:shdep}.
\end{equation}

The energy equation can be expressed as:
\begin{equation}
	\begin{split}
		\frac{\mathrm{d} T}{\mathrm{d} z}=
		\frac{1}{\rho u c_p+\rho_{soot} u f_v 	c_{p,soot}}
		\left[
			-\sum_k h_k
			\left(
			\dot{\omega}_k+\dot{s}_k
			\right) W_k
		\right. \\
		\left.
			+h_{soot}\sum_k \dot{s}_k W_k
			+q^{\prime \prime}\frac{P_c}{A}
		\right],
	\end{split}
	\label{eqn:energypfr}
\end{equation}
\noindent where $q^{\prime \prime}$ is the wall heat flux of the reactor.

\section{Particle dynamics}
\label{sec:particledynamics}
Population balance models rely on the Eulerian description of particles where average properties of particle population such as number density, mass or surface area are treated as continuous quantities and tracked by solving scalar transport equations. Here, we use two particle dynamics models: a monodisperse population balance model (MPBM), which tracks four variables and results in four transport equations, and a fixed sectional population balance model (SPBM), which tracks three variables per section. In the SPBM approach, the total number of transport equations equals the number of tracked variables multiplied by the number of sections. 

Having the total number concentration of agglomerates and primary particles and total carbon content of soot enables the model to describe evolving fractal-like morphology and surface area of soot agglomerates, which is essential to compute their collision frequency~\citep{mulholland1988cluster} as well as oxidation and surface growth rates~\citep{kelesidis2019estimating}. Tracking hydrogen content of soot allows capturing the soot composition, thereby its maturity~\citep{kholghy2016core}, and surface reactivity~\citep{blanquart2009analyzing}.

The common features of the implemented particle dynamics models include the composition (H/C), diffusion coefficient, and morphology of soot particles, all of which are calculated in the same manner for MPBM and for each section in SPBM. Soot morphology is reviewed in Section~\ref{sec:sootmorphology}, and the calculation of the diffusion coefficient is provided in Section~\ref{sec:diffcoef} of the supplementary material. 

However, the particle dynamics models differ in terms of calculating the coagulation partial source term, $I_{coag}$, and processing the contributions from inception, surface growth, oxidation, and coagulation, $I_{\varphi}$, to generate the source terms $S_{\varphi}$ that are incorporated into the transport equations.


\subsection{Soot morphology}
\label{sec:sootmorphology}

\begin{figure}[!htbp]
	\centering
	\includegraphics[height=60mm, ]{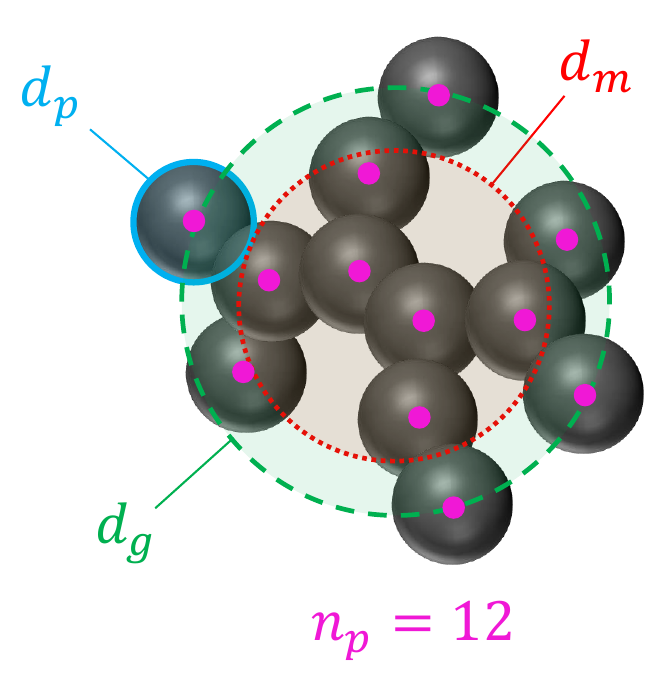}
	\caption{The schematics of a soot agglomerates with 12 primary particles (${n_p=12}$). Primary particle, ${d_p}$, mobility, ${d_m}$, and gyration, ${d_g}$, diameters are shown.}
	\label{fig:Morphology}
\end{figure}

The evolving fractal-like structure of agglomerates is quantified by their mobility diameter normalized by primary particle diameter, $d_m/d_p$, and gyration diameter, $d_m/d_g$, that can be described with power-laws derived from mesoscale simulations.
Incipient soot is initially a sphere formed of PAHs with constant density that grows in size by surface reactions and forms agglomerates by coagulation. The collision frequency of particles depends on their evolving fractal-like structure~\citep{mulholland1988cluster}.
Mobility and gyration diameters are calculated using power-laws developed to describe the morphology of soot from premixed~\citep{abid2008evolution}, diffusion~\citep{yon2015simple} flames, and diesel engines~\citep{rissler2013effective}. Figure~\ref{fig:Morphology} shows a schematic of a soot agglomerate composed of 12 primary particles, with ${d_p}$, ${d_m}$, and ${d_g}$ annotated.

${n^i_p}$ is the number of primary particles per agglomerate of ${i^{th}}$ section that can be obtained by dividing the number concentration of primary particles in ${i^{th}}$ section by that of agglomerates in that section as:

\begin{equation}
	n^i_p = \frac{N^i_{pri}}{N^i_{agg}}
	\label{eqn:n_p}.
\end{equation}

Primary particle diameter, ${d^i_p}$, is determined from total carbon content and number density of primary particles using:

\begin{equation}
	d^i_p = \left(\frac{6}{\pi} \frac{C^i_{tot}\cdot W_{carbon}}{\rho_{soot}} \frac{1}{N^i_{pri}\cdot Av} \right)^{1/3}.
	\label{eqn:d_p}
\end{equation}

The DEM-derived power-laws~\citep{Kelesidis2017} relate ${d^i_m}$ and ${d^i_g}$ to ${d^i_p}$ and ${n^i_p}$ as:

\begin{equation}
	d^i_{m} = d^i_p\cdot {n^i_p}^{0.45}
	\label{eqn:d_m},
\end{equation}

\begin{equation}
	d^i_g = 
	\left\{
	\begin{array}{lr}
		d^i_m/({n^i_p}^{-0.2}+0.4), & \text{if } n^i_p > 1.5\\
		d^i_m/1.29. & \text{if } n^i_p\leq 1.5
	\end{array}
	\right.
	\label{eqn:d_g}
\end{equation}

The collision diameter, ${d^i_c}$, is the maximum of ${d^i_{m}}$ and ${d^i_{g}}$:

\begin{equation}
	d^i_c = \mathrm{max}\left(d^i_m, d^i_g\right).
	\label{eqn:d_c}
\end{equation}

$A^i_{tot}$ (for each section) is defined as the total surface area of soot particles per unit mass of gas mixture obtained as:
\begin{equation}
	A^i_{tot} = N^i_{pri}\cdot Av\cdot \pi {d^i_p}^2
	\label{eqn:Atot}.
\end{equation}

Mass of each agglomerate, $m^i_{agg}$, is expressed as:
\begin{equation}
	m^i_{agg} = \frac{C^i_{tot}\cdot W_{carbon}}{N^i_{agg}\cdot Av}.
	\label{eqn:m_agg}
\end{equation}

\subsection{Sectional Population Balance Model (SPBM)}
\begin{figure}[!htbp]
	\centering
	\includegraphics[height=40mm, ]{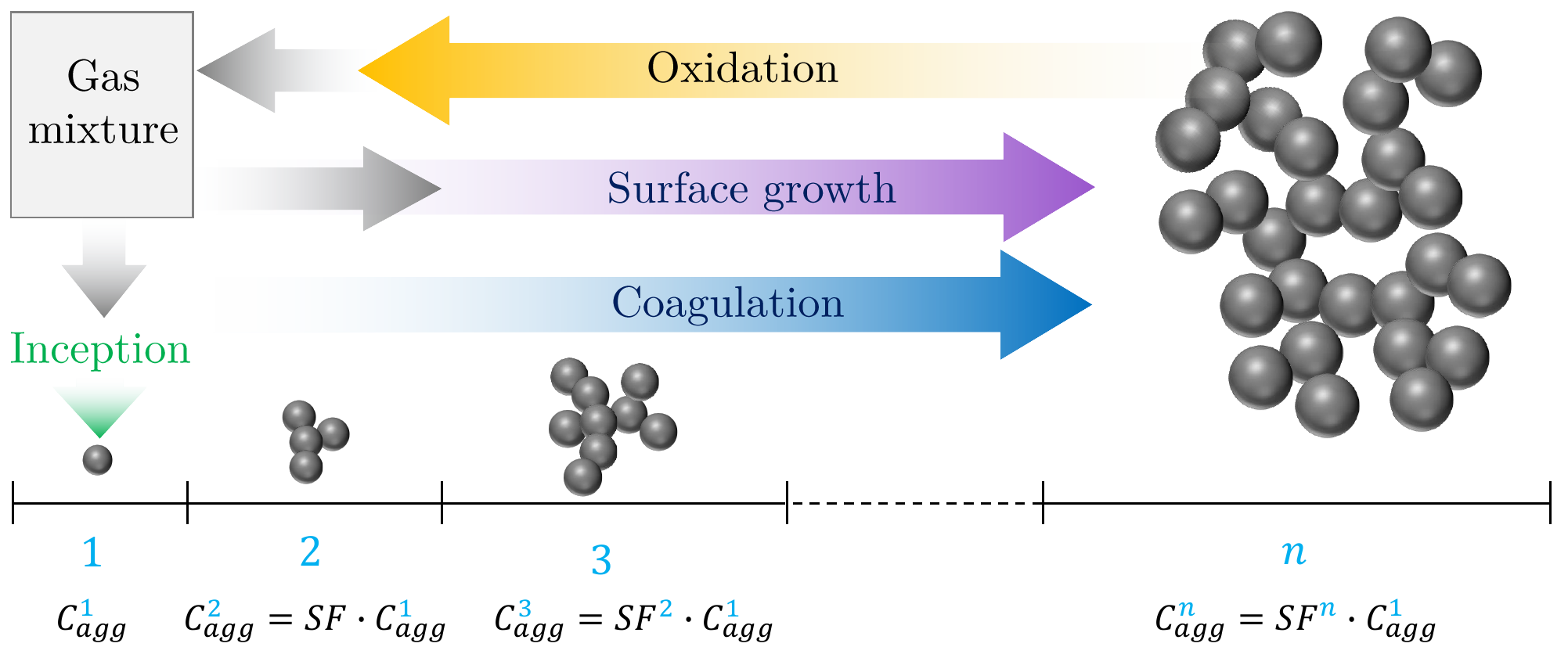}
	\caption{The illustration of mass sections in SPBM that grow progressively by the scale factor of SF. Inception introduces new particles to the first section that propagate to the upper section via coagulation and surface growth and return to lower sections by oxidation. Carbon and hydrogen pass from gas to solid phase through inception and surface growth and goes back via oxidation.}
	\label{fig:sectional}
\end{figure}

A SPBM with fixed pivots is used to describe particle dynamics~\citep{wu1988discrete}. The mass range of particles is divided into discrete sections each of which includes agglomerates of the same mass. Figure~\ref{fig:sectional} illustrates the interaction between the gas phase and mass sections, as well as the mechanisms by which particles move between sections. Inception introduces new particles to the first section with the mass corresponding to the incipient particle. The particles of each section can migrate to upper sections by gaining mass via surface growth and coagulation, and return to lower sections when they lose mass through oxidation without breaking into smaller particles (i.e., fragmentation is not considered). Particles attach at point contact without coalescence. The mass of sections is determined by a geometric progression that has an initial value equal to the mass of incipient soot particle, and a common ratio of SF, known as sectional spacing factor. The mass of each section is approximated by the carbon content of agglomerates in moles as:
\begin{equation}
	C^i_{agg} = \frac{n_{c,min}}{Av}\cdot SF^{(i-1)},
	\label{eqn:Caggsec}
\end{equation}
\noindent where $(i-1)$ represents the exponent of SF. The mass of hydrogen is ignored in the placement of agglomerates in the sections.
The total number density of agglomerates, $N^i_{agg}$, and primary particles, ${N^i_{pri}}$, and the hydrogen content, $H_{tot}$, are tracked for each section. The mass of each section is fixed, so $C_{tot}$, for each section can be easily calculated by knowing the number of agglomerates; that is, $C_{tot} = N^i_{agg} \cdot {Av} \cdot C^i_{agg}$. Morphological parameters for each section are determined according to the equations provided in Section~\ref{sec:sootmorphology}.

New particles formed by coagulation are assigned to an upper section, with a total mass equal to the sum of the masses of the colliding particles. If the mass of resulting particle falls between two consecutive sections, the mass is distributed between them proportionally. In some cases, the mass of the newly formed particle may exceed the upper bound of the last section, causing it to fall outside the tracked mass range. This leads to potential mass loss, which is a known limitation of the fixed pivot sectional model~\citep{zhang2010detailed}. However, this issue can be mitigated by choosing an appropriate number of sections and a suitable spacing factor to ensure the uppermost sections remain unoccupied during the simulation. In the SPBM, the source terms of tracked soot variables (which appear in Equations~\ref{eqn:sootconstuv},~\ref{eqn:sootpsr},~and~\ref{eqn:sootpfr}) are split into four parts showing the contribution of inception, surface growth, oxidation and coagulation. The effect of HACA and PAH adsorption are combined (denoted by the subscript \textit{haca,ads}) because they are similar mass-gaining mechanisms. The source terms for section $i$ are given as:

\begin{equation}
	S^i_{N_{agg}} = 
	\left(S^i_{N_{agg}}\right)_{inc}
	+\left(S^i_{N_{agg}}\right)_{haca, ads}
	+\left(S^i_{N_{agg}}\right)_{ox}
	+\left(S^i_{N_{agg}}\right)_{coag}
	\label{eqn:S_Naggsect},
\end{equation}

\begin{equation}
	S^i_{N_{pri}} = 
	\left(S^i_{N_{pri}}\right)_{inc}
	+\left(S^i_{N_{pri}}\right)_{haca, ads}
	+\left(S^i_{N_{pri}}\right)_{ox}
	+\left(S^i_{N_{pri}}\right)_{coag}
	\label{eqn:S_Nprisect},
\end{equation}

\begin{equation}
	S^i_{H_{tot}} = 
	\left(S^i_{H_{tot}}\right)_{inc}
	+\left(S^i_{H_{tot}}\right)_{haca, ads}
	+\left(S^i_{H_{tot}}\right)_{ox}
	+\left(S^i_{H_{tot}}\right)_{coag}
	\label{eqn:S_Htotsect}.
\end{equation}

The calculation of coagulation source terms is based on the implementation of SPBM by \citet{veshkini2015understanding}, and it is presented in detail along with the other source terms in Section~\ref{sec:sectextra} of the supplementary material.

\subsection{Monodisperse Population Balance Model (MPBM)}
\label{sec:mpbm}
The MPBM used in this study is based on the monodisperse model presented in \citep{kholghy2021surface}, and it tracks $N_{pri}$, $N_{agg}$, $C_{tot}$ and $H_{tot}$. The coagulation model follows the same principles as the SPBM, but the calculations are greatly simplified due to the monodispersity assumption. This approach has been shown to achieve accuracy comparable to that of DEM and SPBM when applied with well-justified assumptions~\citep{Kelesidis2017Flame, kelesidis2019estimating}. The rate of decay of agglomerates is simply described as:

\begin{equation}
	I_{coag} = -\frac{1}{2}\zeta\beta N^2_{agg}
	\label{eqn:Icoag},
\end{equation}
where $\zeta$ is the collision efficiency of agglomerates calculated using Equation~\eqref{eqn:coageff}, and ${\beta}$ is the collision frequency of agglomerates for the free-molecular ($\mathrm{Kn>10}$) to continuum regimes (${Kn<0.1}$). The value of ${\beta}$ in the transition regime (${0.1<Kn<10}$) can be calculated from the harmonic mean of the continuum (${\beta_{cont}}$) and free-molecular (${\beta_{fm}}$) regime values. Additionally, an enhancement factor of \%82 is applied to take into account the effect of polydispersity~\citep{kelesidis2021self} as:
\begin{equation}
	\beta = 1.82\frac{\beta_{fm}\cdot\beta_{cont}}{\beta_{fm}+\beta_{cont}}
	\label{eqn:betahmmono},
\end{equation}
\begin{equation}
	\beta_{fm} = 4\sqrt{\frac{\pi k_B T}{m_{agg}}} d^2_c
	\label{eqn:betafmmono},
\end{equation}
\begin{equation}
	\beta_{cont} = 8\pi d_m D,
	\label{eqn:betacontmono}
\end{equation}
\noindent where $d_c$, $D$, and $m_{agg}$ are calculated using Equations~\eqref{eqn:d_c},~\eqref{eqn:diff},~and~\eqref{eqn:m_agg}, respectively.
Alternatively, $\mathrm{\beta}$ can be obtained using Fuchs interpolation~\citep{fuchs1965mechanics} as:

\begin{equation}
	\beta = \beta_{cont}
	\left(
		\frac{d_c}{d_c+2\sqrt{2}\delta} +
		\frac{8D}{\sqrt{2}c d_c}
	\right)^{-1},
	\label{eqn:betafuchsmono}
\end{equation}
\noindent where $c$ (mean velocity of particles) and $\delta$ (mean stop distance of particles) are calculated using Equations~\eqref{eqn:meanvel}~and~\eqref{eqn:meandist}. The source terms of tracked variables combines the effect of the inception, surface growth, oxidation and coagulation.

\begin{equation}
	S_{N_{agg}} = \frac{I_{N,inc}}{n_{c,min}}+I_{coag},
	\label{eqn:S_N_agg}
\end{equation}
\begin{equation}
	S_{N_{pri}} = \frac{I_{N,inc}}{n_{c,min}},
	\label{eqn:S_N_pri}
\end{equation}
\begin{equation}
	S_{C_{tot}} = I_{C_{tot},inc}+I_{C_{tot},haca}+I_{C_{tot},ads} - I_{C_{tot},ox},
	\label{eqn:S_C_tot}
\end{equation}
\begin{equation}
	S_{H_{tot}} = I_{H_{tot},inc}+I_{H_{tot},haca}+I_{H_{tot},ads} - I_{H_{tot},ox}.
	\label{eqn:S_H_tot}
\end{equation}

\section{PAH growth models}
\label{sec:pahgrowmodel}
Four different PAH growth models are implemented in Omnisoot to describe the conversion of PAHs into incipient particles and their adsorption onto existing agglomerates. In other words, these models explain the calculation of $I_{\varphi, inc}$ and $I^i_{\varphi, ads}$. The implemented models are: Irreversible Dimerization~\cite{frenklach1991detailed}, Reactive Dimerization~\citep{kholghy2018reactive}, Dimer Coalescence~\citep{blanquart2009joint}, and E-Bridge Modified~\citep{frenklach2020mechanism}. All models are based on PAH collision, as supported by ample evidence in the literature~\citep{zhao2003measurement, abid2009quantitative, happold2009soot}, but they differ in terms of reversibility, temperature dependence, and the number of steps involved. The formulation and implementation details of the first three models are provided in Sections~\ref{sec:irrevdim}-\ref{sec:dimcoal} of the supplementary material, while only E-Bridge Modified model is discussed here for brevity.

The collision frequency of gaseous species, including PAH molecules and polymers, depends on their mass and diameter and is calculated as:

\begin{equation}
	\beta_{dim_{jk}}=
	2.2 \cdot d^2_{r} \sqrt{\frac{8 \pi k_B T}{m_{r}}},
	\label{eqn:betadim}
\end{equation}

\noindent where 2.2 is the vdW enhancement factor~\citep{kholghy2018reactive}. ${d_{r}}$ and ${m_{r}}$ are reduced diameter and mass for two PAH molecules/dimers, respectively, calculated as:

\begin{equation}
	d_{r}=
	\frac{d_{{PAH}_k}\cdot d_{{PAH}_j}}{d_{{PAH}_k}+d_{{PAH}_j}},
	\label{eqn:drPAH}
\end{equation}

\begin{equation}
	m_{r}=
		\frac{m_{{PAH}_k}\cdot m_{{PAH}_j}}{m_{{PAH}_k}+ m_{{PAH}_j}},
	\label{eqn:mrPAH}
\end{equation}

\noindent where $m_{PAH_j}$ and $d_{PAH_j}$ are the mass and equivalent diameter of the ${j}^{th}$ PAH molecule, respectively, and are obtained as:

\begin{equation}
	m_{PAH_j}=
	\frac{W_{{PAH}_j}}{Av},
	\label{eqn:mPAH}
\end{equation}

\begin{equation}
	d_{PAH_j}=
	\left(
		\frac{6\cdot m_{{PAH}_j}}{\pi\cdot\rho_{{PAH}_j}}
	\right)^{1/3},
	\label{eqn:dPAH}
\end{equation}

\noindent where $\rho_{PAH_j}$ is the equivalent density of PAH estimated using the relation proposed by \citet{johansson2016formation} as:

\begin{equation}
	\rho_{PAH_j}= 
	171943.5197
	\frac{W_{carbon}\cdot n_{C,{PAH}_j}+W_{hydrogen}\cdot n_{H,{PAH}_j}}
	{n_{C,{PAH}_j}+n_{H,{PAH}_j}},
	\label{eqn:rhoPAH}
\end{equation}

\noindent where ${n_{C,{PAH}_j}}$ and ${n_{H,{PAH}_j}}$ denote the number of carbon and hydrogen atoms in the $j^{th}$ PAH molecule, respectively. The collision frequency of $\mathrm{PAH}_j$ and soot agglomerates in each section can be determined for the entire regime by harmonic mean of the collision frequencies in the free-molecular and continuum regimes as:

\begin{equation}
	\beta^i_{ads_j}=
	\frac{\beta^i_{fm, ads}\cdot \beta^i_{cont, ads}}
	{\beta^i_{fm, ads}+\beta^i_{cont, ads}},
	\label{eqn:betahmads}
\end{equation}

\begin{equation}
	\beta^i_{fm, ads_j}=
	2.2 
	\sqrt{
		\frac{\pi k_B T}{2}\left(\frac{1}{m^i_{agg}}+\frac{1}{m_{PAH_j}}\right)
	}
	\left(d^i_g+d_{PAH}\right)^2,
	\label{eqn:betafmads}
\end{equation}

\begin{equation}
	\beta^i_{cont, ads_j}=
		\frac{2 k_B T}{3 \mu}
		\left[
			\frac{C^i\left(d_m\right)}{d^i_g}+
			\frac{C^i\left(d_{PAH_j}\right)}{d_{PAH_j}}
		\right]
		\left(d_g+d_{PAH_j}\right),
	\label{eqn:betacontads}
\end{equation}
where $C^i$ is the Cunningham correction factor calculated using Equation~\eqref{eqn:cun}.

\subsection{E-Bridge Modified}
\label{sec:ebrimod}
The E-Bridge model was originally proposed by \citet{frenklach2020mechanism} to describe soot inception using a HACA-like scheme that starts with the dehydrogenation of PAH monomers, often pyrene, which forms the monomer radicals and continues with the sequential addition of the radicals that form dimers, trimers and larger polymers until the PAH structure reaches a mass threshold and the clustering process becomes irreversible~\citep{frenklach2020mechanism}. Here, a modified version of this model, called E-Bridge Modified, is used where dimers are considered as incipient soot, and monomer radicals are adsorbed on soot agglomerates. This PAH growth model is described using the following set of pathways:

\reaction[reac:dehyd_ebri]{
	PAH_j + H <-->[$k_{f,d_{j}}$][$k_{r,d_{j}}$] $\dot{\mathrm{PAH}}_j$ + H
}

\reaction[reac:hyd_ebri]{
	$\mathrm{\dot{PAH}}_j$ + H ->[$k_{f,h_{j}}$] PAH_$j$ 
}

\reaction[reac:ebri]{
	$\mathrm{\dot{PAH}}_j$ + $\mathrm{\dot{PAH}}_j$ ->[$k_{inc_j}$] Dimer_{$j$}
}
\reaction[reac:ads_ebri]{
	$\mathrm{\dot{PAH}}_j$ + Soot ->[$k_{ads_j}$] Soot-PAH_{$j$}
}

The rate constants of Reactions~\eqref{reac:dehyd_ebri}~and~\eqref{reac:hyd_ebri} are listed in Table~\ref{tab:Ebridge}. $k_{inc_j}$ and $k_{ads_j}$ are the rate constants of dimer production and PAH adsorption, respectively, which are calculated as:

\begin{equation}
	k_{inc_j}=
	\beta_{jj,PAH}\cdot Av
	\label{eqn:kdim_ebri},
\end{equation}

\begin{equation}
	k^i_{ads_{j}}=
	\beta^i_{ads_j}\cdot Av
	\label{eqn:kads_ebir},
\end{equation}

\noindent where $\beta_{jj,PAH}$ and $\beta^i_{ads_j}$ are obtained using Equations~\eqref{eqn:betadim}~and~\eqref{eqn:betahmads}, respectively. The rate of dimer formation and adsorption is calculated as:

\begin{equation}
	\omega_{dim_j} = k_{inc_{j}} [\cdot{\mathrm{PAH}}_j]^2,
	\label{eqn:wdim_ebri}
\end{equation}

\begin{equation}
	\omega^i_{ads_j} = k^i_{ads_{j}} [\mathrm{soot}^i] [\mathrm{\dot{PAH}}_j].
\end{equation}

\renewcommand{\arraystretch}{1.5}
\begin{table}
	\caption{Rate coefficients for the monomer de-/hydrogenation reaction of E-bridge formation in Arrhenius form $\mathrm{k=AT^n\cdot e^{-E/RT}}$~\citep{frenklach2020mechanism}.}
	\label{tab:Ebridge}
	\centering
	\begin{tabular}{l l l l l}
		\hline
		Reaction & \hspace{0.1cm} & A~$\mathrm{\left[{m^3}/{mol\cdot s} \right]}$ & n & {E}/{R}~[K]  \\
		\hline
		\eqref{reac:dehyd_ebri} & f & $98\times$ $\mathrm{n_{C, PAH_j}}$ & 1.8 & 7563.519 \\
		  & r & $1.6\times 10^{-2}$ & 2.63 & 2145.346\\
		\eqref{reac:hyd_ebri} & f & $4.8658\times10^7
		$ & 0.13 & 0.0\\
		\hline
	\end{tabular}
\end{table}

The calculation of rate of inception and PAH adsorption from $\mathrm{PAH}_j$ requires the concentration of $\mathrm{\dot{PAH}}_j$, which can be determined by applying the steady-state assumption for $\mathrm{\dot{PAH}_j}$, i.e., $d[\mathrm{\dot{PAH}}_j]/dt=0$, resulting in a quadratic equation as:




\begin{equation}
	a_{inc_j}[\mathrm{\dot{PAH}}_j]^2+
	b_{ads_j}[\mathrm{\dot{PAH}}_j] + c_j = 0,
\end{equation}
\begin{equation}
	a_{inc_j}=k_{inc_j}
\end{equation}
\begin{equation}
	b_{ads_j}=k_{r,d_j}[\mathrm{H_2}]+k_{f,h_j}[\mathrm{H}]+\sum_{i=1}^{n_{sec}}k^i_{ads_j}[\mathrm{Soot}]^i
\end{equation}
\begin{equation}
	c_{inc_j}=k_{f,d_j}[\mathrm{PAH}_j][\mathrm{H}]
\end{equation}

Solving the quadratic equation for each PAH gives the concentration of $\mathrm{\mathrm{\dot{PAH}}}_j$ as:
\begin{equation}
	[\mathrm{\mathrm{\dot{PAH}}}_j]=
	\left\{
	\begin{aligned}
		&\frac{-b_{ads_j}+\sqrt{\Delta_j}}{2a_{inc_j}},
		&&
		\text{if } \Delta_j \ge 0
		\\
		& 0 
		&&
		\text{if } \Delta_j < 0
	\end{aligned}
	\right.
	\label{eqn:rad_ebri}
\end{equation}
\begin{equation}
	\Delta_j = b_{ads_j}^2-4a_{inc_j}c_{j}
	\label{eqn:delta_ebri}
\end{equation}

The partial source terms of inception, $I_{\varphi, inc}$, are calculated in the E-Bridge Modified model as: 

\begin{equation}
	I_{N,{inc}} = \frac{1}{\rho}
	\sum_{j=1}^{n_{PAH}}
	2\omega_{inc_{j}}
	n_{PAH_j,C}
	\label{eqn:IN_inc_ebri},
\end{equation}

\begin{equation}
	I_{C_{tot},{inc}} = \frac{1}{\rho}
	\sum_{j=1}^{n_{PAH}}
	2\omega_{inc_{j}} 
	n_{PAH_j,C}
	\label{eqn:ICtot_inc_ebri},
\end{equation}

\begin{equation}
	I_{H_{tot},{inc}} = \frac{1}{\rho}
	\sum_{j=1}^{n_{PAH}}
	\omega_{inc_{j}} 
	\left(
	n_{PAH_j,H}-2
	\right)
	\label{eqn:IHtot_inc_ebri},
\end{equation}

The partial source terms of PAH adsorption, $I^i_{\varphi, ads}$, are calculated for each section in this PAH growth model as: 

\begin{equation}
	I^i_{C_{tot},ads} =
	\frac{1}{\rho}
	\sum_{i=1}^{n_{PAH}}
	\omega^i_{ads_j}
	n_{C,PAH_j}
	\label{eqn:ICtotads_ebri},
\end{equation}

\begin{equation}
	I^i_{H_{tot},ads} =
	\frac{1}{\rho}
	\sum_{i=1}^{n_{PAH}}
	\omega^i_{ads_j}
	\left(n_{H,PAH_j}-2\right)
	\label{eqn:IHtotads_ebri}.
\end{equation}

The rate of removal of each PAH involved in soot inception and PAH adsorption, and release of $\mathrm{H_2}$ to the gas mixture can be expressed as:

\begin{equation}
	\left(
	\frac{\mathrm{d}\left[{\mathrm{PAH}_j}\right]}{\mathrm{d}t}
	\right)_{inc}
	= 
	-2\sum_{k=1}^{n_{sec}}\omega_{inc_{j}},
\end{equation}

\begin{equation}
	\left(
	\frac{\mathrm{d}\left[{\mathrm{PAH}_j}\right]}{\mathrm{d}t}
	\right)_{ads}
	= 
	-\sum_{i=1}^{n_{sec}}\omega^i_{ads_j},
	\label{eqn:PAHscrub_ebri_ads}
\end{equation}

\begin{equation}
	\left(
	\frac{\mathrm{d}\left[{\mathrm{H_2}}\right]}{\mathrm{d}t}
	\right)_{inc}
	= 
	\sum_{i=1}^{n_{sec}}\omega^i_{ads_j}.
	\label{eqn:H2scrub_ebri}
\end{equation}

\section{Surface reactions model}
\label{sec:surfreacmodel}
Heterogeneous surface reactions are described by HACA. The soot growth in HACA scheme is based on a sequential process similar to PAH growth. The hydrogenated armchair sites ($\mathrm{C_{soot}-H}$) on the edge of aromatic rings are dehydrogenated by H abstraction forming $\mathrm{C_{soot\mbox{\textdegree}}}$ that bonds with $\mathrm{C_2H_2}$ resulting in an additional aromatic ring with hydrogenated sites. These sites can also be attacked by $\mathrm{O_2}$ or $\mathrm{OH}$ leading to removal of carbon from soot particles by oxidation. The elementary reactions that describe this sequential process are listed in Table~\ref{tab:HACA}.
The rate of mass growth by HACA is obtained from the reaction of $\mathrm{C_2H_2}$ with dehydrogenated sites as:

\begin{equation}
	\omega^i_{gr} = \alpha^i k_{f4} [\mathrm{C_2H_2}][\mathrm{C}^i_\mathrm{soot\mbox{\textdegree}}]
	\label{eqn:hacaRate},
\end{equation}

\noindent  where ${k_{f4}}$ denotes the forward rate of Reaction~\eqref{reac:haca4} in Table~\ref{tab:HACA}, and $\mathrm{[C^i_{soot\mbox{\textdegree}}]}$ is obtained by multiplying the surface density of dehydrogenated sites, $\mathrm{C}^i_\mathrm{soot\mbox{\textdegree}}$ by the total surface area of soot (per unit of mass of gas mixture) of each section as:

\begin{equation}
	[\mathrm{C}^i_\mathrm{soot\mbox{\textdegree}}] = \frac{\rho}{Av}A^i_{tot}\cdot\chi_{soot\mbox{\textdegree}}
	\label{eqn:csoot0},
\end{equation}

\noindent where $\mathrm{\chi_{soot\mbox{\textdegree}}}$ is the surface density of dehydrogenated sites, and is calculated by assuming the steady-state for $\mathrm{[C_{soot\mbox{\textdegree}}]}$ in the system of reactions in Table~\ref{tab:HACA} as:
\begin{equation}
	\chi_{soot{\mbox{\textdegree}}} = 
	\frac
	{k_{f1}[\mathrm{H}]+k_{f2}[\mathrm{OH}]}
	{k_{r1}[\mathrm{H_2}]+k_{r2}[\mathrm{H_2O}]+k_{f3}[\mathrm{H}]+k_{f4}[\mathrm{C_2H_2}]+k_{f5}[\mathrm{O_2}]} \chi_{soot-{H}},
	\label{eqn:chisoot0}
\end{equation}
\noindent where ${\chi_{soot-{H}}}$ is the surface density of hydrogenated sites estimated based on the assumption that soot ``surface is assumed to be composed of outwardlooking PAH edges with PAH molecular moieties assembled into turbostratic structures"~\citep{frenklach2019new}. Considering the layer spacing of 3.15$\mathrm{\AA}$ and two C–H bonds per benzene ring length, $\chi_{{soot}-H}$ is calculated to be $0.23\:\mathrm{site/\AA^2}=2.3\times10^{19}$ $ \mathrm{site/m^2}$, which gives the maximum theoretical limit of the reaction sites.

In Equation~\eqref{eqn:hacaRate}, $\alpha^i$ is the surface reactivity factor between 0 and 1 that represents the decline of reaction sites from the theoretical limit due to PAH layer orientation, particle aging, growth and maturity~\citep{haynes1982surface, harris1985chemical}, and it has been observed to depend on temperature-time history~\cite{homann1985formation, dasch1985decay}. The value of $\alpha$ has been described using constant target-specific values as well as empirical equations based on particle size and flame temperature. A detailed review of these can be found in Chapter~4 of \citep{veshkini2015understanding}.  Here, the empirical equation proposed by \citet{appel2000kinetic} is used to calculate $\alpha^i$ as:
\begin{equation}
	\alpha^i = \tanh 
	\left(
	\frac{12.56 - 0.00563 T}
	{\mbox{log}_{10}
		\left( \frac{\rho_{soot}\cdot Av}{W_{carbon}} \frac{\pi}{6}{d^i_p}^3 \right) } -1.38+0.00068T
	\right)
	\label{eqn:alpha}.
\end{equation}

Alternatively, $\alpha^i$ can be related to the H/C ratio of soot particles by assuming that all hydrogen atoms reside on the particle surface~\citep{blanquart2009joint} as:

\begin{equation}
	\alpha^i = \frac{H^i_{tot}}{C^i_{tot}}
	\label{eqn:alpha_htoc}.
\end{equation}

The contribution of HACA to growth source terms can be computed from HACA rates considering the number of carbon atoms in $\mathrm{C_2H_2}$ and number of arm-chair and zig-zag hydrogenated sites on soot particle~\cite{blanquart2009analyzing} using

\begin{equation}
	I^i_{C_{tot},haca} = 2\omega^i_{gr}/\rho
	\label{eqn:IiCtotgr},
\end{equation}
\begin{equation}
	I^i_{H_{tot},haca} = 0.25\omega^i_{gr}/\rho
	\label{eqn:IiHtotgr}.
\end{equation}

The rates of concentration change of $\mathrm{C_2H_2}$ and H radical due to HACA are written as:

\begin{equation}
	\left(\frac{d\left[{\mathrm{C_2H_2}}\right]}{dt}\right)_{gr} = -\sum_{i=1}^{n_{sec}}\omega^i_{gr},
	\label{eqn:C2H2rate_gr}
\end{equation}

\begin{equation}
	\left(\frac{d\left[{\mathrm{H}}\right]}{dt}\right)_{gr} = 1.75 \sum_{i=1}^{n_{sec}}\omega^i_{gr}.
	\label{eqn:Hrate_gr}
\end{equation}

\renewcommand{\arraystretch}{1.5}
\begin{table}
	\caption{Arrhenius rate coefficients of the surface reactions in HACA~\citep{appel2000kinetic}, ${k=AT^n\cdot e^{-E/RT}}$.}
	\label{tab:HACA}
	\centering
	\begin{tabular}{l l l l l l}
		\hline
		No. & Reaction & \hspace{0.1cm} & A~$\mathrm{\left[ {m^3}/{mol\cdot s} \right]}$ & n & {E}/{R}~[K]  \\
		\hline
		\refstepcounter{reaction}\label{reac:haca1}\thetag{\thereaction} & \ce{C_{soot-H} + H <-->[$k_{f,1}$][$k_{r,1}$] C_{soot\textdegree} + H_2}  & f & $4.17\times 10^7$ & 0 & 6542.52 \\
		& & r & $3.9\times 10^6$ & 0 & 5535.98 \\
		{\refstepcounter{reaction}\label{reac:haca2}\thetag{\thereaction}} & \ce{C_{soot-H} + OH <-->[$k_{f,2}$][$k_{r,2}$] C_{soot\textdegree} + H_2O} & f & $10^4$ & 0.734 & 719.68\\
		&  & r & 3.68$\times 10^2$ & 1.139 & 8605.94 \\
		\refstepcounter{reaction}\label{reac:haca3}\thetag{\thereaction} & \ce{C_{soot\textdegree} + H ->[$k_{f,3}$] C_{soot-H}} & f & $10^4$ & 0.734 & 719.68\\
		{\refstepcounter{reaction}\label{reac:haca4}\thetag{\thereaction}} & \ce{C_{soot\textdegree} + C_2H_2 ->[$k_{f,4}$] C_{soot-H} + H} & f & 80 & 1.56 & 1912.43\\
		\refstepcounter{reaction}\label{reac:haca5}\thetag{\thereaction} & \ce{C_{soot\textdegree} + O_2 ->[$k_{f,5}$] 2CO + product} & f & 2.2 $\times 10^6$ & 0 & 3774.53\\
		\refstepcounter{reaction}\label{reac:haca6}\thetag{\thereaction} & \ce{C_{soot}-H + OH ->[$k_{f,6}$] CO + product} & f & \multicolumn{3}{c}{$\gamma_{OH}$ = 0.13} \\
		\hline
	\end{tabular}
\end{table}

The carbons on the surface of soot are oxidized via reaction with $\mathrm{O_2}$ (Reaction~\eqref{reac:haca5}) and $\mathrm{OH}$ (Reaction~\eqref{reac:haca6}). As a result, oxidation decreases the total carbon of soot and releases CO and $\mathrm{H_2}$ to the gas mixture. The $\mathrm{O_2}$ and $\mathrm{OH}$ oxidation rates are calculated as:

\begin{equation}
	\omega^i_{ox,O_2} = \alpha^i k_{f5} [\mathrm{O_2}][C^i_{soot\mbox{\textdegree}}],
	\label{eqn:hacaO2Rate}
\end{equation}

\begin{equation}
	\omega^i_{ox,OH} = \gamma_{OH} \beta^i_{OH} Av [\mathrm{OH}][\mathrm{soot}^i],
	\label{eqn:hacaOHRate}
\end{equation}

\noindent where $\gamma_{OH}=0.13$ is the reaction probability of OH radicals with soot particles~\citep{appel2000kinetic}. $\beta_{OH}$ is the collision frequency of OH and soot particles calculated based on the kinetic theory of gases as:

\begin{equation}
	\beta^i_{OH} = 
	\sqrt{
		\frac{\pi k_B T}{2}\left(\frac{1}{m^i_{agg}}+\frac{1}{m_{OH}}\right)
	}
	\left(d^i_c+d_{OH}\right)^2,
	\label{eqn:betaOH}
\end{equation}
\noindent where $m_{OH}=2.824\times10^{-26}$ kg, and $d_{OH}=0.3$ nm~\citep{shepherd2022measurement} are the mass and equivalent diameter of a OH radical, respectively. Then, the oxidation source term is calculated considering the number of carbon atoms removed from soot through each oxidation pathway as:

\begin{equation}
	I^i_{C_{tot},ox} = (2\omega^i_{ox,O_2} + \omega^i_{ox,OH})/\rho
	\label{eqn:ICtot}.
\end{equation}

The rate of change of concentration of CO, $\mathrm{O_2}$ and OH by oxidation is calculates as:

\begin{equation}
	\left(\frac{d\left[{\mathrm{CO}}\right]}{dt}\right)_{ox} = 2\sum_{i=1}^{n_{sec}}\omega^i_{ox,O_2},
	\label{eqn:COrate_ox}
\end{equation}

\begin{equation}
	\left(\frac{d\left[{\mathrm{O_2}}\right]}{dt}\right)_{ox} = -\sum_{i=1}^{n_{sec}}\omega^i_{ox,O_2},
	\label{eqn:O2rate_ox}
\end{equation}

\begin{equation}
	\left(\frac{d\left[{\mathrm{OH}}\right]}{dt}\right)_{ox} = -\sum_{i=1}^{n_{sec}}\omega^i_{ox,OH},
	\label{eqn:Hrate_ox}
\end{equation}

\begin{equation}
	\left(\frac{d\left[{\mathrm{H}}\right]}{dt}\right)_{ox} = \sum_{i=1}^{n_{sec}}\omega^i_{ox,OH}.
	\label{eqn:OHrate_ox}
\end{equation}

\section{Code Validation}
A set of simulations was performed to evaluate the accuracy and reliability of Omnisoot in predicting soot formation. The aerosol dynamics models were validated by comparing the results of the population balance models implemented in Omnisoot with DEM simulations reported in the literature. In addition, carbon and hydrogen mass and energy balances were rigorously assessed to ensure that residuals remain within the bounds of acceptable numerical error.

\subsection{Coagulation}
The collision kernels and their corresponding interpolation schemes implemented in Omnisoot were validated by comparing the computed collision frequencies across a wide range of Knudsen numbers with reference DEM results~\citep{goudeli2015coagulation} shown in Figure~\ref{fig:kernelvalid}.
Additionally, two test cases were set up to validate the coagulation sub-model of MPBM and SPBM in the free-molecular\footnote{\href{https://github.com/mohammadadib-cu/omnisoot-cv/tree/main/validations/coagulation/free_molecular}{https://github.com/mohammadadib-cu/omnisoot-cv/tree/main/validations/coagulation/free\_molecular}} and continuum \footnote{\href{https://github.com/mohammadadib-cu/omnisoot-cv/tree/main/validations/coagulation/continuum}{https://github.com/mohammadadib-cu/omnisoot-cv/tree/main/validations/coagulation/continuum}} regimes. An adiabatic CVR with a volume of 1~$\mathrm{m}^3$ was populated with $2.6261 \times 10^{18}$ spherical particles with the initial diameters of 2~nm and 668~nm for the free-molecular and continuum cases, respectively. The initial temperature in the free-molecular and continuum simulations are 1800~K and 300~K, respectively.
 
\begin{figure}[H]
	\centering
	\begin{subfigure}[t]{0.4\textwidth}
		\begin{tikzpicture}
			\draw (0, 0) node[inner sep=0] 	{\includegraphics[width=1\textwidth]{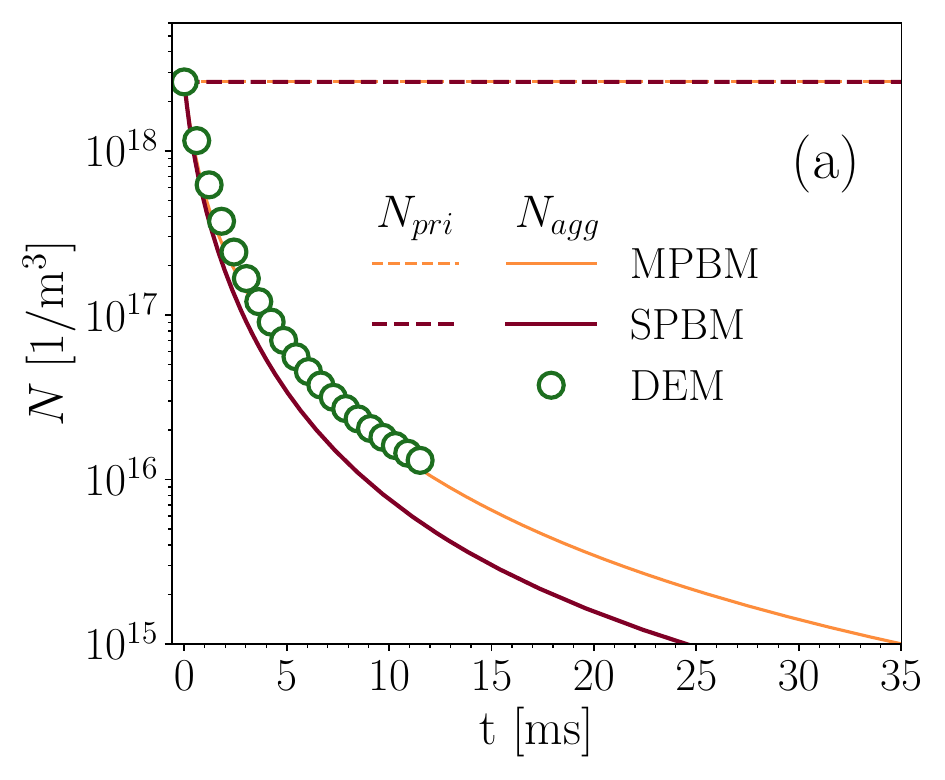}};
			\draw (2.1, 0.03) node {\scriptsize{\cite{kholghy2021surface}}};
		\end{tikzpicture}
	\end{subfigure}
	\begin{subfigure}[t]{0.4\textwidth}
		\begin{tikzpicture}
			\draw (0, 0) node[inner sep=0] 	{\includegraphics[width=1\textwidth]{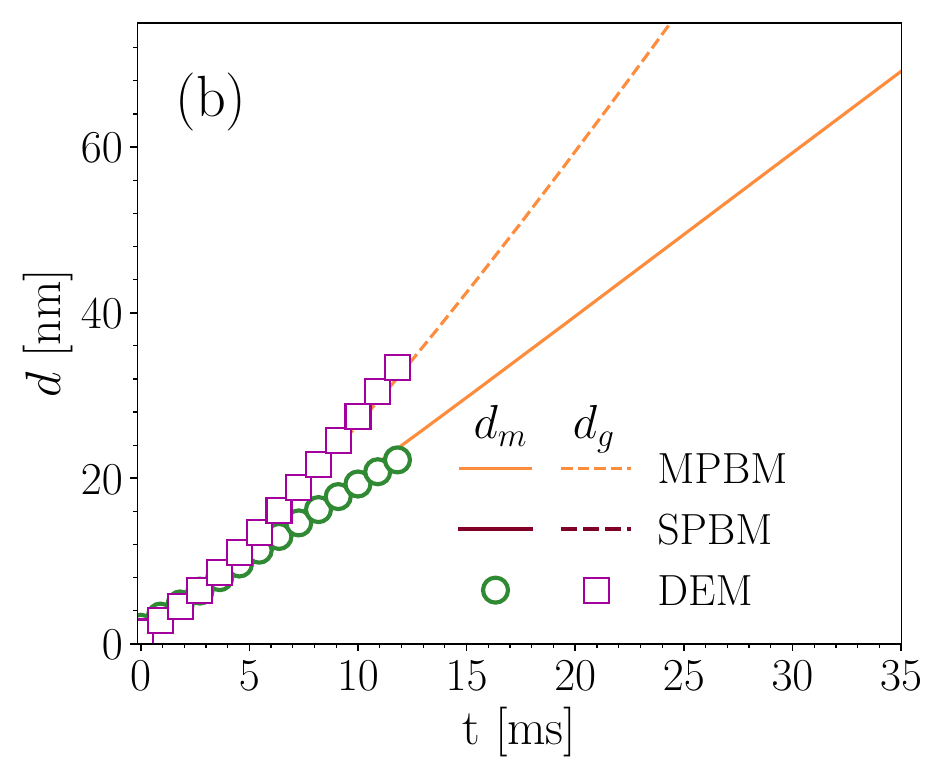}};
			\draw (2.2, -1.22) node {\scriptsize{\cite{kholghy2021surface}}};
		\end{tikzpicture}
	\end{subfigure}

	\caption{The number concentration of agglomerates and primary particles (a), and mobility and gyration diameters (b) obtained by Omnisoot using MPBM and SPBM in the free-molecular regime, which are in close agreement with the DEM results~\citep{kholghy2021surface} indicating the validity of the coagulation sub-models.}
	\label{fig:coagvalid_Nd} 
\end{figure}

The particles collide and grow in size through coagulation in both simulations, without inception, surface growth, or oxidation. Figure~\ref{fig:coagvalid_Nd} demonstrates the evolution of ${N_{agg}}$, ${N_{pri}}$, ${d_m}$, and ${d_g}$ as predicted by MPBM and SPBM in the free-molecular regime, which are in good agreement with DEM results~\citep{kholghy2021surface}. ${N_{pri}}$ is conserved during coagulation, resulting in identical flat trends for both models. In contrast, ${N_{agg}}$ decreases over time, with a faster decay observed in SPBM due to its treatment of agglomerate polydispersity, which leads to a higher collision frequency compared to MPBM. Therefore, $d_m$ and $d_g$  predicted by SPBM, shown in Figure~\ref{fig:coagvalid_Nd}b, are slightly larger than those predicted by MPBM.

\begin{figure}[H]
	\centering
	\begin{subfigure}[t]{0.4\textwidth}
		\begin{tikzpicture}
			\draw (0, 0) node[inner sep=0] 	{\includegraphics[width=1\textwidth]{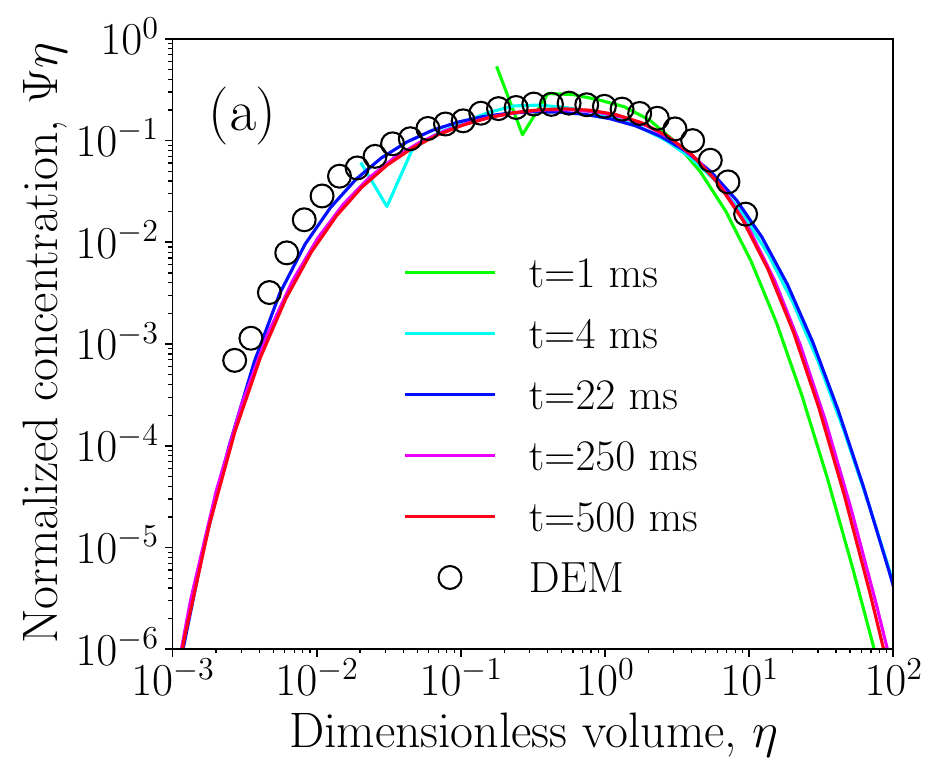}};
			\draw (1.3, -1.2) node {\scriptsize{\cite{goudeli2015coagulation}}};
		\end{tikzpicture}
	\end{subfigure}
	\begin{subfigure}[t]{0.4\textwidth}
		\begin{tikzpicture}
			\draw (0, 0) node[inner sep=0] 	{\includegraphics[width=1\textwidth]{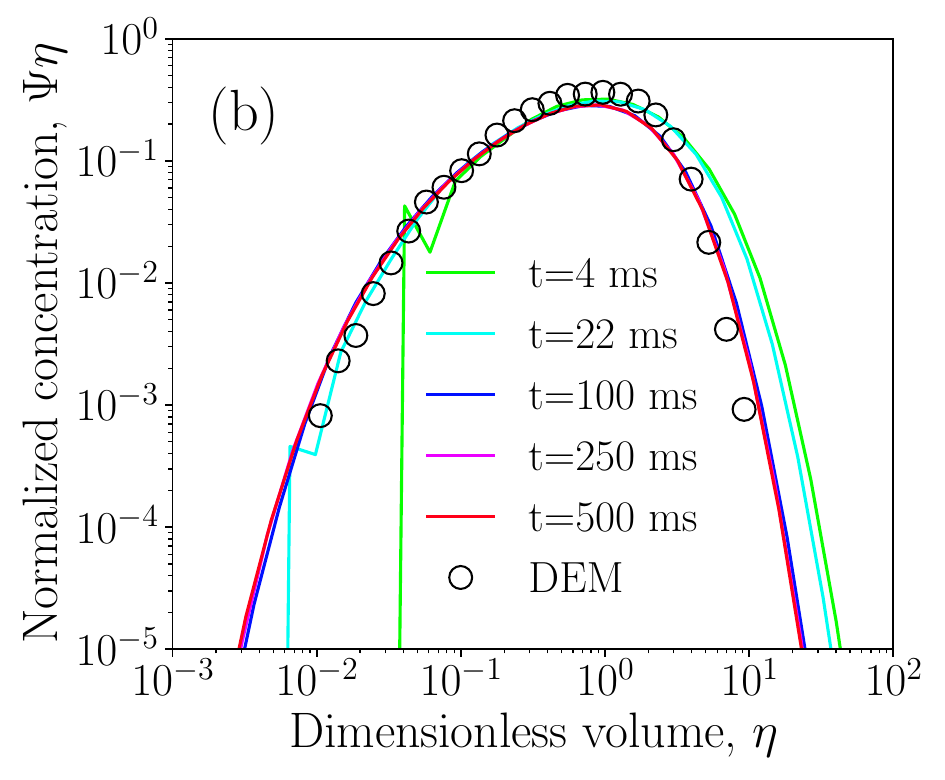}};
			\draw (1.3, -1.22) node {\scriptsize{\cite{kholghy2021surface}}};
		\end{tikzpicture}
	\end{subfigure}
	\caption{The non-dimensional particle size distribution at different residence times in the free-molecular (a) and continuum regimes (b) overlaps after the initial transient phase, indicating the attainment of a self-preserving size distribution, which is also in good agreement with DEM results~\citep{goudeli2015coagulation}.}
	\label{fig:coagvalid_psd} 
\end{figure}

The MPBM model cannot resolve the PSD due to the monodispersity assumption. In contrast, the SPBM tracks the number concentration of particles in discrete sections, which allows the construction of an evolving PSD, and the assessment of the size distribution spread. Figure~\ref{fig:coagvalid_psd} shows the evolution of the non-dimensional PSD from $t=1$~ms to 500 ms for the free-molecular and continuum simulations. The PSD is plotted as a function of the normalized concentration, ${\Psi= \bar{v}n_{agg}(v,t)/N_{agg,\infty}}$ and dimensionless volume, ${\eta= v/ \bar{v}}$, where ${n_{agg}(v,t)}$ is the size distribution function of agglomerate, ${v}$ is the particle volume, ${\bar{v}}$ is the mean particle volume, ${N_{agg,\infty}}$ is the total number concentration of agglomerates. After the initial transient ($t>22$~ms), the PSD rapidly evolves into a full bell curve and remains unchanged, indicating the attainment of a SPSD. This behavior is in good agreement with DEM results and confirms the capability of the SPBM implemented in Omnisoot to capture the SPSD for soot agglomerates in the free-molecular and continuum regimes, a characteristic outcome of Brownian-driven particle coagulation.

Figure~\ref{fig:coagvalid_psd}a shows the standard deviation of the mobility diameter, ${\sigma_g}$, predicted by SPBM in the free-molecular regime, which is in close agreement with DEM results. Initially, ${\sigma_g}=1$, indicating a monodisperse population, and it gradually increases, reaching a final value of 2.03, which corresponds to the characteristic standard deviation for the free-molecular regime~\citep{vemury1995self}.

\begin{figure}[H]
	\centering
	\begin{subfigure}[t]{0.4\textwidth}
		\begin{tikzpicture}
				\draw (0, 0) node[inner sep=0] 	{\includegraphics[width=1\textwidth]{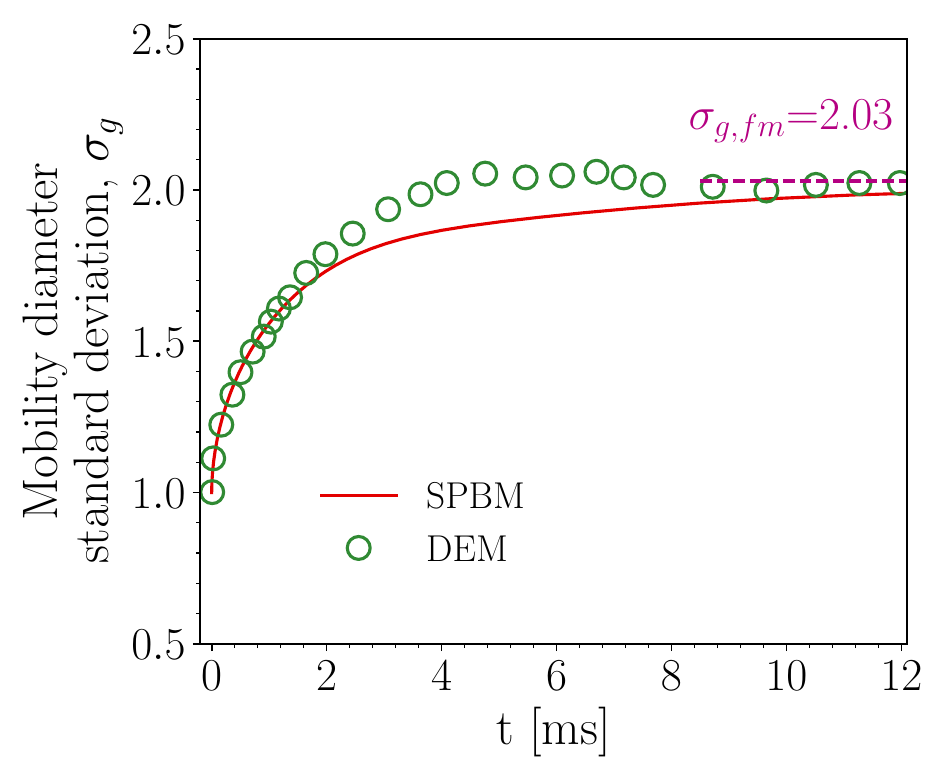}};
				\draw (0.54, -1.02) node {\scriptsize{\cite{kholghy2021surface}}};
			\end{tikzpicture}
	\end{subfigure}
	\caption{Standard deviation of the mobility diameter, $\mathrm{\sigma_g}$, in the free-molecular regime obtained by SPBM is in close agreement with DEM results~\citep{kholghy2021surface}.}
	\label{fig:coagvalid_sigma} 
\end{figure}

\subsection{Elemental and Energy Balance}

\noindent To ensure the accuracy and reliability of the simulations, the conservation of elemental mass and energy was assessed across all reactor models implemented in Omnisoot. The elemental balances of carbon, hydrogen, and the total energy (internal or enthalpy depending on the reactor type) of the gas-particle system were evaluated during soot formation processes under various pyrolysis and combustion conditions.
A set of simulations\footnote{\href{https://github.com/mohammadadib-cu/omnisoot-cv/tree/main/validations/mass_energy}{https://github.com/mohammadadib-cu/omnisoot-cv/tree/main/validations/mass\_energy}} was performed using different combinations of PAH growth and particle dynamics models, and the relative errors in elemental and energy balances were monitored over time or reactor length. The relative error of total carbon, hydrogen and energy in CVR, CPR, PSR, and PFR using different combinations of particle dynamics and inception models are shown in Figures~\ref{fig:constuvvalid}-\ref{fig:pfrvalid}. The residuals for all reactor configurations remained within acceptable limits (typically below $10^{-10}$), confirming that Omnisoot accurately preserves mass and energy during the coupled evolution of gas-phase and soot particles.

\section{Results and Discussion}

\subsection{Uncertainties in gas chemistry}
\label{sec:uncergaschem}
Accurate simulation of soot and carbon black formation remains a major challenge due to the compounded uncertainties introduced when coupling gas-phase chemistry with soot models. Prior studies have shown that the predicted concentrations of large PAHs can vary by orders of magnitude depending on the chosen reaction mechanism~\citep{wang2023systematic}. Moreover, there are not well-established pathways and rate constants for elementary reactions governing PAH dimerization, radical interactions, and surface reactions~\citep{martin2022soot}. Consequently, soot models inherit uncertainty from two sources: (i) the reaction mechanism used to describe fuel pyrolysis and oxidation, and (ii) the choice of inception models and their rate constants. The cumulative effect significantly complicates efforts to produce a reliable prediction of soot mass and morphology. This highlights the importance of coupling between gas-phase chemistry and soot formation models because the reliability of soot and carbon black simulations hinges more on the accuracy of upstream chemical kinetics, inception and surface growth fluxes.

To illustrate the impact of chemical mechanism selection on precursors, a series of simulations\footnote{\href{https://github.com/mohammadadib-cu/omnisoot-cv/tree/main/examples/pressure/mechanism_comparison}{https://github.com/mohammadadib-cu/omnisoot-cv/tree/main/examples/pressure/mechanism\_comparison}} were performed using the CPR model of Omnisoot (with soot formation disabled) for the pyrolysis of 5\%~$\mathrm{CH_4}$-Ar mixture at $\mathrm{T_5}=2203$~K and $\mathrm{P_5}=4.9$~atm~\citep{clack2025}. Six widely used chemical mechanisms were evaluated: ABF~\citep{appel2000kinetic}, Caltech~\citep{blanquart2009chemical}, KAUST~\citep{wang2013pah}, CRECK~\citep{saggese2015kinetic}, ITV~\citep{hellmuth2024role}, and FFCM2~\citep{ZDV2023}.

\begin{figure}[H]
	\centering
	\begin{subfigure}[t]{0.32\textwidth}
		\begin{tikzpicture}
			\draw (0, 0) node[inner sep=0] 	{\includegraphics[width=1\textwidth]{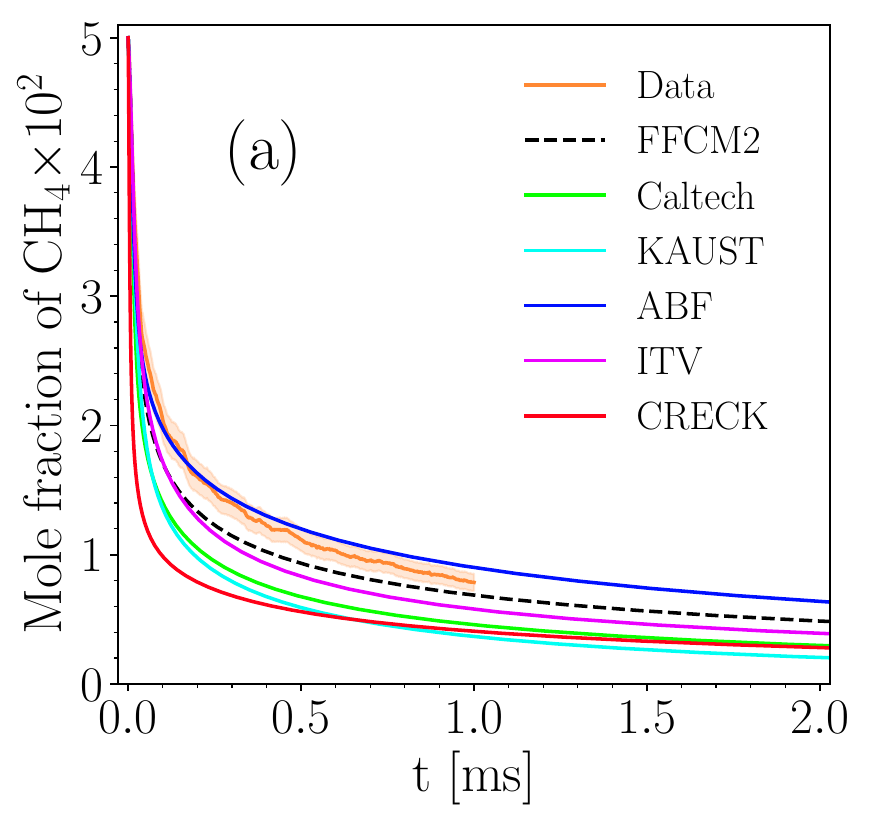}};
			\draw (1.85, 1.82) node {\tiny{\citep{clack2025}}};
		\end{tikzpicture}
	\end{subfigure}
	\begin{subfigure}[t]{0.32\textwidth}
		\includegraphics[width=1\textwidth]{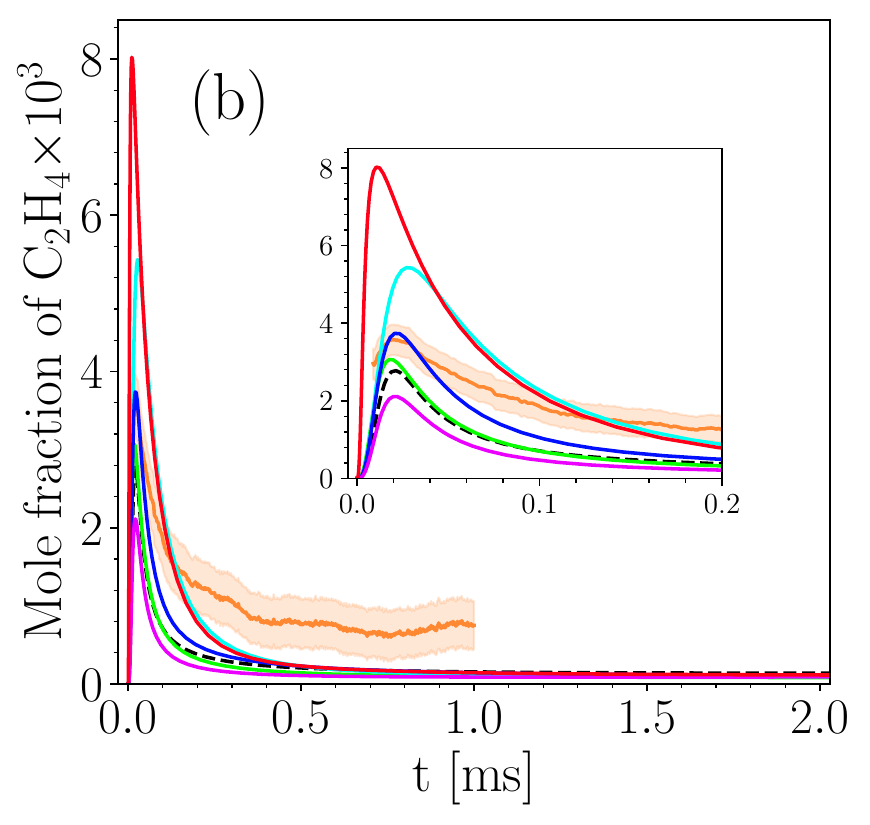}
	\end{subfigure}
	\begin{subfigure}[t]{0.32\textwidth}
		\includegraphics[width=1\textwidth]{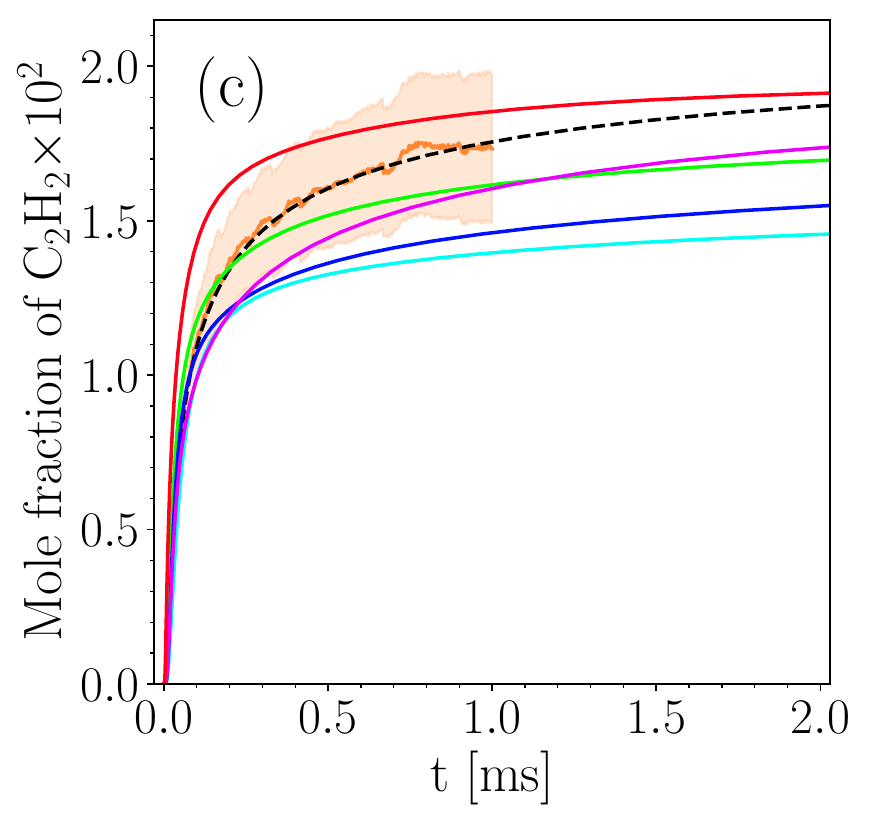}
	\end{subfigure}
	\caption{The mole fraction of $\mathrm{CH_4}$ (a), $\mathrm{C_2H_4}$ (b), and $\mathrm{C_2H_2}$ (c) during pyrolysis of 5\%~$\mathrm{CH_4}$-Ar predicted using different reaction mechanisms and compared with measurements~\citep{clack2025}. The insets provide a zoomed-in view of the early-time behavior and the shaded area represents the uncertainty in the reported experimental data.}
	\label{fig:CH4_C2H2_C2H4_chem} 
\end{figure}

The comparison focuses on the mole fraction of small intermediates and soot volume fraction measured using continuous wave absorption and laser extinction at 632 nm, respectively~\citep{clack2025}. Figure~\ref{fig:CH4_C2H2_C2H4_chem} shows the large variability in the predicted mole fraction of $\mathrm{CH_4}$, $\mathrm{C_2H_4}$, and $\mathrm{C_2H_2}$ across different reaction mechanisms. The mole fraction of $\mathrm{CH_4}$ serves as a useful indicator of carbon flux from the fuel to small and large intermediates. The carbon flux analysis, performed for all the mechanisms, highlighted the major role of $\mathrm{C_2H_4}$ in the production of $\mathrm{C_2H_2}$ via vinyl radical ($\mathrm{C_2H_3}$) as shown in prior studies (Figure~10 in \citep{wang2023systematic}). 

As shown in Figure~\ref{fig:CH4_C2H2_C2H4_chem}a, the CRECK mechanism predicts the lowest $\mathrm{CH_4}$ mole fraction (i.e., the largest conversion) at residence times up to 0.5 ms, whereas the KAUST mechanism yields a lower mole fraction conversion at longer residence times. Both mechanisms significantly underpredict methane mole fraction compared to the measurements. ABF and FFCM2 provide more accurate prediction of the $\mathrm{CH_4}$ considering the uncertainty of the experimental data. As shown in the inset of Figure~\ref{fig:CH4_C2H2_C2H4_chem}b, CRECK and KAUST mechanisms overpredict the initial jump in the mole fraction of  $\mathrm{C_2H_4}$. A similar variability is observed in the prediction of $\mathrm{C_2H_2}$ mole fraction: it is best captured by FFCM2, overpredicted by CRECK, and slightly underpredicted by the ITV and Caltech mechanisms. The KAUST and ABF mechanisms also underpredict the $\mathrm{C_2H_2}$ mole fraction, but their predictions fall outside the uncertainty range of the measurements. Among all, FFCM2 shows the best overall agreement with the measurements; however, this mechanism only includes $\mathrm{C_1}$-$\mathrm{C_4}$ species and excludes benzene and larger PAHs, so it cannot be coupled with the soot models considered in this study.

As mentioned before, there is a large uncertainty in inception and PAH adsorption rates used in PAH growth models. The default values were determined by applying the model to certain targets that vary for each model (some of the references can be found in Sections~\ref{sec:ebrimod} and \ref{sec:irrevdim}–\ref{sec:dimcoal} in which these models are explained). As a results, we do not focus on absolute accuracy of the models and instead use them to provide insight into the effect of chemistry and working parameters such as temperature and composition.
Here, Irreversible Dimerization model with 100\% efficiency of inception and PAH adsorption (i.e., all collisions result in inception and surface growth) is used to compare the maximum possible yield (represented by soot volume fraction) as an upper limit predicted by the selected mechanisms (except for FFCM2) during $5\%$ $\mathrm{CH_4}$ pyrolysis at $\mathrm{T_5}=2203$~K and $\mathrm{P_5}=4.9$~atm.

As shown in Figure~\ref{fig:max_sootfv_chem}, the maximum possible $f_v$ predicted by ABF mechanism is lower than the measured $f_v$ considering the reported uncertainty. The same analysis was performed for other data points with a different $\mathrm{T_5}$ in~\citep{clack2025}, and similar results were observed indicating that ABF mechanism cannot provide enough precursors under pyrolysis conditions in these simulations. In contrast, the maximum possible $f_v$ predicted by other mechanisms is significantly larger than the measurements, which presents the possibility of predicting $f_v$ close to the measurements using more realistic assumptions (i.e., lower dimerization efficiencies). Caltech and KAUST mechanisms are used in the simulations of this study for description of gas chemistry as they provide the highest carbon flux to soot precursors. ITV and CRECK mechanisms can also provide large enough carbon flux to precursors, but they are not used because of their large size that makes parametric studies computationally expensive. Moreover, the selected mechanisms does not change the main findings about the difference between inception models in terms of temperature and composition sensitivity.

In the rest of the results, the inception and PAH adsorption rates of PAH growth models are adjusted by calibrating $\eta_{inc}$ and $\eta_{ads}$ to match predict soot properties with available measurements. Note that, this approach does not resolve the mechanistic uncertainties of soot formation. It allows the model to calculate effective carbon fluxes from the gas phase into the particle phase that are consistent with measured soot characteristics.

\subsection{Methane pyrolysis in shock tube}

The pyrolysis of a 5\% $\mathrm{CH_4}$-Ar mixture was studied using the CPR model over a post-shock temperature range of $\mathrm{T_5} = 1800$-$3000$ K and a pressure range of $\mathrm{P_5} = 4.7$-$7.1$ bar. Since $\mathrm{P_5}$ was not specified for individual experiments (each defined by a given $\mathrm{T_5}$), a linear dependence of $\mathrm{P_5}$ on $\mathrm{T_5}$ was assumed within the stated range~\citep{agafonov2016unified}. The gas-phase chemistry was modeled using the Caltech mechanism. Inception and PAH adsorption rates were adjusted using $\eta_{inc}$ and $\eta_{ads}$, respectively, to match the predicted Carbon Yield (CY) at $t = 1.5$ ms with light extinction measurements at $\lambda = 632$ nm across the studied $\mathrm{T_5}$ and $\mathrm{P_5}$ ranges~\citep{agafonov2016unified}. The experimental CY was extracted from the reported product of CY$\times$E(m). However, this approach introduces uncertainty due to variability in E(m). It is common to assume the E(m) of mature soot, which may not be valid in shock tube experiments with short residence times ($\approx 2$ ms) that involve initial stages of soot formation, as E(m) increases with particle size, composition, and maturity. For instance, during acetylene pyrolysis, E(m) can vary from 0.05 to 0.25 as $d_p$ reaches 20 nm within 1.6 ms. Despite such observations, our quantitative understanding of the evolution of E(m) in soot remains limited. We considered the reported variability of E(m) in the literature, ranging from 0.174~\citep{lee1981optical} to 0.37~\citep{agafonov2011soot}.

A parametric study was conducted on $\eta_{inc}$ and $\eta_{ads}$ using a grid search approach. Each factor was varied across 11 logarithmically spaced values from $10^{-4}$ to 1, resulting in 121 simulation cases per data point and 847 simulations in total. Each $\eta_{inc}$ and $\eta_{ads}$ pair was ranked based on the average relative error of CY across all data points in the studied $\mathrm{T_5}$ range, in order to identify the set of adjustment factors that minimized the CY prediction error. 

To reduce the computational load of the parametric study, simulations were performed using MPBM for all inception models. The comparison of results obtained using SPBM and MPBM shows that CY and $d_p$ are not sensitive to the choice of particle dynamics model, indicating that MPBM results can be reliably used for CY error analysis. However, $N_{agg}$ and $n_p$, shown in Figure~\ref{fig:shockagof_N_agg_n_p_cpr_pdynamics}, differ significantly between SPBM and MPBM because the central assumption of the monodisperse model is the rapid attainment of SPSD, which is not valid at short residence times near 1.5~ms. For example, the non-dimensional PSD at $\mathrm{T_5}=2200$~K, shown in Figure~\ref{fig:shockagof_psd}, significantly changes from 1.0~ms to 2.5~ms for all inception models.

\begin{figure}[H]
	\centering
	\includegraphics[width=0.8\textwidth]{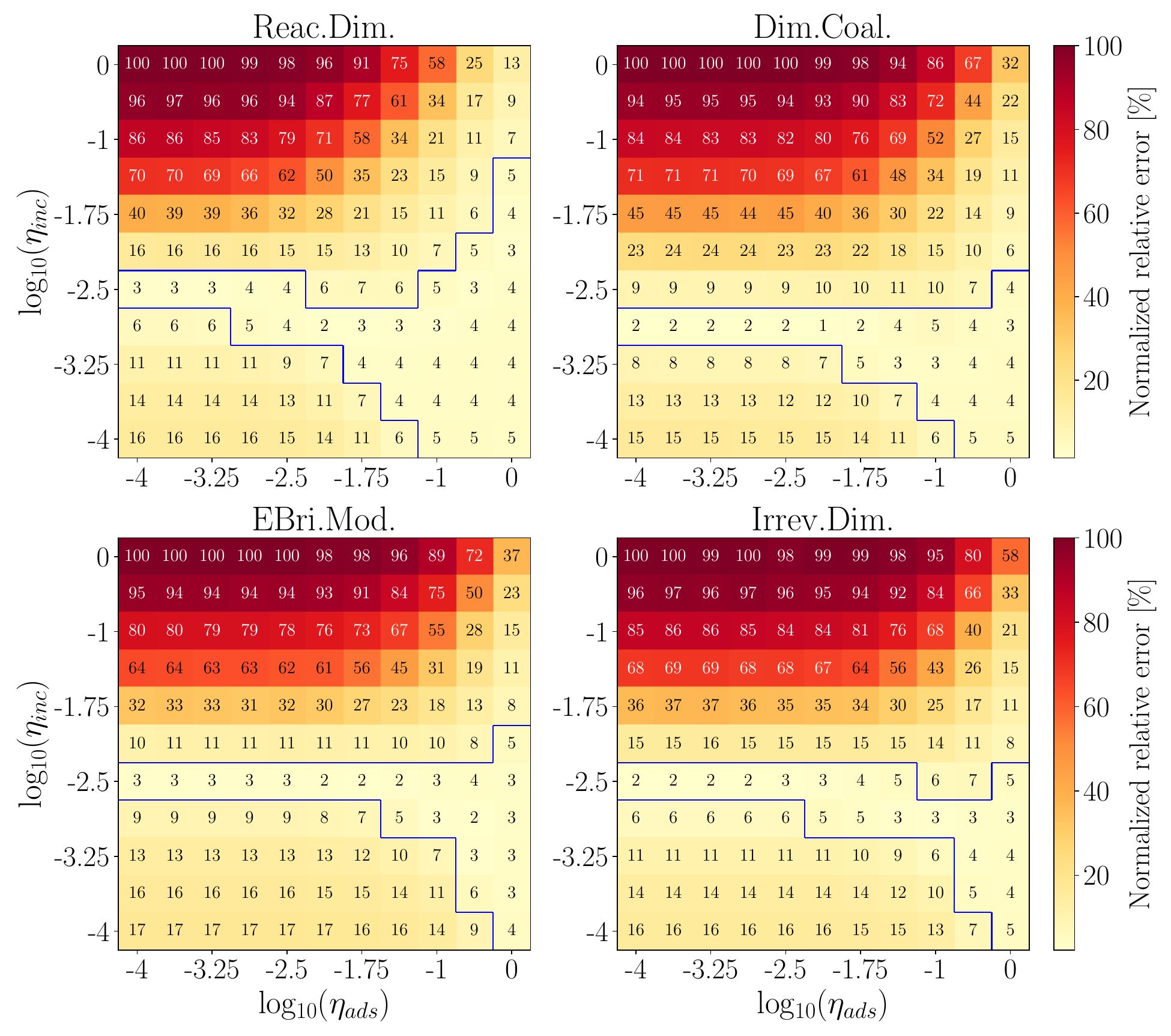}
	\caption{The relative error of CY normalized by the maximum value at $t=1.5$~ms for 5\%~$\mathrm{CH_4}$-Ar obtained using Caltech mechanism and different inception models.}
	\label{fig:shockagof_yielderror_cpr} 
\end{figure}

Figure~\ref{fig:shockagof_yielderror_cpr} shows a heat map of the mean relative error, normalized by the maximum value, for all inception models. The largest error occurs for the combination of maximum inception ($\eta_{inc} = 1$ or $\log_{10}(\eta_{inc}) = 0$) and minimum adsorption ($\eta_{ads} = 10^{-4}$ or $\log_{10}(\eta_{ads}) = -4$). The region of lowest normalized error (less than 5\%), outlined in blue in Figure~\ref{fig:shockagof_yielderror_cpr}, includes a broad set of adjustment factor combinations that enable the model to accurately predict CY. However, accurate prediction of CY does not necessarily imply accurate prediction of other properties, such as primary particle diameter or mobility diameter. The availability of data on other soot parameters or intermediate gaseous species, and including them as optimization targets, could help narrow down the range of optimal adjustment factors.

To highlight the variability in soot morphology across the low-error region of the heat map, we focus on three representative combinations of $\eta_{inc}$ and $\eta_{ads}$: (i) minimum inception flux ($\eta_{inc} = 10^{-4}$), (ii) minimum PAH adsorption rate ($\eta_{ads} = 10^{-4}$), and (iii) equally scaled inception flux and PAH adsorption rate ($\eta_{inc} \approx \eta_{ads}$). It is important to note that soot CY and morphology are closely coupled to gas-phase chemistry under the short residence times studied. Therefore, a set of simulations was performed without soot to isolate gas-phase chemistry and examine the temperature dependence of soot precursors and $\mathrm{C_2H_2}$. Subsequently, inception models were optimized using equal adjustment factors\footnote{\href{https://github.com/mohammadadib-cu/omnisoot-cv/tree/main/examples/pressure/methane_pyrolysis}{https://github.com/mohammadadib-cu/omnisoot-cv/tree/main/examples/pressure/methane\_pyrolysis}} ($\eta_{inc} = \eta_{ads}$), and separate optimizations were carried out for the minimum adsorption case ($\eta_{ads} = 10^{-4}$) and the minimum inception case ($\eta_{inc} = 10^{-4}$), denoted as \textit{Min-Ads} and \textit{Min-Inc}, respectively.

\begin{figure}[H]
	\centering
	\begin{subfigure}[t]{0.4\textwidth}
		\includegraphics[width=1\textwidth]{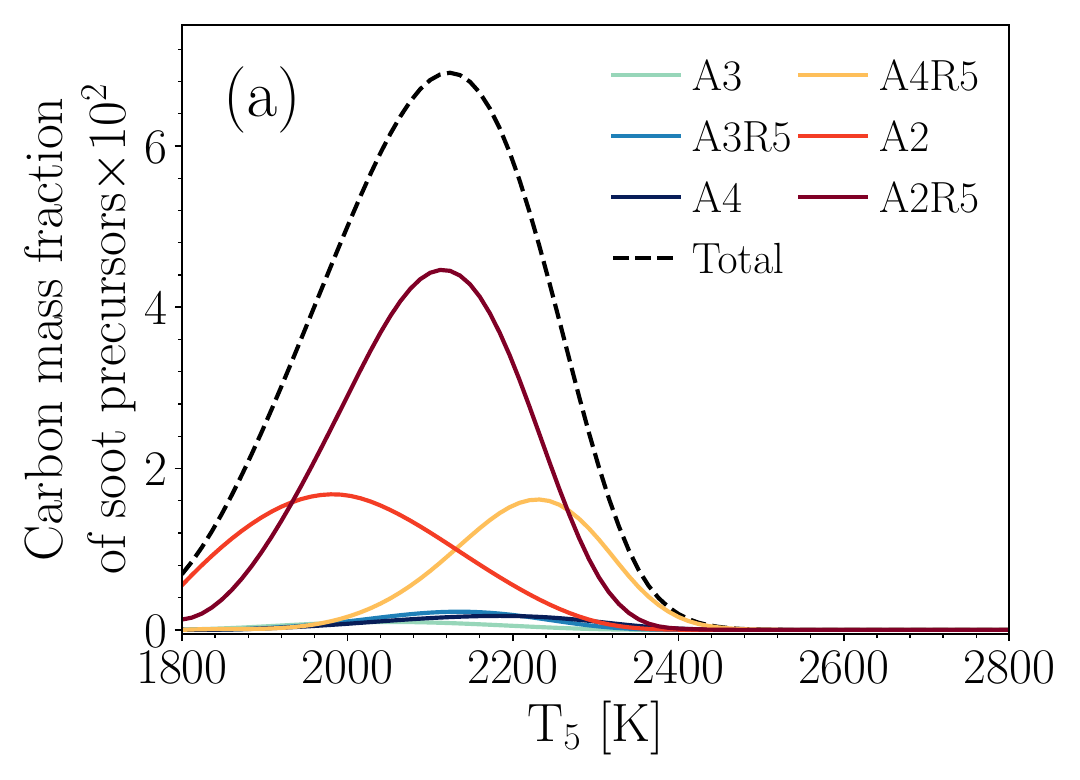}
	\end{subfigure}
	\begin{subfigure}[t]{0.36\textwidth}
		\includegraphics[width=1\textwidth]{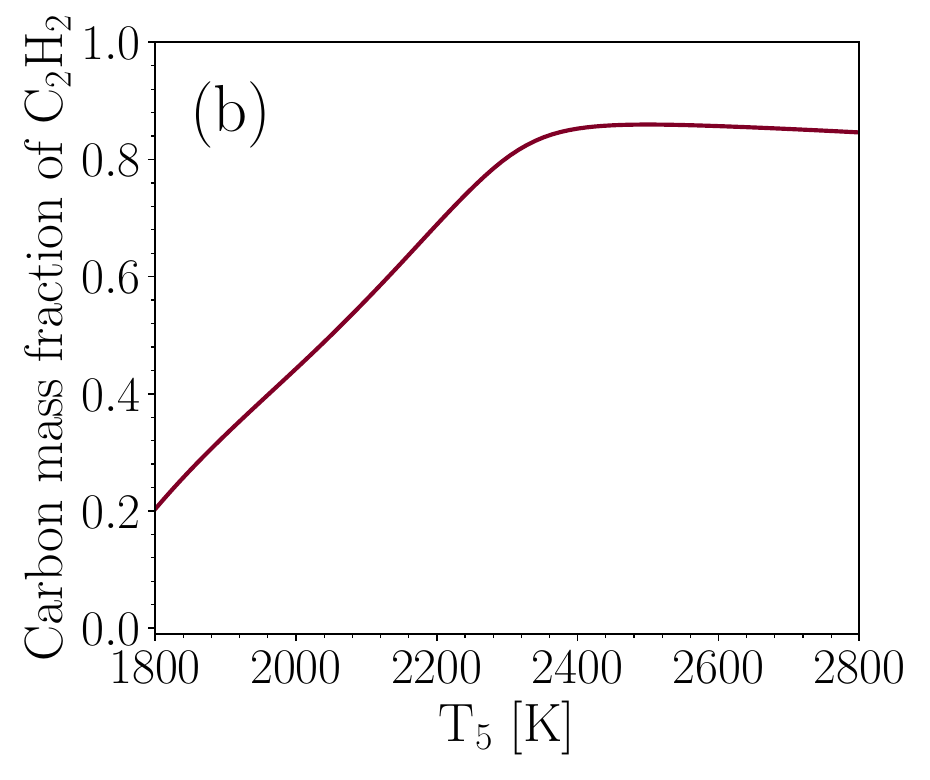}
	\end{subfigure}
	\caption{The bell-shaped temperature profile of carbon mass fraction of soot precursors (A2 and larger) combined (a) and $\mathrm{C_2H_2}$ (b) at $t=1.5$~ms during pyrolysis of 5\%~$\mathrm{CH_4}$-Ar obtained using CPR model with Caltech mechanism without considering soot.}
	\label{fig:SPC_cmf_cpr} 
\end{figure}

Figure~\ref{fig:SPC_cmf_cpr}a shows the CMF of soot precursors (individual PAHs and total) and $\mathrm{C_2H_2}$ at 1.5 ms across the studied temperature range, with the soot model deactivated. The CMF is defined as the ratio of the carbon mass contained in a given species to the total carbon mass in the system. While PAH precursors exhibit a bell-shaped profile, the CMF of $\mathrm{C_2H_2}$ increases linearly with temperature and plateaus near 85\% at $\mathrm{T_5} \approx 2400$ K. Among the considered PAHs, A2, A2R5, and A4R5 account for most of the carbon and are likely major contributors to soot inception and surface growth. The strong variation in PAH CMF highlights the impact of precursor selection on inception flux, surface growth rates, and their temperature dependence. Figure~\ref{fig:shockagof_yieldspc_cpr} illustrates the effect of excluding five-membered ring PAHs (A2R5, A3R5, and A4R5) from the soot precursors on the CY and $d_p$. As expected, excluding these species reduces CY for all inception models, potentially highlighting the significant contribution of five-membered rings PAHs. The majority of carbon mass of PAHs in soot particles over $2000~\mathrm{K}<\mathrm{T_5}<2400$~K originates from A2R5 and A4R5 (Figure~\ref{fig:shockagof_spccont_cpr}).

\begin{figure}[H]
	\centering
	\begin{tikzpicture}
		\draw (0, 0) node[inner sep=0] 	{\includegraphics[width=0.8\textwidth]{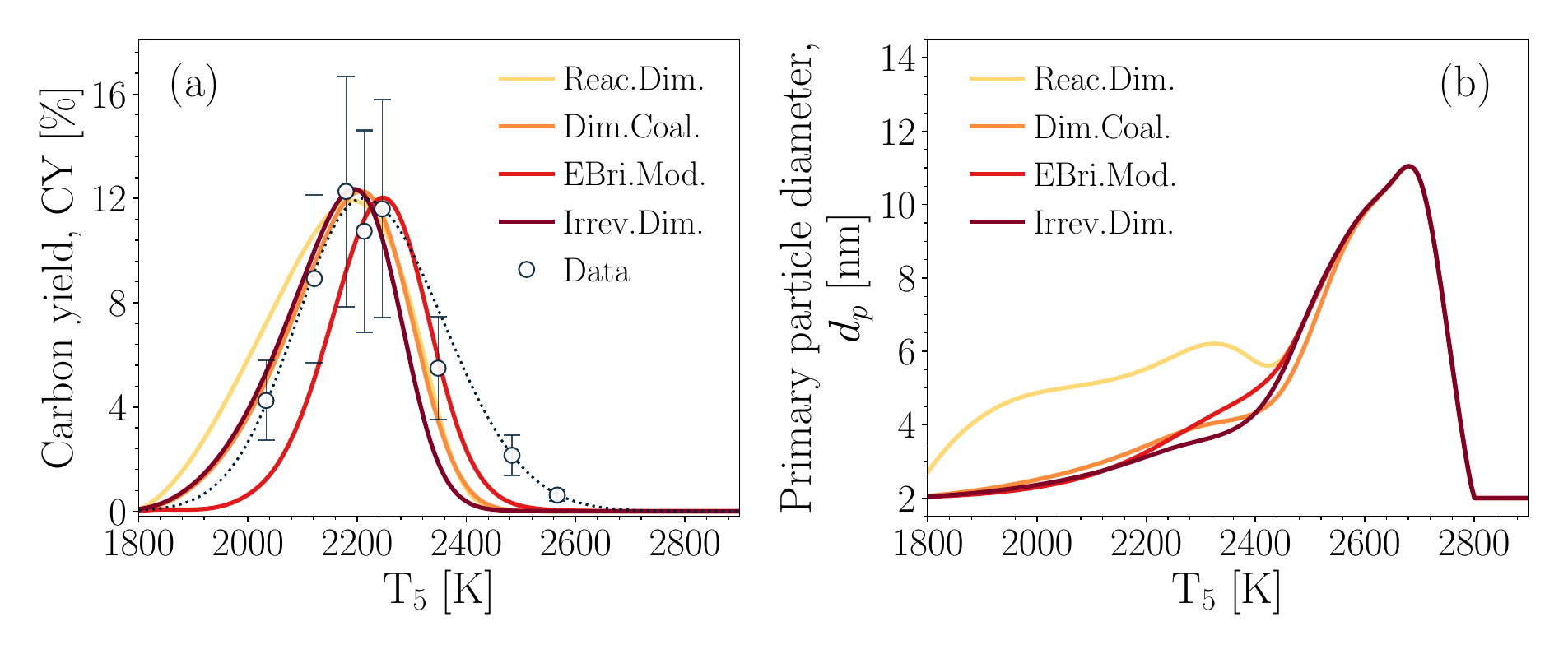}};
		\draw (-0.85, 0.42) node {\scriptsize{\cite{agafonov2016unified}}};
	\end{tikzpicture}
	\caption{The CY (a) and primary particle diameter, $d_p$, (b) at $t=1.5$~ms during the pyrolysis of 5\%~$\mathrm{CH_4}$-Ar obtained using Caltech mechanism, SPBM and different inception models optimized using equal adjustment factors to minimize the prediction error with extinction measurements~\citep{agafonov2016unified}. The dashed line was added to show the trend in the measurements.}
	\label{fig:shockagof_yield_dp_cpr} 
\end{figure}

Next set of simulations were conducted by using equal adjustment factors ($\eta_{inc}=\eta_{ads}$) to minimize the prediction error of CY. Figure~\ref{fig:shockagof_yield_dp_cpr}a compares CY from various inception models with experimental data~\citep{agafonov2016unified}. A skew exponential curve (black dotted line) was fitted to the data to highlight the CY trend and its peak near 12\%. Soot CY follows a bell-shaped profile similar to that of soot precursors, as inception flux and mass growth directly depend on precursor concentrations. Reactive Dimerization model predicts higher CY at lower temperatures, while E-Bridge Modified model shifts the peak to higher temperatures due to a different temperature dependence, which is primarily governed by PAH dehydrogenation via an Arrhenius rate law, in contrast to physical PAH collisions in other models. We also ran simulations using the KAUST mechanism for the equal-adjustment-factors case and compared the model predictions with the CY measurements, as shown in Figure~\ref{fig:shockagof_yield_dp_cpr_kaust}. The bell-shaped profile of CY was captured; however, the $\mathrm{T_5}$ corresponding to the peak CY was overpredicted by all inception models by more than 100~K, in contrast to the Caltech mechanism, which shows closer agreement with the measurements.

We also ran simulations using the KAUST mechanism for the equal-adjustment-factors case and compared the model predictions with the CY measurements, as shown in Figure~\ref{fig:shockagof_yield_dp_cpr_kaust}. The bell-shaped profile of CY was captured; however, the $\mathrm{T_5}$ corresponding to the peak CY is nearly 100~K higher than that of the measurements, whereas the Caltech mechanism provided a better prediction of the peak location.

\begin{figure}[H]
	\centering
	\includegraphics[width=0.8\textwidth]{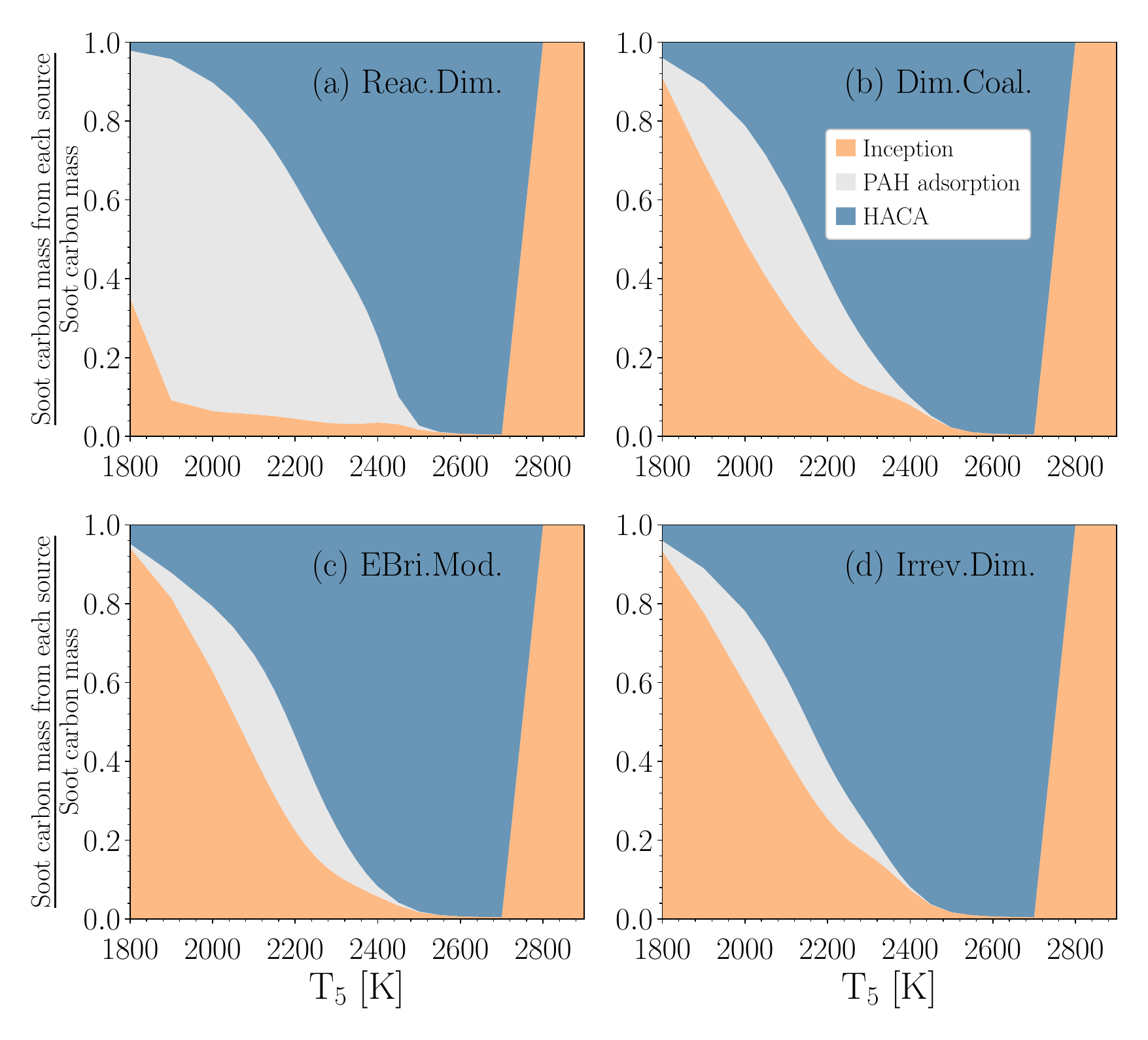}
	\caption{The soot carbon mass from inception, PAH adsorption and HACA normalized by total soot carbon mass at $t=1.5$~ms during the pyrolysis of 5\%~$\mathrm{CH_4}$-Ar obtained using Caltech mechanism, SPBM and different inception models calibrated to minimize the prediction error with extinction measurements~\citep{agafonov2016unified}.}
	\label{fig:shockagof_carbon_map_cpr} 
\end{figure}

As shown in Figure~\ref{fig:shockagof_yield_dp_cpr}b, the predicted $d_p$ increases with $\mathrm{T_5}$, reaching a maximum of 12.5 nm at 2700 K, where yield is very low ($\approx 10^{-7}$), and then drops to the model's minimum allowed value of 2 nm. The differences between inception models in predicting $d_p$ is overall negligible except for Reactive Dimerization model, which predicts a larger $d_p$ in $\mathrm{T_5}<2500$ K. The $d_p$ trends can be better understood by examining Equation~\eqref{eqn:d_p} indicating that $d_p$ is proportional to the third-root of $C_{tot}/N_{pri}$. $C_{tot}$ describes total carbon mass converted to soot through inception and surface growth while $N_{pri}$ is only determined by inception flux. As a result, $d_p$ is controlled by the ratio of total surface growth rate (via HACA and PAH adsorption) to inception flux. The ratio of carbon mass gained up to 1.5 ms by each pathway to the total soot carbon mass was calculated and shown in Figure~\ref{fig:shockagof_carbon_map_cpr}. PAH adsorption is the dominant soot mass growth pathway for Reactive Dimerization model in $\mathrm{T_5}<2000$ K (Figure~\ref{fig:shockagof_carbon_map_cpr}a), which results in larger $C_{tot}/N_{pri}$ and $d_p$ values, but inception is dominant for the other inception models. The contribution of inception decreases with temperature for all inception models while the contribution of HACA becomes dominant up to 2700 K leading to higher $d_p$ (Figure~\ref{fig:shockagof_yield_dp_cpr}b). This is due to the decrease of precursor concentration and the increase of $\mathrm{C_2H_2}$ CMF. Beyond 2700 K, the contribution of inception rapidly increases with temperature leading to the decrease of $d_p$ (Figure~\ref{fig:shockagof_yield_dp_cpr}b).

\begin{figure}[H]
	\centering
	\includegraphics[width=0.8\textwidth]{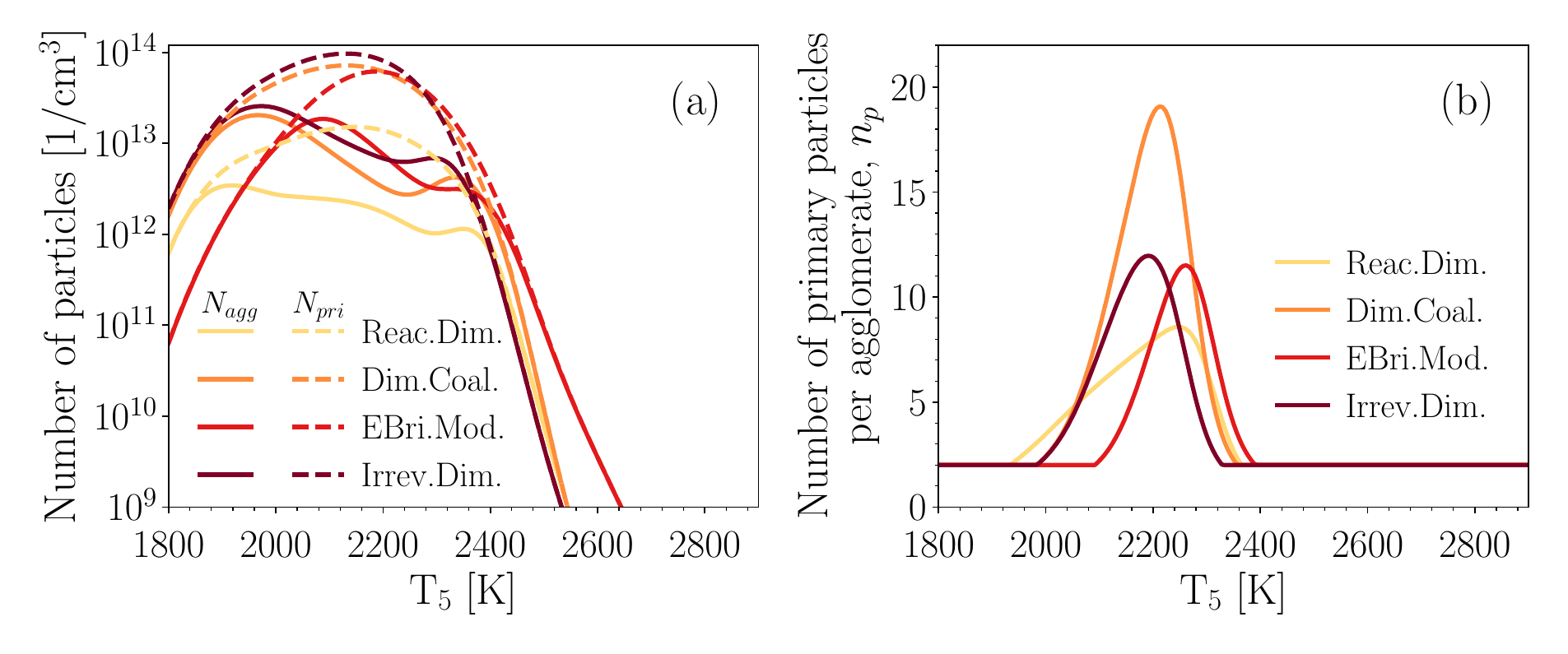}
	\caption{The temperature dependence of total number of agglomerates, $N_{agg}$ (a), number of primary particles per agglomerate, $n_p$ (b) at $t=1.5$~ms during the pyrolysis of 5\%~$\mathrm{CH_4}$-Ar obtained using Caltech mechanism and different inception models optimized using equal adjustment factors to minimize the prediction with extinction measurements~\citep{agafonov2016unified}.}
	\label{fig:shockagof_Nagg_np_cpr} 
\end{figure}

Figure~\ref{fig:shockagof_Nagg_np_cpr}a shows that $N_{pri}$ peaks near 2200 K, aligning with the peak in CY. This is expected because $N_{pri}$ depends solely on the inception flux, which is the highest when PAH concentrations peak. Increased coagulation rates reduce $N_{agg}$ in the $2000$-$2500$ K range. Reactive Dimerization model yields the fewest particles ($N_{pri}$ and $N_{agg}$) because more precursors are directed toward surface growth. Consequently, $n_p$ is lowest for Reactive Dimerization model. Differences between inception models are more evident in $n_p$ than in $d_p$, influencing predicted particle morphology.

\begin{figure}[H]
	\centering
	\begin{tikzpicture}
		\draw (0, 0) node[inner sep=0] 	{\includegraphics[width=0.8\textwidth]{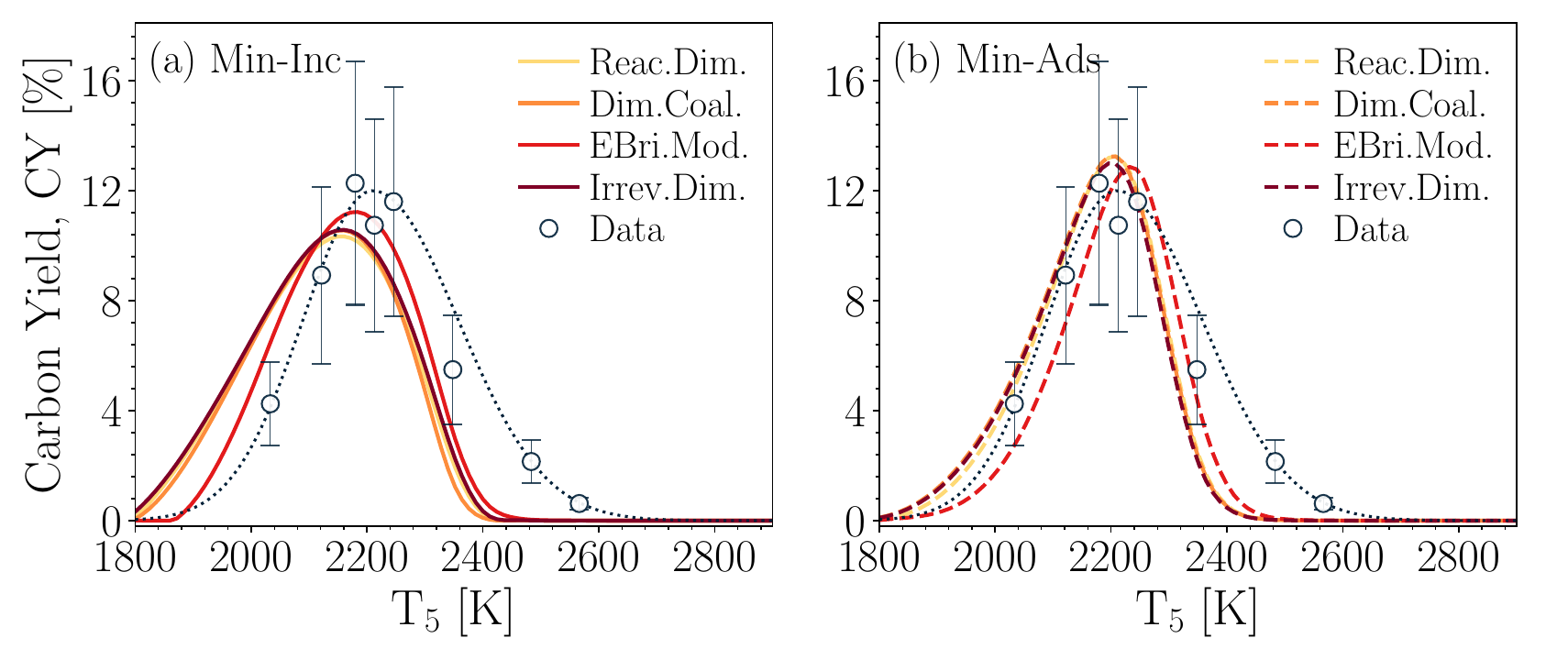}};
		\draw (-0.48, 0.75) node {\scriptsize{\cite{agafonov2016unified}}};
		\draw (5.33, 0.75) node {\scriptsize{\cite{agafonov2016unified}}};
	\end{tikzpicture}
	\caption{The comparison of CY at $t=1.5$~ms when the minimum inception (a) and the minimum PAH adsorption (b) adjustment factors were applied to minimize the prediction error compared with the measurements~\citep{agafonov2016unified} for 5\%~$\mathrm{CH_4}$-Ar obtained using Caltech mechanism, SPBM and different inception models.}
	\label{fig:shockagof_yield_maxincads_cpr} 
\end{figure}

\begin{figure}[H]
	\centering
	\includegraphics[width=0.5\textwidth]{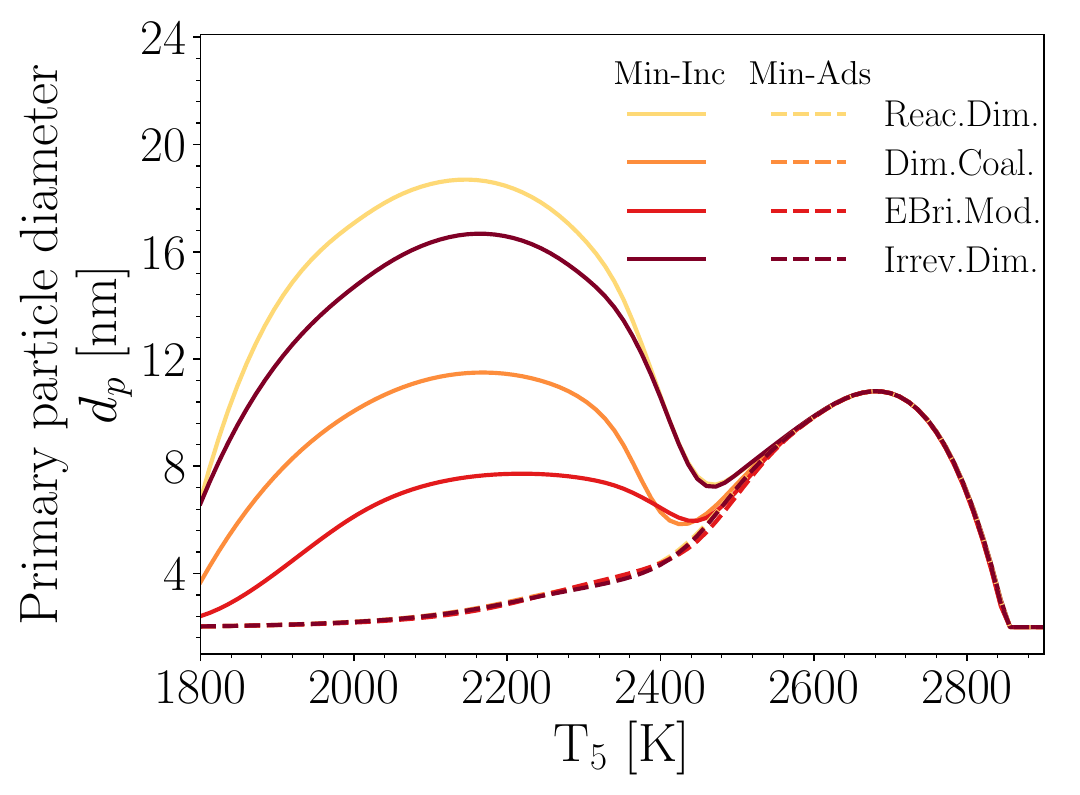}
	\caption{The comparison of mean primary particle, $d_p$, at $t=1.5$~ms when the minimum inception (solid lines) and adsorption (dashed lines) were applied to minimized the prediction error compared with measurements~\citep{agafonov2016unified} for 5\%~$\mathrm{CH_4}$-Ar obtained using Caltech mechanism, SPBM and different inception models.}
	\label{fig:shockagof_dp_maxincads_cpr} 
\end{figure}

Figure~\ref{fig:shockagof_yield_maxincads_cpr} shows that the optimized models with minimum $\eta_{inc}$ and $\eta_{ads}$ yield similar CY across inception models, except E-Bridge Modified model, which shifts slightly toward higher temperatures. The minimum adsorption case has a higher peak and narrower profile compared with the minimum inception case. As shown in Figure~\ref{fig:shockagof_dp_maxincads_cpr}, $d_p$ predicted in Min-Ads mode is smaller and less sensitive to the selection of inception model compared with Min-Inc mode. Two peaks can be seen in Min-Inc mode for all inception models: The first peak occurs due to high PAH adsorption rates at the temperature range close to the peak of CMF of soot precursors (Figure~\ref{fig:SPC_cmf_cpr}a); the second peak which occurs around $\mathrm{T_5}=2700$~K and $d_p=12$~nm, which is very similar to $d_p$ trends predicted using equal adjustment factors (Figure~\ref{fig:shockagof_yield_dp_cpr}b). Reactive Dimerization model has the largest variation at the peak from 2 to 19~nm between Min-Ads and Min-Inc modes. The existence of such a variation highlight the need for characterizing primary particle size in the experiment to reduce the uncertainty in the soot model and restrict the range of expected inception and surface growth rate required to accurately predict soot yield and morphology.

\subsection{Ethylene pyrolysis in a flow reactor}
The PFR model was used to simulate\footnote{\href{https://github.com/mohammadadib-cu/omnisoot-cv/tree/main/examples/plug_flow/sectional}{https://github.com/mohammadadib-cu/omnisoot-cv/tree/main/examples/plug\_flow/sectional}} the pyrolysis of 0.6\% $\mathrm{C_2H_4}$-$\mathrm{N_2}$ in a 1.4 m long, and 16 mm diameter flow reactor. Figure~\ref{fig:pfr_temp} shows the axial temperature profile used in the simulations. The maximum temperature, $\mathrm{T_{max}} = 1673$ K, was imposed on the model based on thermocouple measurements reported by~\citet{mei2019quantitative} for various flow rates along the reactor centerline. The temperature increases from 300 K at the reactor inlet to the hot zone, where it reaches a value within 10\% of $\mathrm{T_{max}}$. The length of the hot zone varies with flow rate due to advection, ranging from 0.71 m at $\mathrm{Q}=8$ L/min to 0.76 m at $\mathrm{Q}=12$ L/min. Near the reactor outlet, the temperature drops to approximately 650 K.

\begin{figure}[H]
	\centering
	\begin{tikzpicture}
		\draw (0, 0) node[inner sep=0] 	{\includegraphics[width=1\textwidth]{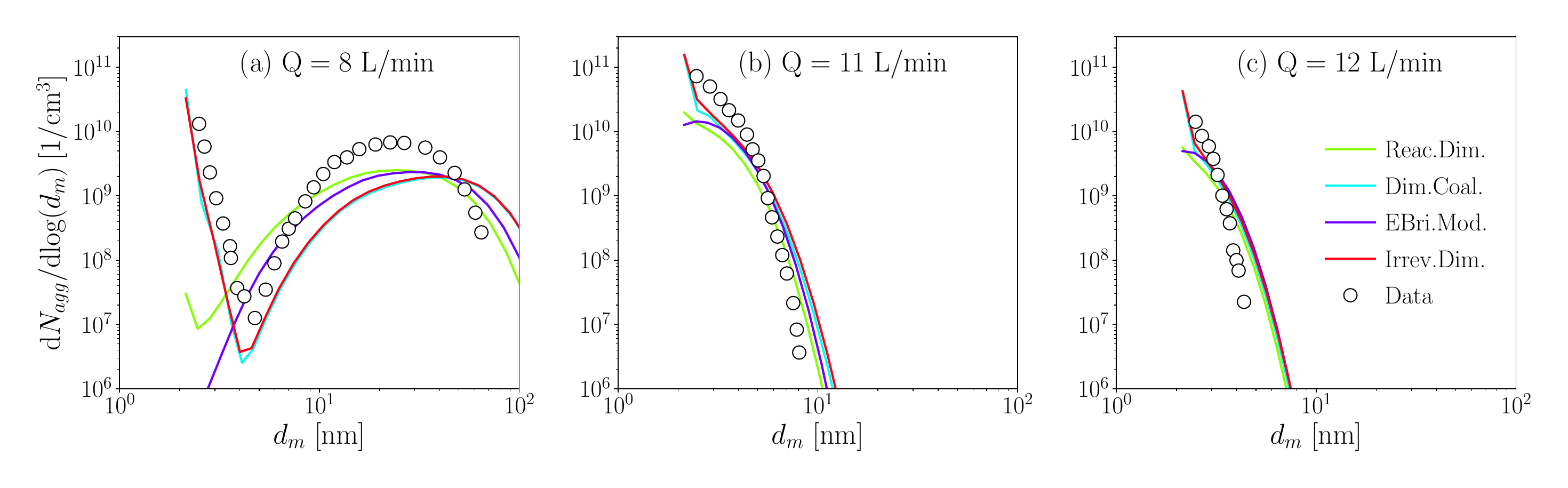}};
		\draw (6.63, -0.51) node {\scriptsize{\cite{mei2019quantitative}}};
	\end{tikzpicture}
	\caption{The particle size distribution at the end of PFR for $\mathrm{Q}=8$ (a), 11 (b), and 12 L/min (c) obtained using KAUST mechanism, SPBM and different inception models calibrated to match the predictions with measurement~\citep{mei2019quantitative}.}
	\label{fig:pfr_psd} 
\end{figure}

The inception and PAH adsorption adjustment factors were varied to match the predicted PSD with Scanning Mobility Particle Sizer
(SMPS) measurements conducted on centerline at the reactor exit for the flow rates of 8, 11, and 12 L/min~\citep{mei2019quantitative}, as shown in Figure~\ref{fig:pfr_psd}. The parametric analysis revealed that each inception model requires a unique set of adjustment factors to minimize the PSD prediction error across different flow rates. The KAUST mechanism~\citep{wang2013pah} was used to describe gas-phase chemistry, as it provided better agreement with the measured PSDs than the Caltech mechanism, shown in Figure~\ref{fig:pfr_psd_caltech}.

A bimodal size distribution can be observed for $\mathrm{Q}=8$ L/min (Figure~\ref{fig:pfr_psd}a) that was attributed to the continuous inception~\citep{zhao2003measurement}. The comparison of simulations with measurements reveals different temperature dependence of inception models. Irreversible Dimerization and Dimer Coalescence models capture the bimodality of PSD in good agreement with the measurements, but the other two models predict a nearly unimodal PSD. Specifically, the number concentration of the first section is lower by more than three orders of magnitude for the Reactive Dimerization and E-Bridge Modified models indicating the lack of active inception at the sampling location due to the temperature drops near the end of reactor that suppresses soot inception.

All inception models capture the disappearance of the PSD shoulder at flow rates of 11 and 12 L/min. In these cases, the shorter particle residence time limits coagulation, preventing formation of a second peak. However, Reactive Dimerization and E-Bridge Modified models still underpredict the number concentration of nascent particles by approximately one order of magnitude compared to the irreversible models (Irreversible Dimerization and Dimer Coalescence).

\begin{figure}[H]
	\centering
	\begin{tikzpicture}
		\draw (0, 0) node[inner sep=0] 	{\includegraphics[width=1\textwidth]{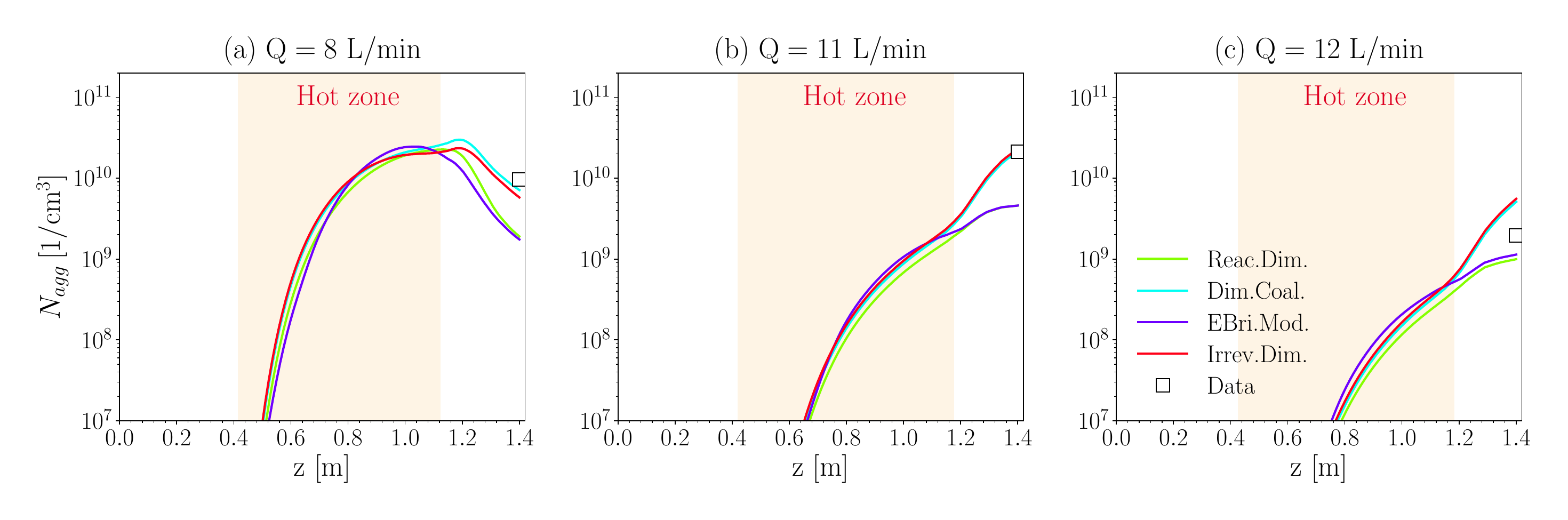}};
		\draw (4.85, -1.25) node {\tiny{\cite{mei2019quantitative}}};
	\end{tikzpicture}
	\caption{The total number of agglomerates along the PFR for $\mathrm{Q}=8$ (a), 11 (b), and 12 L/min (c) obtained using KAUST mechanism, SPBM and different inception models compared with data~\citep{mei2019quantitative}. The yellow area represents the hot zone ($\mathrm{T}>0.9\mathrm{T_{max}}$).}
	\label{fig:pfr_Nagg} 
\end{figure}

Figure~\ref{fig:pfr_Nagg} presents the axial evolution of $N_{agg}$. All inception models predict nearly the same $N_{agg}$ up to $z \approx 1.1$ m, corresponding to the end of the hot zone. Beyond this point, $N_{agg}$ continues to increase for the Irreversible Dimerization and Dimer Coalescence models, which allows soot inception to proceed in cooler downstream regions. These models agree well with experimental data at 8 and 11 L/min but tend to overpredict $N_{agg}$ at 12 L/min.

\begin{figure}[H]
	\centering
	\includegraphics[width=1\textwidth]{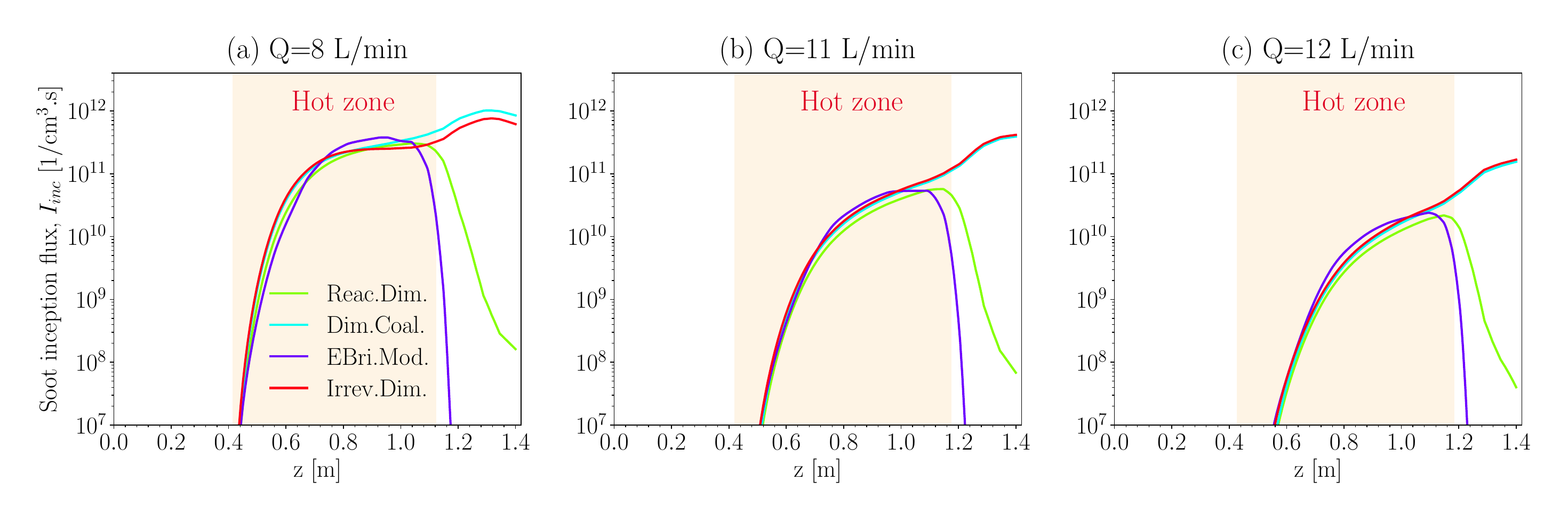}
	\caption{The soot inception flux along the PFR for $\mathrm{Q}=8$ (a), 11 (b), and 12 L/min (c) obtained using KAUST mechanism, SPBM and different inception models. The yellow area represents the hot zone ($\mathrm{T}>0.9\mathrm{T_{max}}$).}
	\label{fig:pfr_Iinc} 
\end{figure}

\begin{figure}[H]
	\centering
	\includegraphics[width=0.8\textwidth]{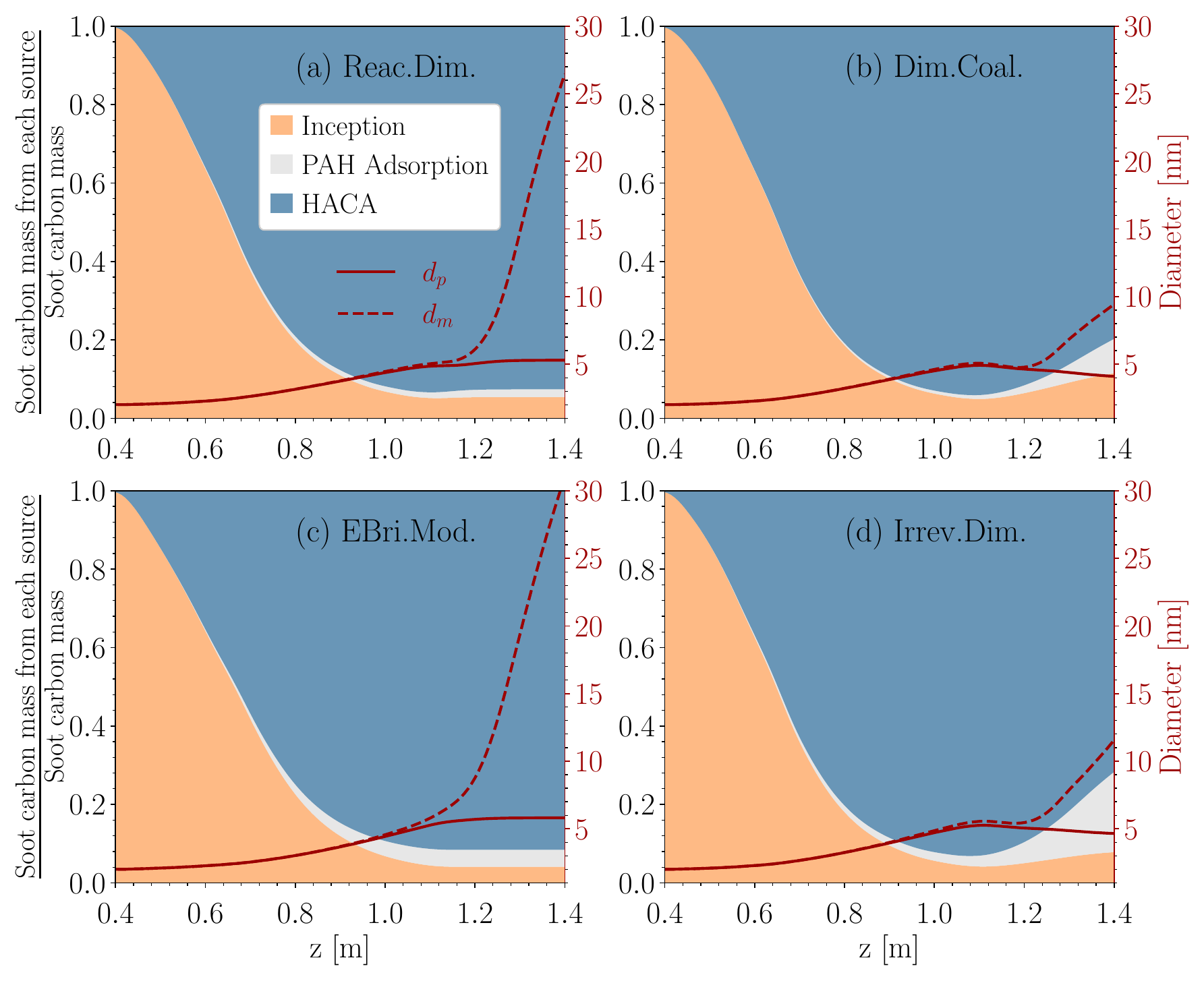}
	\caption{The soot carbon mass from inception, PAH adsorption and HACA normalized by total soot carbon mass for $\mathrm{Q}=8$ L/min along with $d_p$ and $d_m$ obtained using KAUST mechanism, SPBM and different inception models.}
	\label{fig:pfr_cmap} 
\end{figure}

As shown in Figure~\ref{fig:pfr_Iinc}, soot inception flux rises sharply as the gas enters the hot zone. The axial location where the flux exceeds $10^7~\mathrm{cm^{-3}~s^{-1}}$ is similar across all inception models, as it is primarily determined by PAH chemistry. However, this location shifts downstream with increasing flow rate due to reduced residence time. Within the hot zone, differences among the inception models are minimal. Outside the hot zone, Irreversible Dimerization and Dimer Coalescence models maintain high inception flux, while Reactive Dimerization and E-Bridge Modified models predict a rapid decrease of more than three orders of magnitude due to cooling. Figure~\ref{fig:pfr_cmap} demonstrates the contribution of each pathway to total carbon mass of soot along the reactor, which elucidates the link between particle morphology and the balance of inception and surface growth. The relative contribution of inception decreases from 100\% to less than 5\% by $z=1.1$ m for all inception models leading to a gradual increase in $d_p$ and $d_m$. Irreversible Dimerization and Dimer Coalescence models predict an increase in the contribution of inception beyond $z>1.1$~m resulting in the decline of $d_p$ while Reactive Dimerization and E-Bridge Modified models predict a nearly constant $d_p$ close to 5 nm. The final increase in $d_m$ beyond the end of the hot zone near $z = 1.1$~nm is more pronounced for the Reactive Dimerization and E-Bridge Modified models. This is because the inception rate (Figure~\ref{fig:pfr_Iinc}) and surface growth rate (Figure~\ref{fig:pfr_surfacegrowth}) rapidly decrease in these models, and soot morphology becomes primarily governed by coagulation, which conserves $d_p$ and increases $d_m$.


\subsection{Ethylene combustion in a perfectly stirred reactor}
The PSR model of Omnisoot\footnote{\href{https://github.com/mohammadadib-cu/omnisoot-cv/tree/main/examples/psr_pfr}{https://github.com/mohammadadib-cu/omnisoot-cv/tree/main/examples/psr\_pfr}} was used to simulate ethylene combustion at different equivalence ratios, $\phi = 1.9$, 2.0, and 2.1 in a perfectly stirred reactor connected to a flow reactor~\citep{manzello2007soot}. PSD measurements were conducted at the end of the flow reactor using a Nano-Differential Mobility Analyzer (Nano-DMA). The volume and nominal residence time of the PSR are 250~ml and 11~ms, respectively~\citep{manzello2007soot}. The flow reactor is 70~cm long with an inner diameter of 5.1~cm. The reactants enter the adiabatic PSR at 300~K with an inlet mass flow rate of $\dot{m}_{in} = \rho V / \tau$, where $\rho$ was calculated at the reactor temperature of 1723~K~\citep{lenhert2009effects}. The PSR simulation was continued until a steady-state condition was reached for all the solution variables. The final gas composition, temperature, and soot properties from the PSR were used as inlet conditions for the PFR simulations, yielding an average axial velocity of 14.5~m/s in the PFR, which is consistent with the values reported by~\citet{manzello2007soot}.

The KAUST mechanism was used to describe the gas-phase chemistry. The inception and PAH adsorption adjustment factors were varied for each PAH growth model to match the predicted PSD with the measurements as closely as possible for all three equivalence ratios. A similar match could not be reached using the Caltech mechanism because the number concentration of particles was underestimated by at least four orders of magnitude even with $\eta_{inc}=\eta_{ads}=1$. 

\begin{figure}[H]
	\centering
	\begin{tikzpicture}
		\draw (0, 0) node[inner sep=0] 	{\includegraphics[width=1\textwidth]{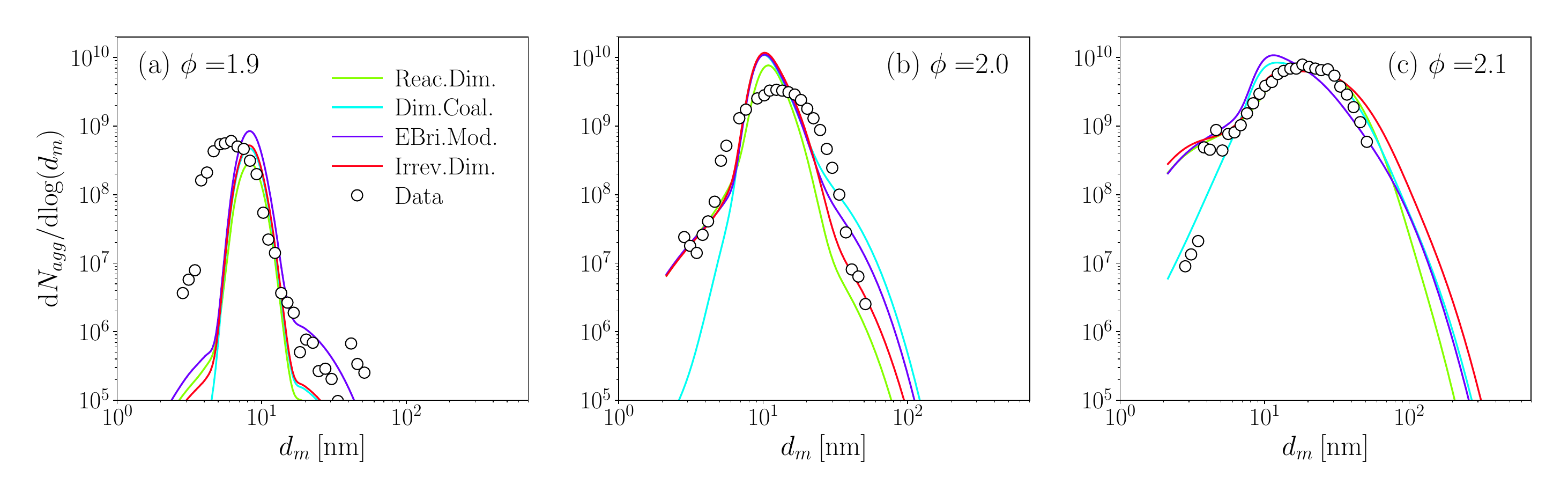}};
		\draw (-3.05, 0.50) node {\tiny{\cite{manzello2007soot}}};
	\end{tikzpicture}
	\caption{The particle size distribution at the end of PFR for $\phi=1.9$ (a), 2.0 (b), and 2.1 (c) obtained using KAUST mechanism, SPBM and different inception models calibrated to match the predictions with the PSD measurements~\citep{manzello2007soot}.}
	\label{fig:psrpfr_psd} 
\end{figure}

As shown in Figure~\ref{fig:psrpfr_psd}, all inception models capture the peak number concentration and the unimodal shape of the PSD, indicating that particle inception has ceased and coagulation is the dominant mechanism for agglomerate size growth. As reported in Table~\ref{tab:psrpfr_morpcomp}, the geometric mean mobility diameter, $d_{m,g}$, and the geometric mobility standard deviation, $\sigma_{m,g}$, obtained using all inception models are in good agreement with the values calculated from the measured PSD~\citep{manzello2007soot}. As shown in Figures~\ref{fig:psrpfr_psd}a and~\ref{fig:psrpfr_psd}b, the spread of the predicted PSD is narrower than that of the measurements for $\phi = 1.9$ and 2.0, corresponding to an underprediction of $\sigma_{m,g}$ for these equivalence ratios.

\begin{figure}[H]
	\centering
	\includegraphics[width=1\textwidth]{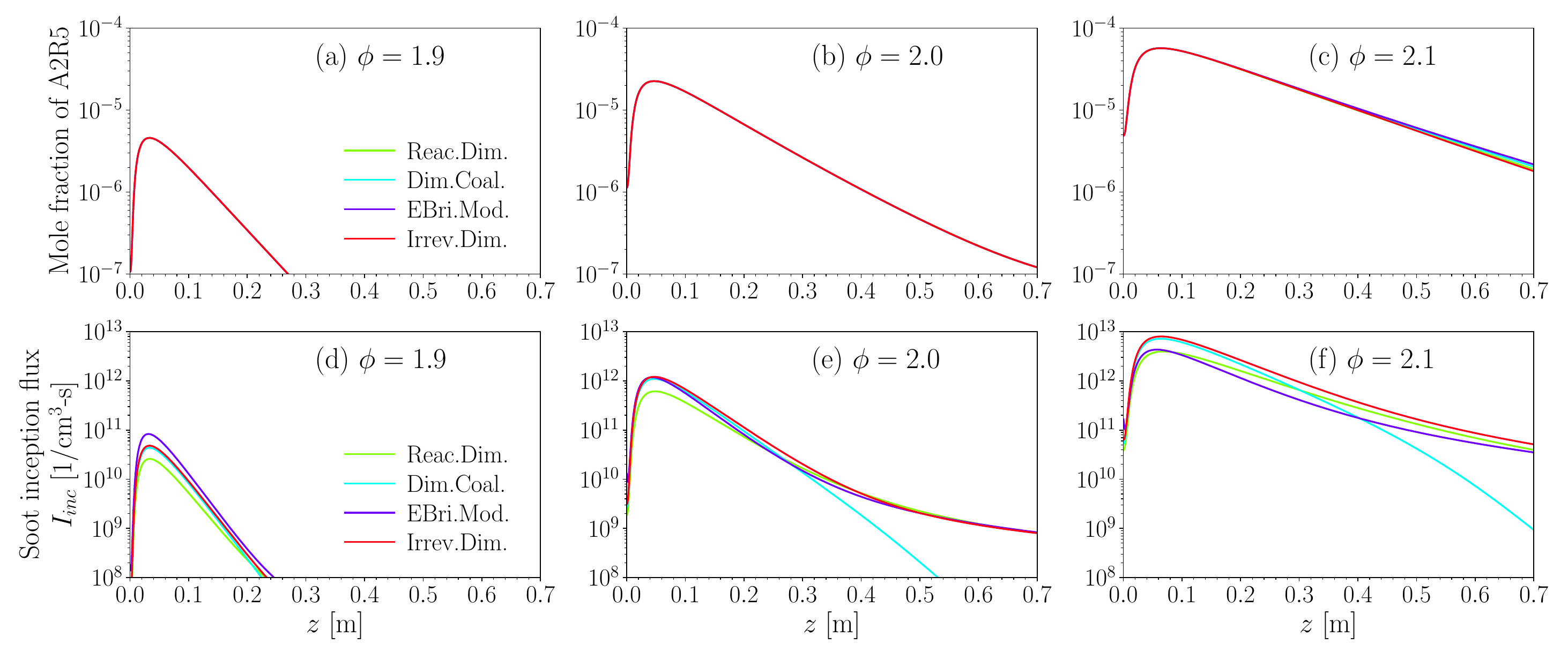}
	\caption{The soot inception flux, $I_{inc}$ and acenaphthylene (A2R5) mole fraction along the PFR for $\phi=1.9$ (a,d), 2.0 (b,e), and 2.1 (c,f) obtained using KAUST mechanism, SPBM and different inception models calibrated to match the predictions with the PSD measurements~\citep{manzello2007soot}.}
	\label{fig:psrpfr_Iinc_PAH} 
\end{figure}

All inception models underpredict the number concentration of small particles ($d_m < 6$~nm) at $\phi = 1.9$. The discrepancy decreases at $\phi = 2.0$, and the model predictions align well with the measurements except for Dimer Coalescence model that underpredicts the number concentration of small particles ($d_m<5$~nm). In the simulation results, the number concentration of the smallest particles with a diameter of 2~nm is strongly affected by the equivalence ratio, as the precursor production rate increases with $\phi$. This behavior can be better understood by comparing the acenaphthylene (A2R5) mole fraction at different $\phi$ values, as shown in Figures~\ref{fig:psrpfr_Iinc_PAH}a-c. The inlet mole fraction of A2R5 at $z = 0$~m corresponds to the value obtained from the PSR, which increases by factors of 10 and 4 when $\phi$ is increased incrementally from 1.9 to 2.0, and from 2.0 to 2.1, respectively. This relative increase is nearly the same across all inception models. As shown in Figures~\ref{fig:psrpfr_Iinc_PAH}d-e, the jump in A2R5 mole fraction leads to a substantial increase in the inception flux, which explains the sensitivity of the smallest particle concentrations to the equivalence ratio. The strong similarity between the A2R5 mole fraction profiles and the soot inception profiles along the reactor highlights the dominant role of A2R5 in soot inception. Another important observation is that the A2R5 mole fraction is not affected by the choice of inception model.

\begin{figure}[H]
	\centering
	\includegraphics[width=1\textwidth]{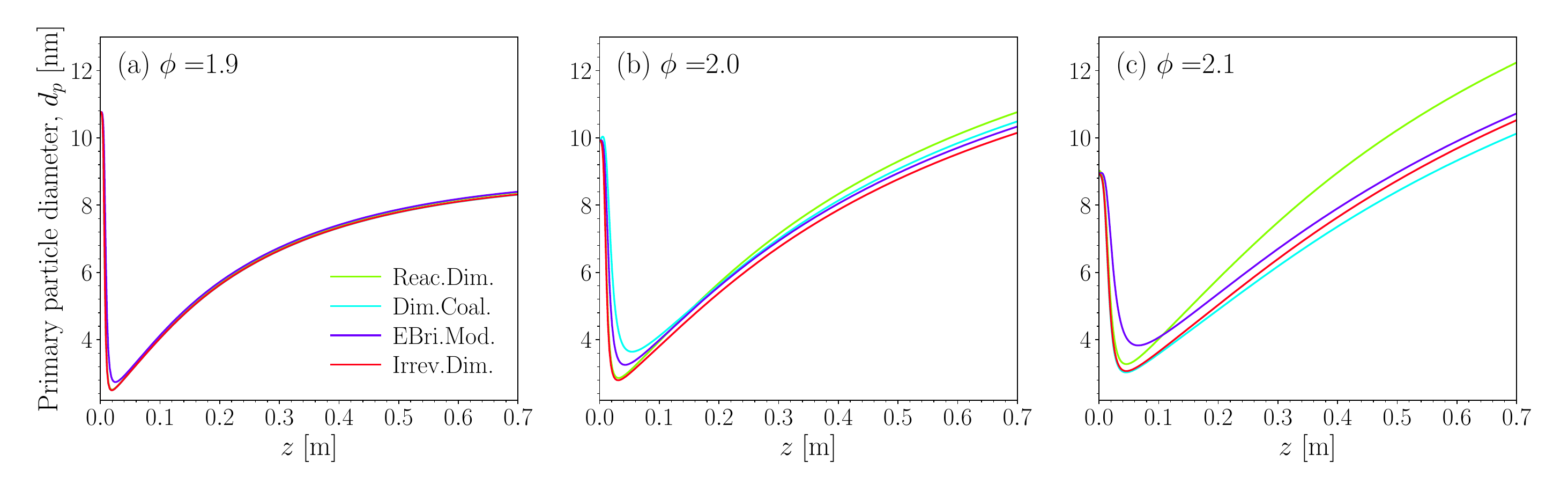}
	\caption{The primary particle diameter, $d_p$, along the PFR for $\phi=1.9$ (a), 2.0 (b), and 2.1 (c) obtained using KAUST mechanism, SPBM and different inception models calibrated to match the predictions with the PSD measurements~\citep{manzello2007soot}.}
	\label{fig:psrpfr_dp} 
\end{figure}

As shown in Figure~\ref{fig:psrpfr_dp}, $d_p$ along the PFR is nearly identical for all inception models at $\phi = 1.9$, but the differences become more noticeable at higher $\phi$ values. The Reactive Dimerization model predicts the largest final $d_p$ because it channels more PAH mass into surface growth rather than inception, which is consistent with the lower peak number concentration observed in Figure~\ref{fig:psrpfr_psd}. The value of $d_p$ decreases rapidly at the beginning of the reactor due to a surge in the production of incipient particles (spherical particles with a diameter of 2~nm), which lowers the mean $d_p$. It then increases toward the end of the reactor as surface growth becomes dominant. Although the final $d_p$ increases slightly with $\phi$, the overall range remains similar across $\phi$ values, indicating a comparable balance between inception and surface growth. 

\begin{figure}[H]
	\centering
	\includegraphics[width=1\textwidth]{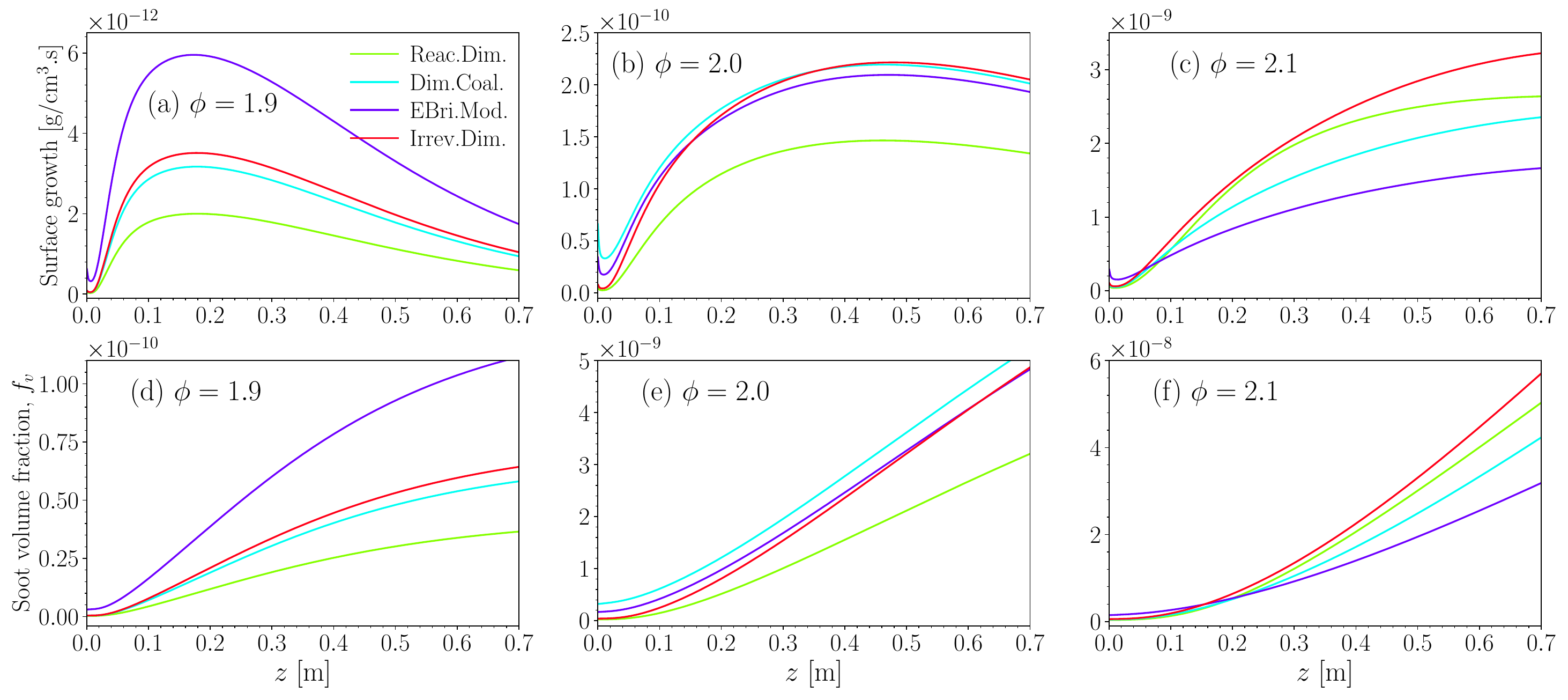}
	\caption{The soot volume fraction, $f_v$, and total surface growth rate by HACA and PAH adsorption along the PFR for $\phi=1.9$ (a, d), 2.0 (b, d), and 2.1 (c, e) obtained using KAUST mechanism and different inception models calibrated to match the predictions with the PSD measurements~\citep{manzello2007soot}.}
	\label{fig:psrpfr_fv} 
\end{figure}

\begin{figure}[H]
	\centering
	\begin{subfigure}[t]{0.43\textwidth}
		\includegraphics[width=1\textwidth]{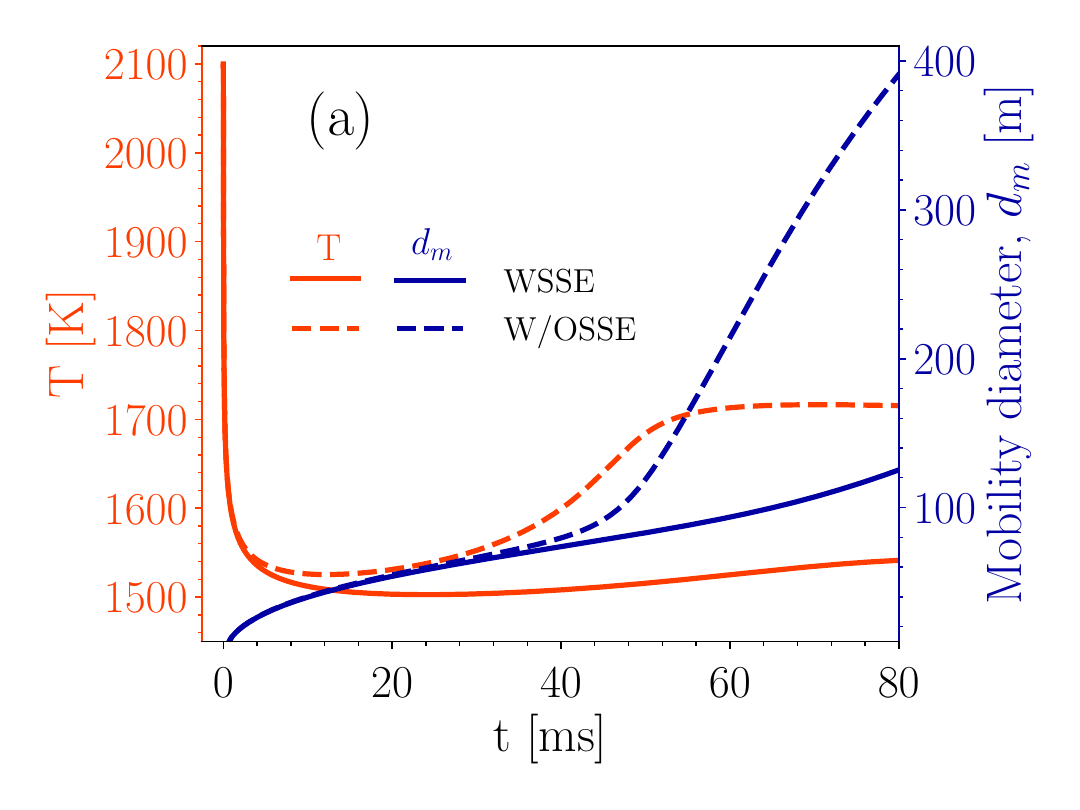}
	\end{subfigure}
	\begin{subfigure}[t]{0.4\textwidth}
		\includegraphics[width=1\textwidth]{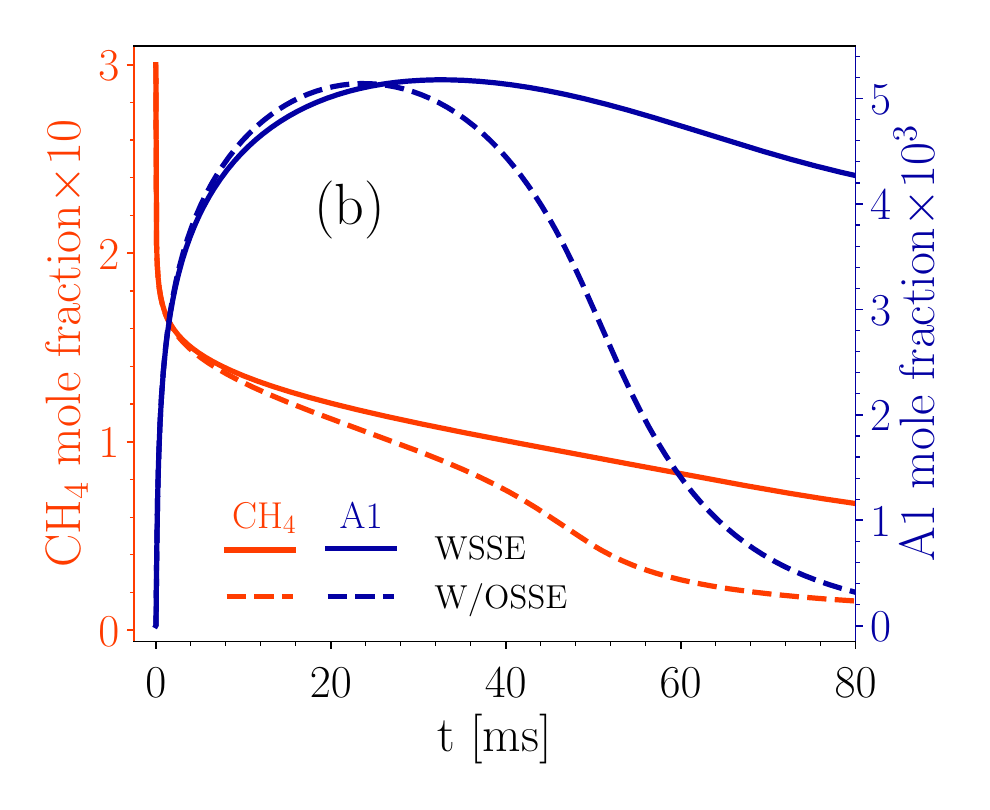}
	\end{subfigure}
	\caption{The comparison of temperature and soot mobility diameter, $d_m$ (a) and the mole fraction of methane, $\mathrm{CH_4}$, and benzene, A1, in the simulation of the pyrolysis of 30\%~$\mathrm{CH_4}$-Ar when soot sensible energy is considered (labeled as ``WSSE") and neglected (labeled as ``W/OSSE"). The CVR was used along with Caltech mechanism~\citep{blanquart2009chemical}, Reactive Dimerization and MPBM.}
	\label{fig:sseeffect}
\end{figure}

As shown in Figure~\ref{fig:psrpfr_fv}, the surface growth rate (HACA and PAH adsorption combined) and the soot volume fraction, however, are highly sensitive to $\phi$, which can be attributed to the strong influence of $\phi$ on precursor production, inception flux, and the number of particles. An analysis of surface growth rates at the end of the PFR revealed that more than 95\% of the soot mass is acquired through the HACA mechanism. The apparent difference in $f_v$ between the inception models originates from the different surface growth rate predicted by the inception models. For example, at $\phi = 1.9$, the E-Bridge Modified model predicts the highest surface growth rate and soot volume fraction due to a higher number of agglomerates (as shown in Figure~\ref{fig:psrpfr_psd}a), which leads to a larger total surface area, which is the key factor in calculating the HACA growth rate (Equation~\ref{eqn:csoot0}).

\section{Conclusion}

This work presents Omnisoot as a robust computational tool built on Cantera to simulate soot formation coupled with gas-phase chemistry across various combustion and pyrolysis scenarios. Omnisoot integrates four reactor models, two particle dynamics models, and four inception and surface growth models, providing a comprehensive platform for analyzing uncertainties in predicting soot yield and morphology originating from chemical kinetics, inception processes, and surface growth rates. The tool was employed to elucidate differences among inception models in shock tubes, flow reactors, and perfectly stirred reactors.

Global adjustment factors were applied to inception and surface growth rates to accurately reproduce experimental data for each target scenario across a range of parameters, including temperature, pressure, composition, and flow rates. The performance of inception models was evaluated based on soot yield, morphology, size distribution, inception and surface growth rates, and carbon mass source analysis. Although different inception models predicted similar carbon yields that closely matched experimental observations, they produced varying morphological characteristics. This highlights the importance of characterizing soot morphology to better constrain inception flux and surface growth rate predictions.

Simulations of ethylene pyrolysis in a flow reactor demonstrated the temperature sensitivity of inception models. Irreversible models (Irreversible Dimerization and Dimer Coalescence) effectively captured the bimodal PSD resulting from high inception flux in the lower-temperature region near the reactor exit. However, the Reactive Dimerization and E-Bridge models generated nearly unimodal distributions due to their strong temperature dependence, causing a significant drop in inception flux in the same region.

The PSD sampled at the outlet of a flow reactor downstream of a perfectly stirred reactor showed good agreement with experimental measurements across three equivalence ratios. Simulations indicated that active inception and surface growth primarily occur at high temperatures within the perfectly stirred reactor and at the entrance region of the flow reactor, far upstream of the sampling location. Under these conditions, the particle size distribution, morphology, and inception flux predictions were largely consistent across all inception models, although notable differences were observed in soot volume fraction due to variations in predicted HACA growth rate.
\section{Acknowledgements}

The research leading to these results has received funding from the Canada Research Chair Program (Grant \#CRC-2019-232527) and the Natural Sciences and Engineering Research Council of Canada (Discovery Grant \#RGPIN-2019-06330 and -Early Career Supplemental Award \#DGECR-2019-00220  and Alliance \#576739-2022), MITACS (\#IT28731).
\clearpage
\bibliography{main.bib}
\clearpage
\section*{Supplementary Material}

\begin{center}
	\vspace*{1cm} 
	\textbf{\LARGE 		Omnisoot: an object-oriented process design package for gas-phase synthesis of carbonaceous nanoparticles} 
\end{center}

\begin{center}
	Mohammad Adib$^{1,*}$, Sina Kazemi$^1$, M. Reza Kholgy$^{1,*}$ \\
	{\small *Corresponding author} \\
	$^1$ Department of Mechanical and Aerospace Engineering, Carleton University, 1125 Colonel By Dr, Ottawa, ON K1S 5B6, Canada
\end{center}

\beginsupplement

\section{The Effect of formation and sensible energy of soot}

\section{Diffusion of soot particles}
\label{sec:diffcoef}
The diffusion coefficient of soot particle, $D^i$, is calculated as:

\begin{equation}
	D^i = \frac{k_B T}{f^i}
	\label{eqn:diff},
\end{equation}
\noindent where $f^i$ is the friction factor of particles in gas and is calculated as:

\begin{equation}
	f^i = \frac{3\pi\mu d^i_m}{C^i(d^i_m)},
	\label{eqn:fraction}
\end{equation}

\noindent where ${C^i}$ is the Cunningham correction factor that accounts for non-continuum effects in the transition and free-molecular regimes. ${C^i}$ is calculated for a given diameter, $d$, as: 
\begin{equation}
	C^i(d) = 1+\frac{2\lambda}{d}
	\left(
	1.21+0.4\cdot\mathrm{exp}(\frac{-0.78d}{\lambda})
	\right)
	\label{eqn:cun},
\end{equation}
\noindent  where $\lambda$ is the mean free path of gas given as:
\begin{equation}
	\lambda = \frac{\mu}{\rho}\sqrt{\frac{\pi \cdot W_{gas}}{2k_B \cdot Av \cdot T}}
	\label{eqn:lambda}.
\end{equation}
Note that $\lambda$ is a property of the gas mixture that does not depend on particle mass and morphology. 

\section{Sectional Population Balance Model}
\label{sec:sectextra}

\subsection{Coagulation Source Term}
\label{sec:sectcoagsource}
Coagulation redistributes the total number of agglomerates and primary particles as well as hydrogen atoms among the sections. The partial coagulation source terms for ${N^i_{agg}}$, ${N^i_{pri}}$ and ${H^i_{tot}}$ can be calculated as:

\begin{equation}
	\left(S^i_{N_{agg}}\right)_{coag}
	=
	\sum_{k=1}^{n_{sec}}\sum_{j=k}^{n_{sec}}
	\left(
	1-\frac{\delta_{jk}}{2}
	\right)
	\eta_{ijk}\zeta^{jk}\beta^{jk}N^j_{agg}N^k_{agg}
	-
	N^i_{agg}
	\sum_{m=1}^{n_{sec}}\zeta^{im}\beta^{im}N^m_{agg},
	\label{eqn:IcoagNaggsect}
\end{equation}

\begin{equation}
	\left(S^i_{N_{pri}}\right)_{coag}
	=
	\sum_{k=1}^{n_{sec}}\sum_{j=k}^{n_{sec}}
	\left(
	1-\frac{\delta_{jk}}{2}
	\right)
	\eta_{p,ijk}\eta_{ijk}\zeta^{jk}\beta^{jk}N^j_{agg}N^k_{agg}
	-
	N^i_{pri}
	\sum_{m=1}^{n_{sec}}\zeta^{im}\beta^{im}N^m_{agg},
	\label{eqn:IcoagNprisect}
\end{equation}

\begin{equation}
	\left(S^i_{H_{tot}}\right)_{coag}
	=
	\sum_{k=1}^{n_{sec}}\sum_{j=k}^{n_{sec}}
	\left(
	1-\frac{\delta_{jk}}{2}
	\right)
	\eta_{h,ijk}\eta_{ijk}\zeta^{jk}\beta^{jk}N^j_{agg}N^k_{agg}
	-
	H^i_{tot}
	\sum_{m=1}^{n_{sec}}\zeta^{im}\beta^{im}N^m_{agg}.
	\label{eqn:IcoagHtotsect}
\end{equation}
\noindent where ${\delta_{jk}}$ is the Kronecker delta defined as:

\begin{equation}
	\delta_{jk}=
	\left\{
	\begin{array}{lr}
		1, & \text{if } j = k\\
		0. & \text{if } j \neq k
	\end{array}
	\right.
	\label{eqn:deltakronecker}
\end{equation}

The collision frequency between sections $j$ and $k$ ($\beta^{jk}$) can be obtained from the harmonic mean of the values in the continuum ($\beta_{cont}^{jk}$) and free molecular ($\beta_{fm}^{jk}$) regimes as:

\begin{equation}
	\beta^{jk} = 				       \frac{\beta^{jk}_{fm}\cdot\beta^{jk}_{cont}}{\beta^{jk}_{fm}
		+\beta^{jk}_{cont}}
	\label{eqn:betahmsect},
\end{equation}

\begin{equation}
	\beta^{jk}_{fm} =
	\sqrt{
		\frac{\pi k_B T}{2}
		\left(
		\frac{1}{m^j_{agg}}+
		\frac{1}{m^k_{agg}}
		\right)
	} 
	\left(
	d^j_c+d^k_c
	\right)^2
	\label{eqn:betafmsect},
\end{equation}
\begin{equation}
	\beta^{ij}_{cont} = \frac{2k_BT}{3\mu}
	\left(
	\frac{C^j}{d^j_m}+
	\frac{C^k}{d^k_m}
	\right)
	\left(
	d^j_c+d^k_c
	\right)^2
	\label{eqn:betacontsect}.
\end{equation}

The collision frequency can also be determined from the Fuchs interpolation as:

\begin{equation}
	\beta^{jk}=
	\beta^{ij}_{cont}
	\left[
	\frac{d^j_c+d^k_c}{d^j_c+d^k_c+2+\delta^{jk}_r}+
	\frac{8\left(D^j+D^k\right)}
	{\bar{c}^{jk}_r\left(d^j_c+d^k_c\right)}
	\right]^{-1},
	\label{eqn:betafuchssect}
\end{equation}
\noindent where ${\delta^{jk}_r}$ and ${\bar{c}^{jk}_r}$ are the mean square root of mean distance and velocity of particles, respectively, and they are obtained as:

\begin{equation}
	\delta^{jk}_r=
	\sqrt{
		{\delta^j_a}^2+{\delta^k_a}^2
	}
	\label{eqn:sqrtmeandist},
\end{equation}

\begin{equation}
	\bar{c}^{jk}_r=
	\sqrt{
		{c^j}^2+{c^k}^2
	}
	\label{eqn:sqrtmeanvel}.
\end{equation}

The mean velocity, ${c^i}$, and mean stop distance of particles, ${\lambda^i_a}$, can be calculated as:

\begin{equation}
	c^i = \sqrt{\frac{8k_B T}{\pi m^i_{agg}}},
	\label{eqn:meanvel}
\end{equation}

\begin{equation}
	\delta^i_a=\frac{1}{d^i_c\lambda^i_a}
	\left[
	\left(
	d^i_c+\lambda^i_a
	\right)^3
	-\left(
	{d^i_c}^2+{\lambda^i_a}^2
	\right)^{3 / 2}
	\right]
	-d^i_{c},
	\label{eqn:meandist}
\end{equation}
\noindent $\lambda^i_a $is the agglomerate stopping distance defined as:
\begin{equation}
	\lambda^i_a = \frac{8D^i}{\pi c^i}
	\label{eqn:stopdist}.
\end{equation}

In Equation~\eqref{eqn:IcoagNaggsect}, $\mathrm{\eta_{ijk}}$ assigns newly formed agglomerates to the two consecutive sections in order to conserve mass during coagulation~\citep{park2005aerosol}.

\begin{equation}
	\eta_{ijk}=
	\left\{
	\begin{aligned}
		&\frac{C^{i+1}_{agg}-C^{jk}_{agg}}{C^{i+1}_{agg}-C^i_{agg}},
		&&
		\text{if } C^i_{agg} \le C^{jk}_{agg} < C^{i+1}_{agg}
		\\
		&\frac{C^{i}_{agg}-C^{jk}_{agg}}{C^{i}_{agg}-C^{i-1}_{agg}}, 
		&&
		\text{if } C^{i-1}_{agg} \le C^{jk}_{agg} < C^{i}_{agg}
		\\
		&0,
		&&\text{else}
	\end{aligned}
	\right.
	\label{eqn:etacoag}
\end{equation}
\noindent where ${C^{jk}_{agg}=C^{j}_{agg}+C^{k}_{agg}}$. Similarly, $\eta_{p,ijk}$ in Equation~\eqref{eqn:IcoagNprisect} and $\eta_{h,ijk}$ in Equation~\eqref{eqn:IcoagHtotsect} adjust the number of primary particles and hydrogen atoms added to consecutive sections based on their mass, respectively.

\begin{equation}
	\eta_{p,ijk}=
	\frac{C^i_{agg}}{C^{jk}_{agg}}
	\left(
	n^j_p + n^k_p
	\right),
	\label{eqn:etapcoag}
\end{equation}

\begin{equation}
	\eta_{h,ijk}=
	\frac{C^i_{agg}}{C^{jk}_{agg}}
	\left(
	H^j_{agg} + H^k_{agg}
	\right).
	\label{eqn:etahcoag}
\end{equation}

In Equation~\eqref{eqn:IcoagNaggsect}, $\zeta^{jk}$ is the coagulation efficiency of soot particles in sections $j$ and $k$. A value of $\zeta^{jk} = 1$ indicates that every collision between two soot particles successfully results in the formation of a new agglomerate. However, numerical models~\citep{narsimhan1985brownian} and experimental evidence~\citep{d2005surface} have shown that coagulation efficiency drastically decreases for particles smaller than 10~nm in the free-molecular regime ($\mathrm{Kn}\gg10$), due to their high kinetic energy exceeding the magnitude of attractive forces~\citep{wang1991filtration}. The coagulation efficiency between two colliding particles can be described by~\citep{narsimhan1985brownian}:

\begin{equation}
	\zeta^{ij} = 1 - 
	\left(1 + \frac{\Phi^{ij}_0}{k_BT} \right)
	\mathrm{exp}\left(-\frac{\Phi^{ij}_0}{k_BT}\right),
	\label{eqn:coageff}
\end{equation}

\noindent where $\Phi_0$ is the potential well depth, i.e., the minimum interaction energy between two colliding particles. \citet{hou2020coagulation} calculated $\Phi_0$ for soot particles ranging from 1 to 15 nm by considering the attractive and repulsive interactions between constituent carbon and hydrogen atoms, and proposed an equation based on the reduced diameter, $d^{jk}_r$, of colliding particles as:

\begin{equation}
	\Phi^{ij}_0 = -6.6891\times10^{-23} (d^{jk}_r)^3 + 1.1244\times10^{-21} (d^{jk}_r)^2 + 1.1394\times10^{-20} d^{jk}_r - 5.5373\times10^{-21},
	\label{eqn:coageffphi}
\end{equation}

\begin{equation}
	d^{jk}_r = \frac{d^i_c\cdot d^j_c}{d^i_c+d^j_c}.
	\label{eqn:coageffredcueddia}
\end{equation}

Equation~\eqref{eqn:coageffphi} is valid for $d^{jk}_r$ between 1 and 7~nm, and $\zeta^{ij}$ is assumed as unity for particles with a reduced diameter larger than 7~nm.

\subsection{Other Source terms}
\label{sec:sectothersource}

Inception introduces equal number of agglomerates and primary particles to the first section.

\begin{equation}
	\begin{aligned}
		\left(S^i_{N_{agg}}\right)_{inc} =
		&\frac{1}{Av}\frac{I_{N, inc}}{C^i_{agg}}, && i=1.
	\end{aligned}
	\label{eqn:S_Nagg_incsect}
\end{equation}

\begin{equation}
	\begin{aligned}
		\left(S^i_{N_{pri}}\right)_{inc} =
		&\frac{1}{Av}\frac{I_{N, inc}}{C^i_{agg}}, && i=1.
	\end{aligned}
	\label{eqn:S_Npri_incsect}
\end{equation}

\begin{equation}
	\begin{aligned}
		\left(S^i_{H_{tot}}\right)_{inc} =
		&I_{H, inc}, && i=1.
	\end{aligned}
	\label{eqn:S_Htot_incsect}
\end{equation}
Surface growth and PAH adsorption increase both the carbon mass and hydrogen content of agglomerates, transferring them to higher sections. The rate at which agglomerates are removed from the original section and added to the target section is calculated to ensure mass conservation. Specifically, it is determined by dividing the mass growth rate by the difference in mass between adjacent sections as:

\begin{equation}
	\left(S^i_{N_{agg}}\right)_{haca, ads}=
	\frac{1}{Av}
	\left\{
	\begin{aligned}
		&-\frac{I^i_{C_{tot},haca}+I^i_{C_{tot},ads}}{C^{i+1}_{agg}-C^{i}_{agg}},
		&&
		\text{if } i = 1
		\\
		&\frac{I^{i-1}_{C_{tot},haca}+I^{i-1}_{C_{tot},haca}}{C^{i}_{agg}-C^{i-1}_{agg}}
		-\frac{I^{i}_{C_{tot},haca}+I^{i}_{C_{tot},ads}}{C^{i+1}_{agg}-C^{i}_{agg}},
		&&
		\text{if } 1 < i < n_{sec}
		\\
		&\frac{I^{i-1}_{C_{tot},haca}+I^{i-1}_{C_{tot},ads}}{C^{i}_{agg}-C^{i-1}_{agg}}.
		&&\text{if } i=n_{sec}
	\end{aligned}
	\right.
	\label{eqn:S_Nagg_gradssect}
\end{equation}

As agglomerates move up/down through sections, they carry the number of primary particles as well as hydrogen atoms, so the transfer rate of agglomerates is multiplied by ${n^i_p}$ and ${H^i_{agg}}$, respectively. 

\begin{equation}
	\left(S^i_{N_{pri}}\right)_{haca, ads}=
	\frac{1}{Av}
	\left\{
	\begin{aligned}
		&-\frac{I^i_{C_{tot},haca}+I^i_{C_{tot},ads}}{C^{i+1}_{agg}-C^{i}_{agg}},
		&&
		\text{if } i = 1
		\\
		&\frac{I^{i-1}_{C_{tot},haca}+I^{i-1}_{C_{tot},ads}}{C^{i}_{agg}-C^{i-1}_{agg}}n^{i-1}_p
		-\frac{I^{i}_{C_{tot},haca}+I^{i}_{C_{tot},ads}}{C^{i+1}_{agg}-C^{i}_{agg}}n^{i}_p,
		&&
		\text{if } 1 < i < n_{sec}
		\\
		&\frac{I^{i-1}_{C_{tot},haca}+I^{i-1}_{C_{tot},ads}}{C^{i}_{agg}-C^{i-1}_{agg}}n^{i-1}_p,
		&&\text{if } i=n_{sec}
	\end{aligned}
	\right.
	\label{eqn:S_Npri_gradssect}
\end{equation}

\begin{equation}
	\left(S^i_{H_{tot}}\right)_{haca, ads}=
	\frac{1}{Av}
	\left\{
	\begin{aligned}
		&-\frac{I^i_{C_{tot},haca}+I^i_{C_{tot},ads}}{C^{i+1}_{agg}-C^{i}_{agg}}H^{i}_{agg} 
		+ I^{i}_{H_{tot}, haca} + I^{i}_{H_{tot}, ads},
		&&
		\text{if } i = 1
		\\
		&\frac{I^{i-1}_{C_{tot},haca}+I^{i-1}_{C_{tot},ads}}{C^{i}_{agg}-C^{i-1}_{agg}}H^{i-1}_{agg}
		-\frac{I^{i}_{C_{tot},haca}+I^{i}_{C_{tot},ads}}{C^{i+1}_{agg}-C^{i}_{agg}}H^{i}_{agg}
		+ I^{i}_{H_{tot}, haca} + I^{i}_{H_{tot}, ads},
		&&
		\text{if } 1 < i < n_{sec}
		\\
		&\frac{I^{i-1}_{C_{tot},haca}+I^{i-1}_{C_{tot},ads}}{C^{i}_{agg}-C^{i-1}_{agg}}H^{i-1}_{agg}
		+ I^{i}_{H_{tot}, haca} + I^{i}_{H_{tot}, ads}.
		&&\text{if } i=n_{sec}
	\end{aligned}
	\right.
	\label{eqn:S_Htot_gradssect}
\end{equation}

Similarly, the agglomerates lose carbon mass by oxidation, and descend to the lower sections carrying primary particle and hydrogen.

\begin{equation}
	\left(S^i_{N_{agg}}\right)_{ox}=
	\frac{1}{Av}
	\left\{
	\begin{aligned}
		&\frac{I^{i+1}_{C_{tot},ox}}{C^{i+1}_{agg}-C^{i}_{agg}}
		-
		\frac{I^{i}_{C_{tot},ox}}{C^{i}_{agg}},
		&&
		\text{if } i = 1
		\\
		&\frac{I^{i+1}_{C_{tot},ox}}{C^{i+1}_{agg}-C^{i}_{agg}}
		-
		\frac{I^{i}_{C_{tot},ox}}{C^{i}_{agg}-C^{i-1}_{agg}},
		&&
		\text{if } 1 < i < n_{sec}
		\\
		&
		-
		\frac{I^{i}_{C_{tot},ox}}{C^{i}_{agg}-C^{i-1}_{agg}},
		&&\text{if } i=n_{sec}
	\end{aligned}
	\right.
	\label{eqn:S_Nagg_oxsect}
\end{equation}

\begin{equation}
	\left(S^i_{N_{pri}}\right)_{ox}=
	\frac{1}{Av}
	\left\{
	\begin{aligned}
		&\frac{I^{i+1}_{C_{tot},ox}}{C^{i+1}_{agg}-C^{i}_{agg}}n^{i+1}_p
		-
		\frac{I^{i}_{C_{tot},ox}}{C^{i}_{agg}},
		&&
		\text{if } i = 1
		\\
		&\frac{I^{i+1}_{C_{tot},ox}}{C^{i+1}_{agg}-C^{i}_{agg}}n^{i+1}_p
		-
		\frac{I^{i}_{C_{tot},ox}}{C^{i}_{agg}-C^{i-1}_{agg}}n^{i}_p,
		&&
		\text{if } 1 < i < n_{sec}
		\\
		&
		-
		\frac{I^{i}_{C_{tot},ox}}{C^{i}_{agg}-C^{i-1}_{agg}}n^{i}_p,
		&&\text{if } i=n_{sec}
	\end{aligned}
	\right.
	\label{eqn:S_Npri_oxsect}
\end{equation}

\begin{equation}
	\left(S^i_{H_{tot}}\right)_{ox}=
	\frac{1}{Av}
	\left\{
	\begin{aligned}
		&\frac{I^{i+1}_{C_{tot},ox}}{C^{i+1}_{agg}-C^{i}_{agg}}H^{i+1}_{agg}
		-
		\frac{I^{i}_{C_{tot},ox}}{C^{i}_{agg}}H^{i}_{agg}
		+ I^{i}_{H_{tot}, ox},
		&&
		\text{if } i = 1
		\\
		&\frac{I^{i+1}_{C_{tot},ox}}{C^{i+1}_{agg}-C^{i}_{agg}}H^{i+1}_{agg}
		-
		\frac{I^{i}_{C_{tot},ox}}{C^{i}_{agg}-C^{i-1}_{agg}}H^{i}_{agg}
		+ I^{i}_{H_{tot}, ox},
		&&
		\text{if } 1 < i < n_{sec}
		\\
		&
		-
		\frac{I^{i}_{C_{tot},ox}}{C^{i}_{agg}-C^{i-1}_{agg}}H^{i}_{agg}
		+ I^{i}_{H_{tot}, ox}.
		&&\text{if } i=n_{sec}
	\end{aligned}
	\right.
	\label{eqn:S_Htot_oxsect}
\end{equation}

\section{PAH growth models}

\subsection{Irreversible Dimerization}
\label{sec:irrevdim}

The Irreversible Dimerization is based on the collision of a pair of PAH molecules forming a dimer. The sequential growth continues leading formation of trimers and tetramers until the PAH cluster mass reaches a threshold that can be considered a solid particle. For practical purposes, a dimer is usually considered as an incipient particle that grows by surface growth and coagulation. A single-step collision of two similar PAHs forms a new dimer as:

\reaction[react:irrevdiminc]{
	$\mathrm{PAH}_j$ + $\mathrm{PAH}_j$ ->[$k_{f,dim_j}$] $\mathrm{Dimer}_j$
}
Similarly, the adsorption of each PAH molecule on soot particles is described by the irreversible collision of soot and $\mathrm{PAH}_j$ as:
\reaction[react:irrevdimads]{
	$\mathrm{PAH}_j$ + Soot ->[$k_{f,ads_j}$] Soot-$\mathrm{PAH}_j$
}
The forward rate of dimerization, ${k_{f,dim_j}}$, and adsorption, $k_{f,ads_j}$, in Reactions~\eqref{react:irrevdiminc} and \eqref{react:irrevdimads} are calculated as:

\begin{equation}
	k_{f,dim_j}=
	\gamma_{inc}\cdot\beta_{jj,PAH}\cdot Av
	\label{eqn:kfdim},
\end{equation}

\begin{equation}
	k^i_{f,ads_j}=
	\gamma_{ads_j}\cdot\beta^i_{j,ads}\cdot Av
	\label{eqn:kfads},
\end{equation}

\noindent where $\beta_{jk,\mathrm{PAH}}$ and $\beta^i_{j,\mathrm{ads}}$ are computed using Equations~\eqref{eqn:betadim} and~\eqref{eqn:betahmads}, respectively, and $\gamma_{\mathrm{inc}}$ and $\gamma_{\mathrm{ads}}$ are the collision efficiencies for dimerization and adsorption, respectively. Their values range from $\mathrm{10^{-7}}$ to 1 and are typically chosen to match the predicted soot properties with experimental data. The dimerization rate of $\mathrm{PAH}_j$ is then calculated as:

\begin{equation}
	\omega_{dim_j} = \eta_{inc} k_{f,dim_{j}} [\mathrm{PAH}_j] [\mathrm{PAH}_j].
	\label{eqn:irrevdim_wdim}
\end{equation}

The partial source terms of inception are calculated as:

\begin{equation}
	I_{N,inc} =\frac{1}{\rho} \sum_{j=1}^{n_{PAH}} 2\omega_{dim_j} n_{PAH_j,C},
	\label{eqn:INinc}
\end{equation}
\begin{equation}
	I_{C_{tot},inc} = \frac{1}{\rho}\sum_{j=1}^{n_{PAH}} 2\omega_{dim_j} n_{PAH_j,C}.
	\label{eqn:ICtotinc}
\end{equation}
\begin{equation}
	I_{H_{tot},inc} =\frac{1}{\rho} \sum_{j=1}^{n_{PAH}} 2\omega_{dim_j} n_{PAH_j,H}.
	\label{eqn:IHtotinc}
\end{equation}
The rate of PAH adsorption for each section is obtained as:
\begin{equation}
	\omega^i_{ads_j} = \eta_{ads} k^i_{f,ads_{j}} [\mathrm{soot}^i] [\mathrm{PAH}_j].
	\label{eqn:adsrate_irrevdim}
\end{equation}

The contribution of PAH adsorption to the source terms are expressed as:

\begin{equation}
	I^i_{C_{tot},ads} = \frac{1}{\rho}\sum_{j=1}^{n_{PAH_j}} \omega^i_{ads_j} n_{PAH_j,C},
	\label{eqn:ICtotads}
\end{equation}
\begin{equation}
	I^i_{H_{tot},ads} =\frac{1}{\rho} \sum_{j=1}^{n_{PAH_j}} \omega^i_{ads_j} (n_{PAH_j,H}-2).
	\label{eqn:IHtotads}
\end{equation}

Note that PAH adsorption is a mass growth phenomenon that only changes $C_{tot}$ and $H_{tot}$ but does not affect number of $N_{agg}$ and $N_{pri}$. Each PAH molecule loses one H atom becoming a radical that forms bonds with a dehydrogenated site on soot surface, so two H atoms are released during adsorption that is taken into account in Equation~\eqref{eqn:IHtotads}.

The formation of a dimer consumes two PAH molecules, and during adsorption one PAH molecule is removed from the gas mixture, so the total rate of removal of $\mathrm{PAH}_j$ by Irreversible Dimerization is obtained as:

\begin{equation}
	\left(
	\frac{\mathrm{d}\left[{\mathrm{PAH}_j}\right]}{\mathrm{d}t}
	\right)_{inc}
	= 
	-2\omega_{dim_j},
	\label{eqn:PAHscrub_irrevdim_inc}
\end{equation}

\begin{equation}
	\left(
	\frac{\mathrm{d}\left[{\mathrm{PAH}_j}\right]}{\mathrm{d}t}
	\right)_{inc}
	= 
	-\sum_{i=1}^{n_{sec}}\omega^i_{ads_j}.
	\label{eqn:PAHscrub_irrevdim_ads}
\end{equation}

Moreover, one $\mathrm{H_2}$ is released to the gas mixture as a result of the adsorption process.
\begin{equation}
	\left(
	\frac{\mathrm{d}\left[{\mathrm{H_2}}\right]}{\mathrm{d}t}
	\right)_{ads}
	= 
	\sum_{i=1}^{n_{sec}}\omega^i_{ads_j}
	\label{eqn:H2scrub_irrevdim}.
\end{equation}

\subsection{Reactive Dimerization}
\label{sec:reacvdim}

This model was built on Irreversible Dimerization with a main difference: The first step of dimerization and adsorption is reversible forming physically bonded dimers followed by a irreversible carbonization that leads to chemical bond formation in dimers \citep{kholghy2018reactive}. This approach allows the formation of homo- and hetero-dimers. The dimerization of $\mathrm{PAH}_j$ and $\mathrm{PAH}_k$ is described as:
\reaction[reac:phydim_reacdim]{
	PAH_j + PAH_k <-->[$k_{f,dim_{jk}}$][$k_{r,dim_{jk}}$] Dimer^*_{$ij$}
}
\reaction[reac:chemdim_reacdim]{
	Dimer^*_{$jk$} ->[$k_{reac}$] Dimer_{$jk$}
}
\noindent where $\mathrm{Dimer^*}_{jk}$ and $\mathrm{Dimer}_{jk}$ are physically and chemically bonded dimers, respectively, from $\mathrm{PAH}_j$ and $\mathrm{PAH}_k$. The forward rate of physical dimerization, ${k_{f,dim_{jk}}}$, is calculated as:

\begin{equation}
	k_{f,dim_{jk}}=
	p^{''}\cdot\beta_{jk,PAH}\cdot Av
	\label{eqn:kfphydim_reacdim},
\end{equation}

\noindent where $\beta_{jk,\mathrm{PAH}}$ is calculated using Equation~\eqref{eqn:betadim}, and $p^{\prime\prime} = 0.1$ accounts for the probability of PAH–PAH collisions occurring in the "FACE" configuration that result in successful van der Waals (vdW) bond formation~\citep{miller1984intermolecular}. The reverse rate of physical dimerization, $k_{r,\mathrm{dim}_{jk}}$, is obtained from the dimerization equilibrium constant~\citep{miller1991kinetics} as:

\begin{equation}
	k_{r,dim_{jk}} = \frac{k_{f,dim_{jk}}}{K_{eq}} 
	\label{eqn:krphydim_reacdim},
\end{equation}

\begin{equation}
	\mathrm{log}_{10}K_{eq}=
	a\frac{\epsilon_{jk}}{RT}+b
	\label{eqn:keq_reacdim},
\end{equation}

\begin{equation}
	\epsilon_{jk} = cW_{jk} -d
	\label{eqn:epsilon_reacdim},
\end{equation}

\begin{equation}
	W_{jk} = \frac{W_j\cdot W_k}{W_j+W_k}
	\label{eqn:Wjk_reacdim},
\end{equation}
\noindent where $a=0.115$ (obtained from pyrere dimerization data~\citep{sabbah2010exploring}) and $b=1.8$~\citep{kholghy2018reactive}, $c=933420$~J/kg, and $d=34053$~J/mol~\cite{kholghy2018reactive}. 

The rate constant of chemical bond formation, $k_{reac}$, is defined in the Arrhenius form~\cite{naseri2022simulating} as
\begin{equation}
	k_{reac} = 5\times10^6\cdot e^{(-96232/RT)}
	\label{eqn:kc_reacdim}.
\end{equation}

Assuming a steady state condition for physical dimers, i.e., $\mathrm{\partial [Dimer^*_{jk}]/\partial t=0}$, the rate of formation of chemically-bonded dimers can be obtained as:

\begin{equation}
	\omega_{dim_{jk}} = k_{reac}\frac{k_{f,dim_{jk}}[\mathrm{PAH}_j][\mathrm{PAH}_k]}
	{k_{r,dim_{jk}}+k_{reac}}
	\label{eqn:chemdimer_reacdim}.
\end{equation}

The contribution of dimer formation to the partial source terms is evaluated by looping over all combinations of PAH precursors as:

\begin{equation}
	I_{N,{inc}} = 
	\frac{1}{\rho}
	\sum_{j=1}^{n_{PAH}} \sum_{k=j}^{n_{PAH}}  \omega_{dim_{kj}} 
	\left(
	n_{PAH_j,C}+n_{PAH_k,C}
	\right),
	\label{eqn:IN_inc}
\end{equation}

\begin{equation}
	I_{C_{tot},{inc}} = 
	\frac{1}{\rho}
	\sum_{j=1}^{n_{PAH}} \sum_{k=j}^{n_{PAH}}  \omega_{dim_{kj}} 
	\left(
	n_{PAH_j,C}+n_{PAH_k,C}
	\right),
	\label{eqn:ICtot_inc}
\end{equation}

\begin{equation}
	I_{H_{tot},{inc}} = 
	\frac{1}{\rho}
	\sum_{j=1}^{n_{PAH}} \sum_{k=j}^{n_{PAH}}  \omega_{dim_{kj}} 
	\left(
	n_{PAH_j,H}+n_{PAH_k,H}
	\right).
	\label{eqn:IHtot_inc}
\end{equation}

Similarly, PAH adsorption is described by a two-step process where the collision of $\mathrm{PAH}_j$ with soot agglomerates leads to a physically bonded Soot$-$$\mathrm{PAH}^*_j$ that is carbonized and forms a chemically bonded Soot$-$$\mathrm{PAH}_j$ on soot surface.

\reaction[reac:physootPAH_reacdim]{
	PAH_j + Soot <-->[$k_{f,ads_j}$][$k_{r,ads_j}$] Soot-PAH^*_j
}

\reaction[reac:chemsootPAH_reacdim]{
	Soot-PAH^*_j ->[$k_{c,ads}$] Soot-PAH_j
}

The forward and reverse rate of PAH-soot collision are calculated as:

\begin{equation}
	k^i_{f,ads_j}=\beta^i_{j,ads}\cdot Av,
	\label{eqn:kfads_reacdim}
\end{equation}

\begin{equation}
	k^i_{r,ads_j}=k^i_{f,ads_j}\cdot10^{-b}e^{-a\epsilon^i_{soot,j} \mathrm{ln}(10)/(RT)},
	\label{eqn:krads_reacdim}
\end{equation}

\begin{equation}
	\epsilon^i_{soot,j} = cW^i_{soot,j} - d,
	\label{eqn:epsilonads_reacdim}
\end{equation}

\noindent where $\beta^i_{j,ads}$ is calculated using Equation~\eqref{eqn:betahmads}. The values of $a$, $b$, $c$, and $d$ are the same as those explained in the inception part. Computing ${\epsilon^i_{soot,j}}$ also requires the equivalent soot molecular weight, ${W^i_{soot}}$, which is estimated from carbon mass of each agglomerate as:

\begin{equation}
	W^i_{soot}=\frac{C^i_{tot}W_{carbon}}{N^i_{agg}}.
\end{equation}

The rate constant of carbonization of Soot$-$$\mathrm{PAH}^*_j$ is defined as in an Arrhenius form similar to the inception formulation (Equation~\eqref{eqn:kc_reacdim}). The pre-exponential factor is adjusted by matching the numercial PSD~\citep{naseri2022simulating} with measurements in the ethylene pyrolysis in a flow reactor~\cite{araki2021effects}. 

\begin{equation}
	k_{c,ads} = 2\times10^{10}\cdot e^{(-96232/RT)}
	\label{eqn:kcads_reacdim}.
\end{equation}

The total adsorption rate can be calculated assuming a steady-state concentration for the physically adsorbed PAH on soot, i.e., $\partial$[Soot$-\mathrm{PAH}^*_j$]/$\partial t=0$, calculated in a similar way to the inception flux (Equation~\eqref{eqn:chemdimer_reacdim}) as:

\begin{equation}
	\omega^i_{ads_j} = k_{c,ads}\frac{k_{f,ads_j}[\mathrm{soot}^i][\mathrm{PAH}_j]}{k_{r,ads_j}+k_{c,ads_j}}
	\label{eqn:wads_reacdim},
\end{equation}

The contribution of PAH adsorption rate to the partial source terms can be expressed as:

\begin{equation}
	I^i_{C_{tot},ads} =
	\frac{1}{\rho}
	\sum_{i=1}^{n_{PAH}}
	\omega^i_{ads_j}
	n_{C,PAH_j}
	\label{eqn:ICtotads_reacdim},
\end{equation}

\begin{equation}
	I^i_{H_{tot},ads} =
	\frac{1}{\rho}
	\sum_{i=1}^{n_{PAH}}
	\omega^i_{ads_j}
	\left(n_{H,PAH_j}-2\right)
	\label{eqn:IHtotads_reacdim}.
\end{equation}

The rate of removal of PAH from the gas mixture due to inception and PAH adsorption is given as:

\begin{equation}
	\left(
	\frac{\mathrm{d}\left[{\mathrm{PAH}_j}\right]}{\mathrm{d}t}
	\right)_{inc}
	= 
	-\sum_{k=1}^{n_{PAH}}\omega_{dim_{jk}},
	\label{eqn:PAHscrub_inc_reacdim}
\end{equation}

\begin{equation}
	\left(
	\frac{\mathrm{d}\left[{\mathrm{PAH}_j}\right]}{\mathrm{d}t}
	\right)_{ads}
	= -\sum_{i=1}^{n_{sec}}\omega^i_{ads_j}.
	\label{eqn:PAHscrub_ads_reacdim}
\end{equation}

Additionally, during the PAH adsorption process one $\mathrm{H_2}$ is released to the gas mixture changing the concentration of $\mathrm{H_2}$ as:
\begin{equation}
	\left(
	\frac{\mathrm{d}\left[{\mathrm{H_2}}\right]}{\mathrm{d}t}
	\right)_{ads}
	= 
	\sum_{i=1}^{n_{sec}}\omega^i_{ads_j}
	\label{eqn:H2scrub_reacdim}.
\end{equation}

\subsection{Dimer Coalescence}
\label{sec:dimcoal}
The Dimer Coalescence model is a multi-step irreversible model proposed by \citet{blanquart2009joint} where self-collision of PAH molecules forms dimers, which are an intermediate state between gaseous PAH molecules and solid soot particles. The dimers can either form incipient soot particles through self-coalescence or adsorb on the surface of existing soot particles and contribute to their surface growth. The following equations describing the inception and surface growth in Dimer Coalescence adopted from the work of \citet{sun2021modelling}.

\reaction[reac:dim_dimcoal]{
	PAH_j + PAH_j ->[$k_{dim_{j}}$] Dimer_{j}
}
\reaction[reac:chemdim_reacdim]{
	Dimer_{j} + Dimer_{j} ->[$k_{inc_j}$] Tetramer_{j}
}
\reaction[reac:chemdim_reacdim]{
	Dimer_{j} + Soot ->[$k_{ads_j}$] Soot-PAH_{j}
}

The rate constant of dimerization, ${k_{dim_{j}}}$, and inception, ${k_{inc_{j}}}$, are calculated from the collision rate of PAHs as:

\begin{equation}
	k_{dim_{j}}=
	\gamma_{dim_j}\cdot\beta_{jj,PAH}\cdot Av
	\label{eqn:kdim_dimcoal},
\end{equation}

\begin{equation}
	k_{inc_{j}}=
	\beta_{jj,dimer}\cdot Av
	\label{eqn:kinc_dimcoal},
\end{equation}

\noindent where $\beta_{jj,PAH}$, $\beta_{jj,dimer}$ are both calculated using Equation~\eqref{eqn:betadim}. $\gamma_{dim_j}$ is the dimerization efficiency assumed to scale with the fourth power of the PAH molecular weight~\cite{blanquart2009analyzing} as:

\begin{equation}
	\gamma_{dim_j}=
	C_{N,j}\cdot W_{PAH_j}^4.
	\label{eqn:gamma_dimcoal}
\end{equation} 

\citet{blanquart2009joint} estimated the constant ${C_{N,j}}$ in Equation~\eqref{eqn:gamma_dimcoal} by comparing the profiles of several PAH species with the experimental measurements in a single premixed benzene flame~\citep{tregrossi1999combustion}, and provide ${C_{N,j}}$ values for various PAHs, which are listed in Table 1 in~\citep{blanquart2009analyzing}. The rate of dimer collision is expressed as:

\begin{equation}
	\omega_{dim_j} = \eta_{inc} k_{inc_{j}} [\mathrm{Dimer}_j] [\mathrm{Dimer}_j].
	\label{eqn:wdim_dimcoal}
\end{equation}

Similarly, the rate of adsorption of dimers on soot particles is obtained as:

\begin{equation}
	\omega^i_{ads_j} = \eta_{ads} k^i_{ads_{j}} [\mathrm{soot}^i] [\mathrm{Dimer}_j],
\end{equation}

\begin{equation}
	k^i_{ads_{j}}=
	\beta^i_{ads_j}\cdot Av.
	\label{eqn:kads_dimcoal}
\end{equation}

Assuming fast dimer consumption leads to the steady-state concentration of dimers, which can be determined by solving a quadratic equation~\citep{blanquart2009analyzing} as:
\begin{equation}
	a_{inc_j}[\mathrm{Dimer}_j]^2+b_{ads_j}[\mathrm{Dimer}_j]=\omega_{dim,j},
	\label{eqn:quad_dimcoal}
\end{equation}
\begin{equation}
	[\mathrm{Dimer}_j]=
	\left\{
	\begin{aligned}
		&\frac{-b_{ads_j}+\sqrt{\Delta_j}}{2a_{inc_j}},
		&&
		\text{if } \Delta_j \ge 0
		\\
		& 0 
		&&
		\text{if } \Delta_j < 0
	\end{aligned},
	\right.
	\label{eqn:dimer_dimcoal}
\end{equation}
\begin{equation}
	\Delta_j = b_{ads_j}^2+4a_{inc_j}\omega_{dim,j},
	\label{eqn:delta_dimcoal}
\end{equation}

\noindent where ${a_{inc_j} = k_{inc_{j}}}$ and ${b_{ads_j}}$ is calculated by summing the adsorption rate of dimers for all sections as:

\begin{equation}
	b_{ads_j} = \sum_{i=1}^{n_{sec}} k^i_{ads_{j}} [\mathrm{soot}^i]
\end{equation}


After determining the concentration of each dimer, the contributions of inception and PAH adsorption to the source terms of the tracked soot variables can be calculated in a manner similar to previous inception models, by accounting for the number of carbon and hydrogen atoms involved in the process as:

\begin{equation}
	I_{N,{inc}} = \frac{1}{\rho}
	\sum_{j=1}^{n_{PAH}}
	4\omega_{inc_{j}} 
	n_{PAH_j,C}
	\label{eqn:IN_inc_dimcoal},
\end{equation}

\begin{equation}
	I_{C_{tot},{inc}} = \frac{1}{\rho}
	\sum_{j=1}^{n_{PAH}}
	4\omega_{inc_{j}} 
	n_{PAH_j,C}
	\label{eqn:ICtot_inc_dimcoal},
\end{equation}

\begin{equation}
	I_{H_{tot},{inc}} = \frac{1}{\rho}
	\sum_{j=1}^{n_{PAH}}
	4\omega_{inc_{j}} 
	n_{PAH_j,H}
	\label{eqn:IHtot_inc_dimcoal},
\end{equation}

\begin{equation}
	I^i_{C_{tot},ads} =
	\frac{1}{\rho}
	\sum_{i=1}^{n_{PAH}}
	2\omega^i_{ads_j}
	n_{C,PAH_j}
	\label{eqn:ICtotads_dimcoal},
\end{equation}

\begin{equation}
	I^i_{H_{tot},ads} =
	\frac{1}{\rho}
	\sum_{i=1}^{n_{PAH}}
	2\omega^i_{ads_j}
	\left(n_{H,PAH_j}-1\right)
	\label{eqn:IHtotads_dimcoal}.
\end{equation}

The rate of removal of PAHs and release of $\mathrm{H_2}$ molecule due to PAH adsorption is calculated as:

\begin{equation}
	\left(
	\frac{\mathrm{d}\left[{\mathrm{PAH}_j}\right]}{\mathrm{d}t}
	\right)_{inc}
	= 
	-4\sum_{k=1}^{n_{PAH}}\omega_{inc_{j}},
	\label{eqn:PAHscrub_dimcoal_inc}
\end{equation}

\begin{equation}
	\left(
	\frac{\mathrm{d}\left[{\mathrm{PAH}_j}\right]}{\mathrm{d}t}
	\right)_{ads}
	= 
	-2\sum_{i=1}^{n_{sec}}\omega^i_{ads_j},
	\label{eqn:PAHscrub_dimcoal_ads}
\end{equation}

\begin{equation}
	\left(
	\frac{\mathrm{d}\left[{\mathrm{H_2}}\right]}{\mathrm{d}t}
	\right)_{ads}
	= 
	\sum_{i=1}^{n_{sec}}\omega^i_{ads_j}
	\label{eqn:H2scrub_dimcoal}.
\end{equation}

\section{Mass and energy balance}

\subsection{Constant Volume Reactor}
\begin{figure}[H]
	\centering
	\includegraphics[width=0.8\textwidth]{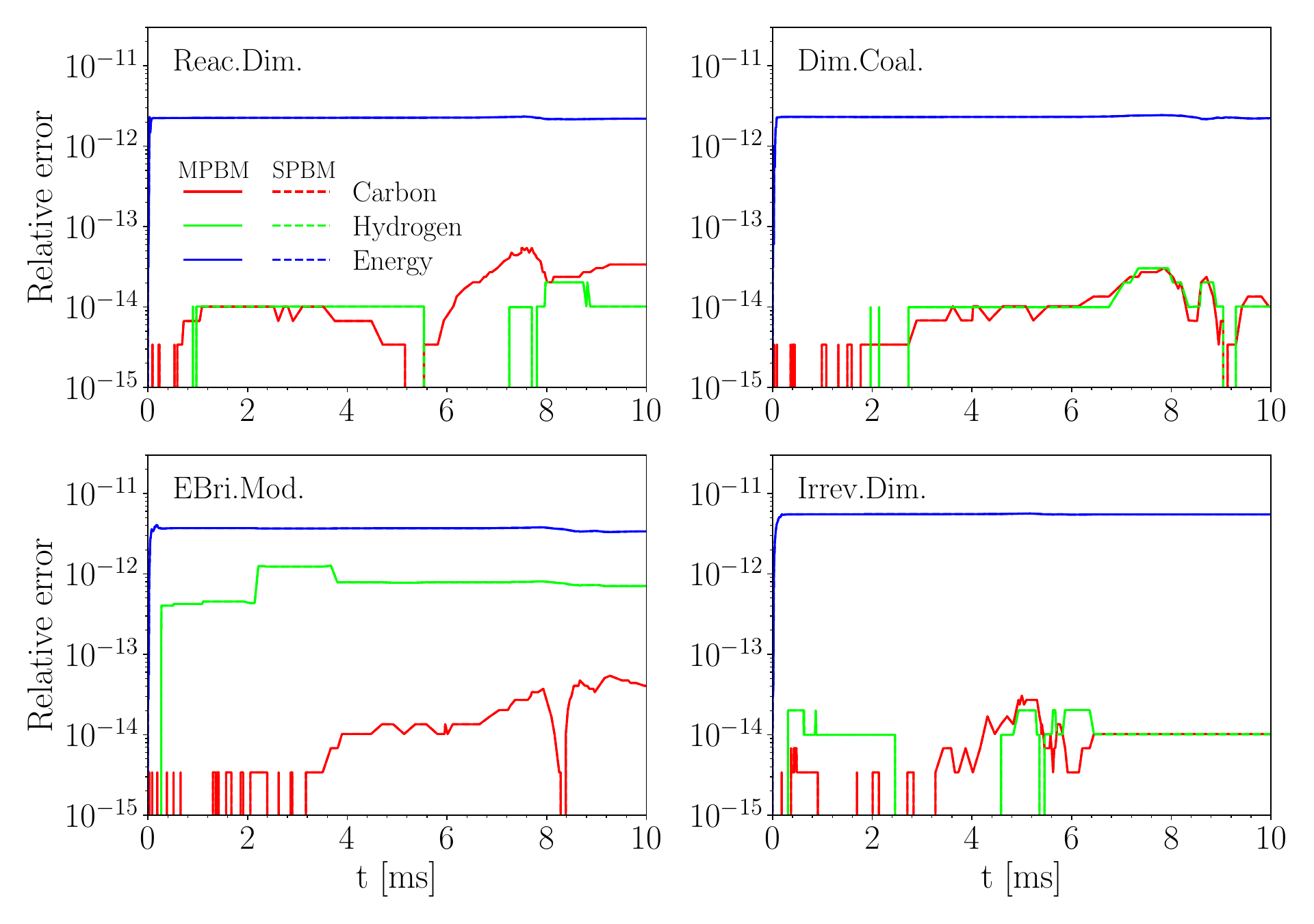}
	\caption{The relative error of total carbon (red line) and hydrogen (green line) mass, and total internal energy residual of gas and soot (blue line) plotted against residence time during pyrolysis of 30\% $\mathrm{CH_4}$-$\mathrm{N_2}$ at 2455 K and 3.47 atm in the constant volume reactor simulated using different PAH growth models along with MPBM (solid line) and SPBM (dashed line).}
	\label{fig:constuvvalid}
\end{figure}

The pyrolysis of 30\% $\mathrm{CH_4}$ diluted in $\mathrm{N_2}$ with the initial temperature and pressure of 2455 K and 3.47 atm, respectively, was simulated using CVR with a residence time of 40 ms. The combination of available PAH growth and particle dynamics models results in eight different cases, which were simulated to verify conservation of mass and energy. Here, we focus on the total elemental balance of carbon and hydrogen, as they are key elements involved in soot formation.
Figure~\ref{fig:constuvvalid} shows the relative errors in total carbon, hydrogen, and energy for the different PAH growth and particle dynamics models. In all cases, the errors fall below $\mathrm{10^{-10}}$, confirming that CVR satisfies mass and energy conservation.

\subsection{Constant Pressure Reactor}
\begin{figure}[H]
	\centering
	\includegraphics[width=0.8\textwidth]{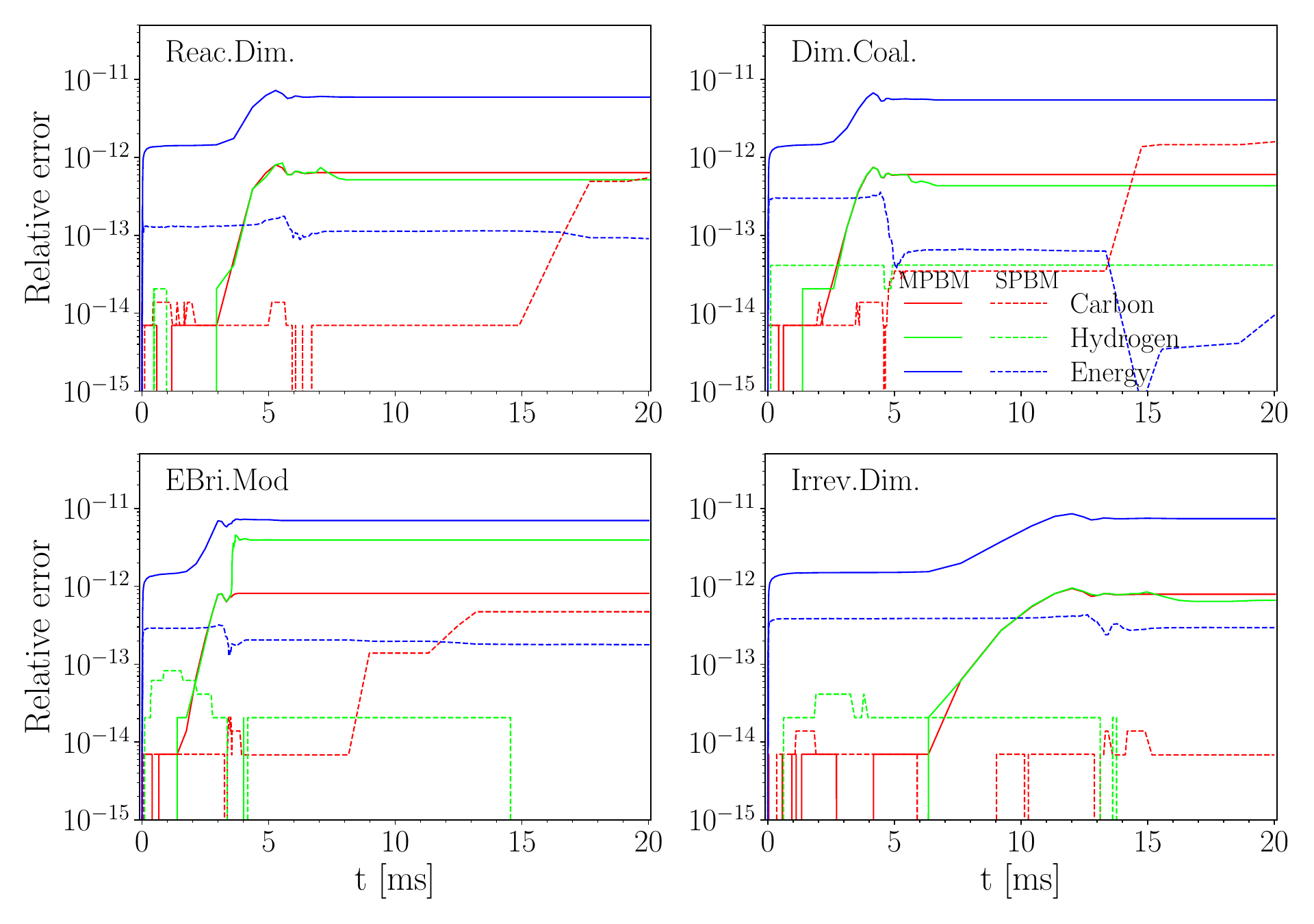}
	\caption{The relative error of total carbon (red line) and hydrogen (green line) mass, and total internal energy residual of gas and soot (blue line) plotted against residence time during pyrolysis of 5\% $\mathrm{CH_4}$-Ar at 2355 K and 4.64 atm simulated using CPR with different combinations of PAH growth models and particle dynamics models: MPBM (solid line) and SPBM (dashed line).}
	\label{fig:cprvalid}
\end{figure}

The pyrolysis of 5\% $\mathrm{CH_4}$-Ar in a shock-tube with post-reflected-shock temperature and pressure of 2355 K and 4.64 atm, respectively, was simulated using CPR model. Figure~\ref{fig:cprvalid} shows the relative error of total carbon, hydrogen and energy of system for different PAH growth and particle dynamics models in the constant pressure that falls below $\mathrm{10^{-10}}$ for all parameters confirming the validity of model in satisfying the mass and energy balance in the constant pressure reactor using all models.

\subsection{Perfectly Stirred Reactor}
\label{sec:psrvalid}
\begin{figure}[H]
	\centering
	\includegraphics[width=0.8\textwidth]{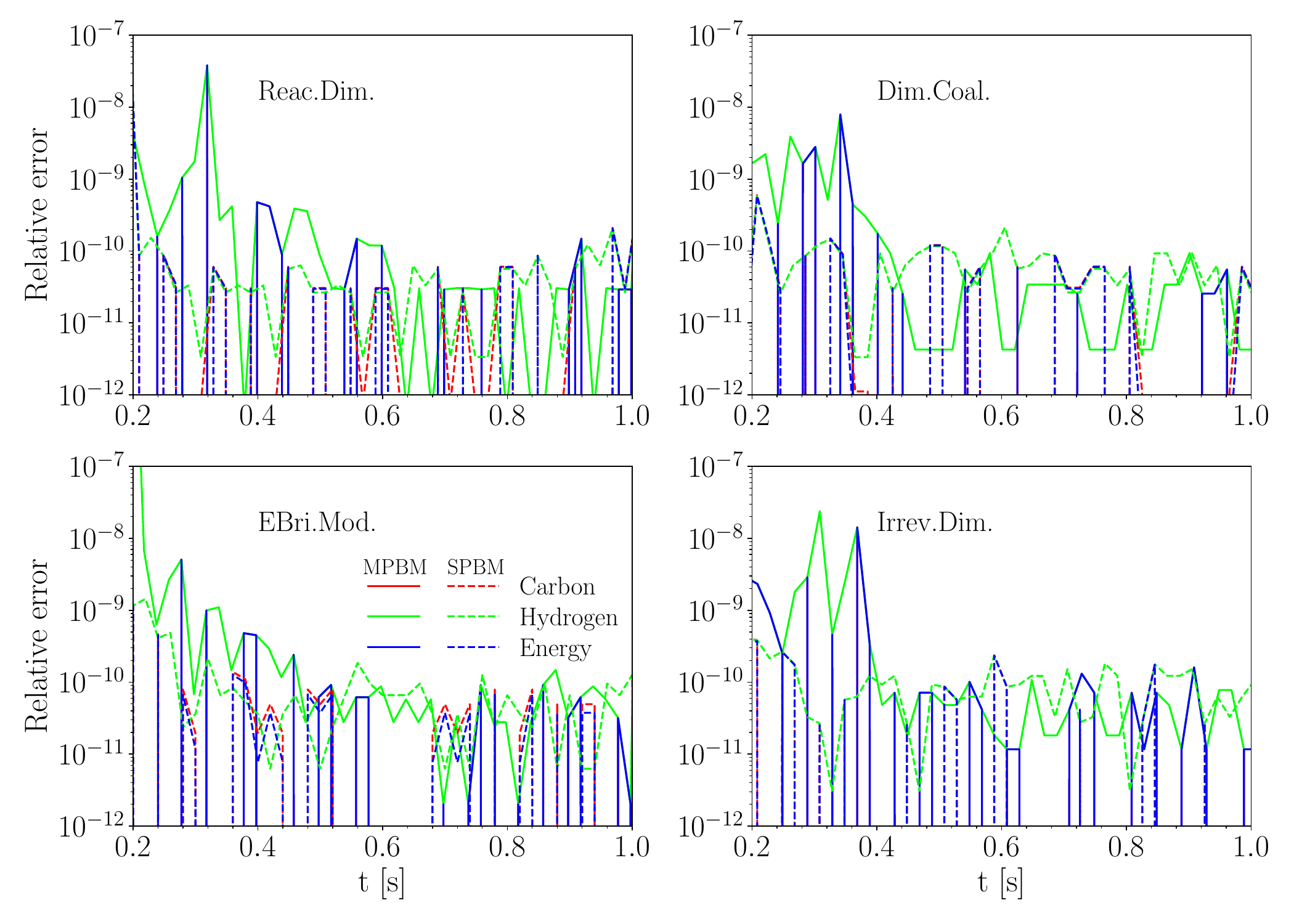}
	\caption{The relative error of total carbon (red line) and hydrogen (green line) mass, and total internal energy residual of gas and soot (blue line) plotted in simulation time during adiabatic combustion of $\mathrm{C_2H_4}$-air with $\phi=2$ at 1 atm simulated using different combinations of PAH growth models and particle dynamics models: MPBM (solid line) and SPBM (dashed line).}
	\label{fig:psrvalid}
\end{figure}

The mass and energy balance are investigated for soot formation during ethylene-air oxidation at equivalence ratio of $\phi=2$ in a perfectly stirred reactor. The simulation conditions were chosen based on the combustor implemented and utilized by \citet{stouffer2002combustion}. The reactants enter the reactor with the volume of 250 ml at 300 K and atmospheric pressure. The simulation is initialized from a high temperature ($\approx2000$~K) to avoid trivial solution (cold reactant leaving the reactor with no chemical reactions) and to ensure the model captures a sustained combustion. The residence time of products in the reactor is 8.5 ms. Figure~\ref{fig:psrvalid} shows the relative error of total elemental carbon and hydrogen mass and total enthalpy of gas and soot, which is less than $10^{-6}$ for all combinations of particle dynamics and PAH growth models.

\subsection{Plug Flow Reactor}
\begin{figure}[H]
	\centering
	\includegraphics[width=0.8\textwidth]{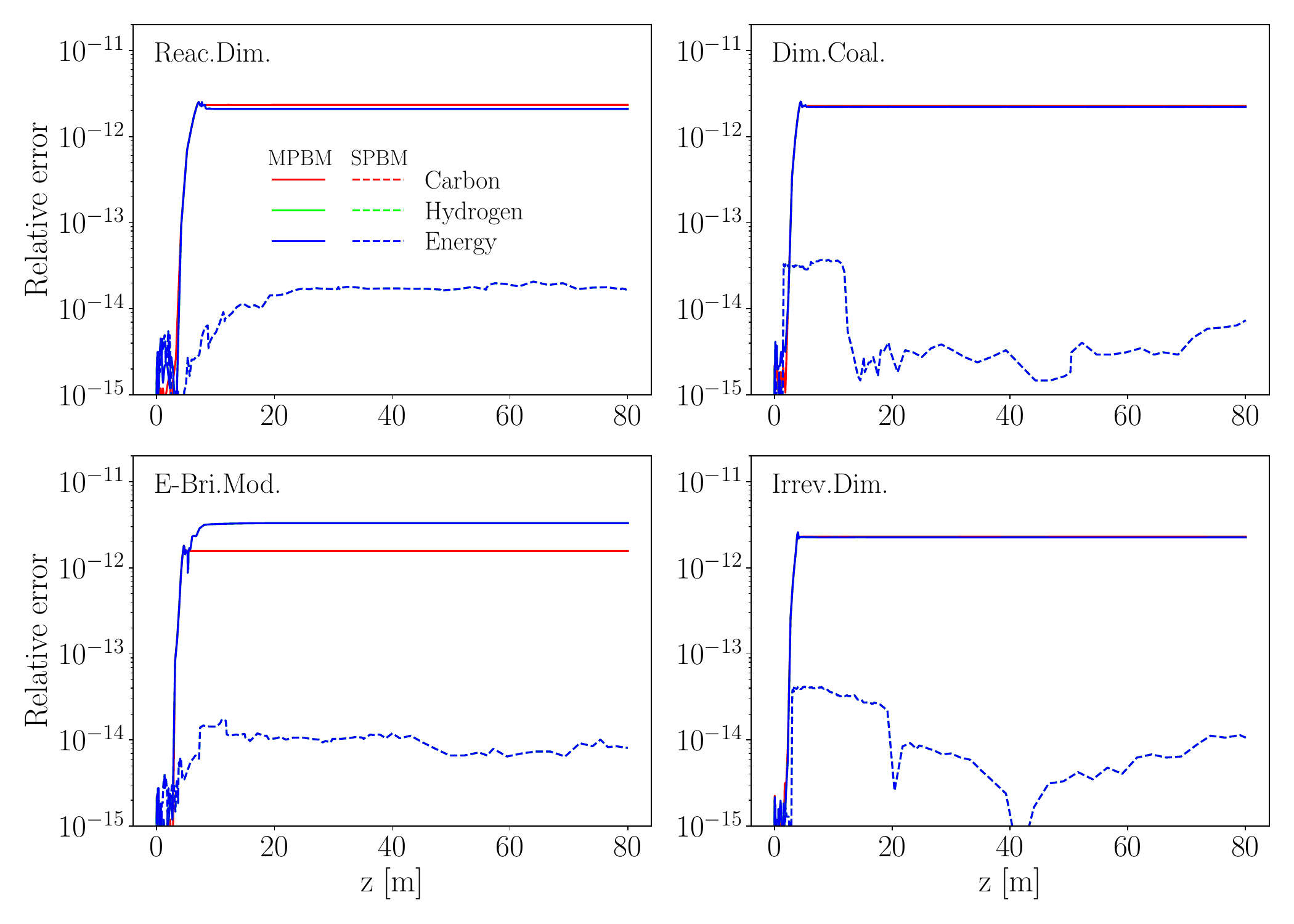}
	\caption{The relative error of total carbon (red line), hydrogen (green line) mass, and total internal energy residual of gas and soot (blue line) plotted against reactor length (cm) in the adiabatic flow reactor during pyrolysis of 30\%~$\mathrm{CH_4}$-$\mathrm{N_2}$ at 2100 K and 1 atm simulated using different combinations of PAH growth models and particle dynamics models: MPBM (solid line) and SPBM (dashed line).}
	\label{fig:pfrvalid}
\end{figure}

Methane pyrolysis in an adiabatic PFR was used to assess the conservation of elemental carbon, hydrogen, and energy. The inlet stream 30\% $\mathrm{CH_4}$ diluted in $\mathrm{N_2}$ with an initial temperature of 2100 K and pressure of 1 atm. Figure~\ref{fig:pfrvalid} shows the residuals of total elemental carbon, hydrogen, and energy along the reactor length, up to 40 cm, for all combinations of PAH growth and particle dynamics models. The residuals remain below $10^{-11}$, confirming that the PFR model in Omnisoot satisfies mass and energy conservation.

\section{Validation of Collision Frequency}
\label{sec:validcolfreq}
The collision frequency function determines the rate at which two particles collide, which results in the reduction of the total number of agglomerates and the increase in particle size. In the absence of strong flow shear or external forces, Brownian motion is the main driving force for particle coagulation. As explained in Sections~\ref{sec:sectextra} and \ref{sec:mpbm}, Omnisoot employs harmonic mean and Fuchs interpolations to calculate collision frequency of agglomerates from free-molecular ($\mathrm{Kn}\ge10$) to continuum ($\mathrm{Kn}\le0.1$) regimes based on the gas mean free path and the particle size.

\begin{figure}[H]
	\centering
	\begin{tikzpicture}
		\draw (0, 0) node[inner sep=0] 	{\includegraphics[width=0.45\textwidth]{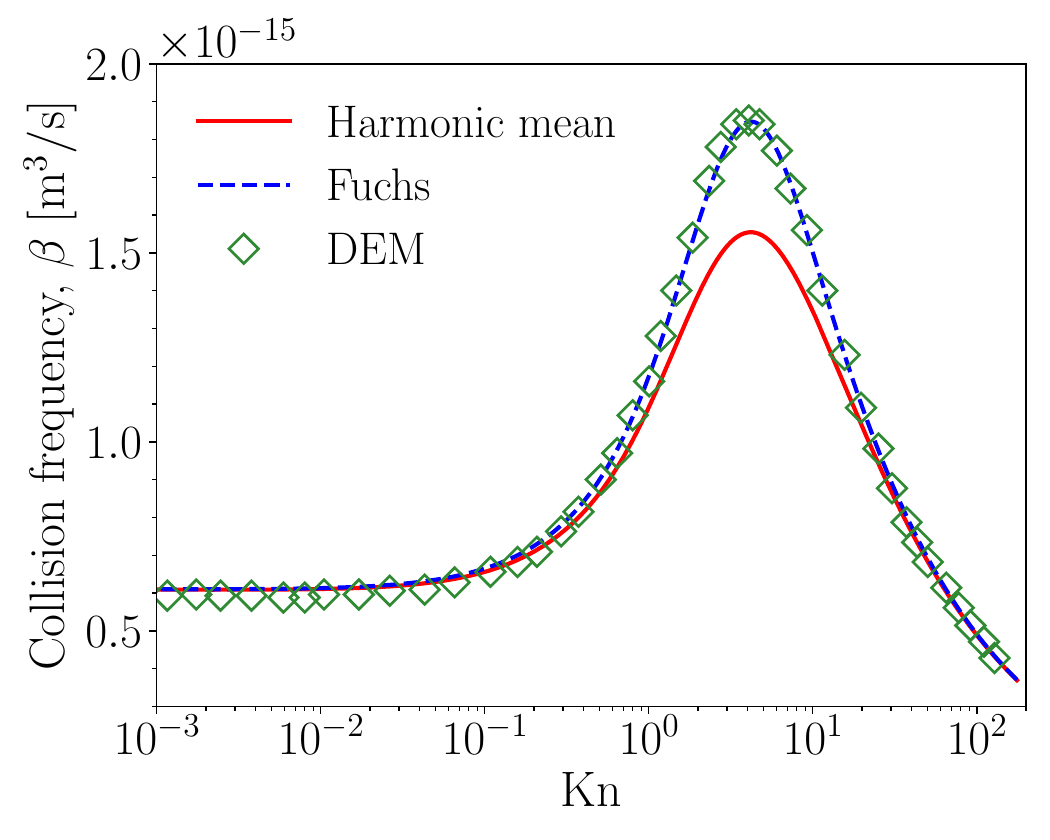}};
		\draw (-0.33, 1.11) node {\footnotesize{\cite{goudeli2015coagulation}}};
	\end{tikzpicture}
	\caption{The comparison of collision frequency, $\beta$, obtained by Omnisoot using harmonic mean (red solid line) and Fuchs interpolations (blue dashed line) with DEM results (symbols)~\citep{goudeli2015coagulation}.}
	\label{fig:kernelvalid} 
\end{figure}

\begin{figure}[H]
	\centering
	\begin{tikzpicture}
		\draw (0, 0) node[inner sep=0] 	{\includegraphics[width=0.35\textwidth]{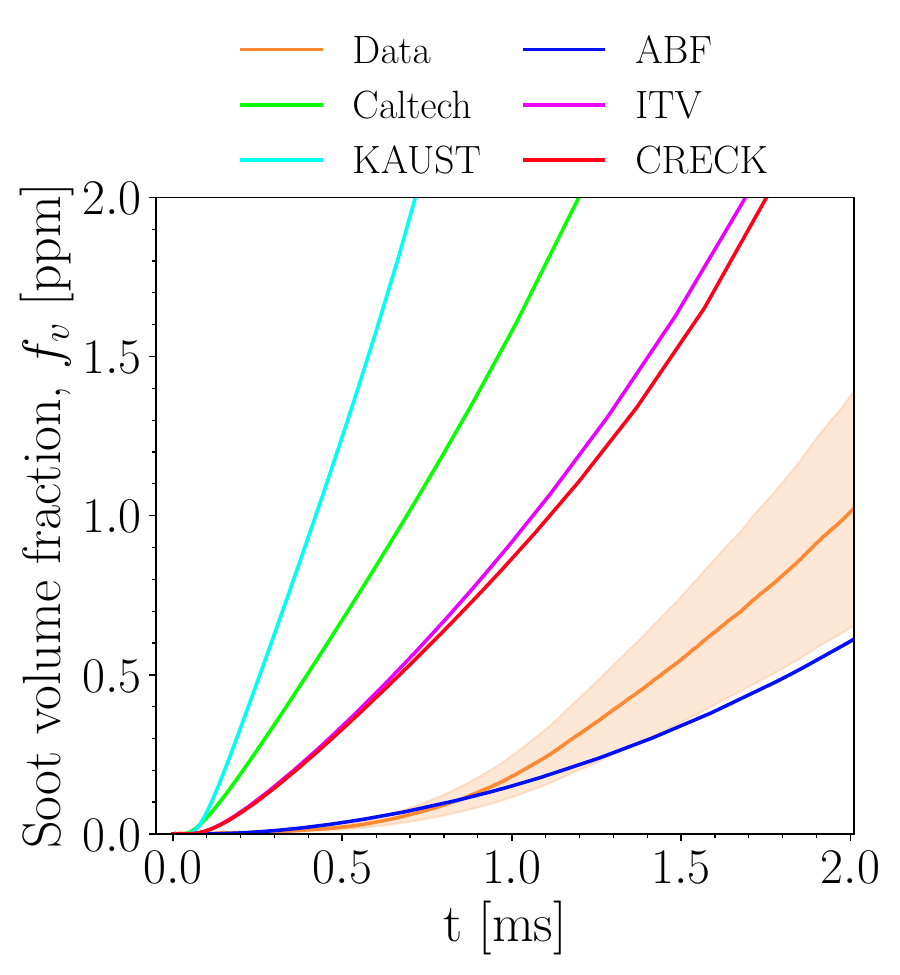}};
		\draw (0.14, 2.58) node {\tiny{\citep{clack2025}}};
	\end{tikzpicture}
	\caption{The maximum soot volume fraction predicted using Irreversible Dimerization and different reaction mechanisms during pyrolysis of 5\%~$\mathrm{CH_4}$-Ar was compared with measurements~\citep{clack2025}. The shaded area represents the uncertainty in the reported experimental data.}
	\label{fig:max_sootfv_chem} 
\end{figure}

The test case for validation of collision frequency is based on the DEM simulation of 2000 monodisperse spherical particles with the density of 2200 $\mathrm{kg/m^3}$ in
a cubic cell with the constant temperature of 298 K and pressure of 1 atm~\citep{goudeli2015coagulation}. Figure~\ref{fig:kernelvalid} depicts the collision frequency plotted against Knudsen number ($\mathrm{Kn}=2\lambda/d_m$) obtained by Omnisoot using harmonic mean (red solid line) and Fuchs interpolation (blue dashed line) and DEM results of \citet{goudeli2015coagulation}. The Fuchs interpretation perfectly matches DEM data over the free-molecular regime to the continuum regime. Harmonic mean is also in good agreement with the DEM results in the free-molecular and continuum regimes, but slightly underpredicts the collision frequency in the transition regime with relative errors less than 16\%.

\section{The comparison of maximum possible soot volume fraction predicted using different reaction mechanisms}

\section{The comparison of Monodisperse and Sectional population balance models}

\begin{figure}[H]
	\centering
	\includegraphics[width=0.8\textwidth]{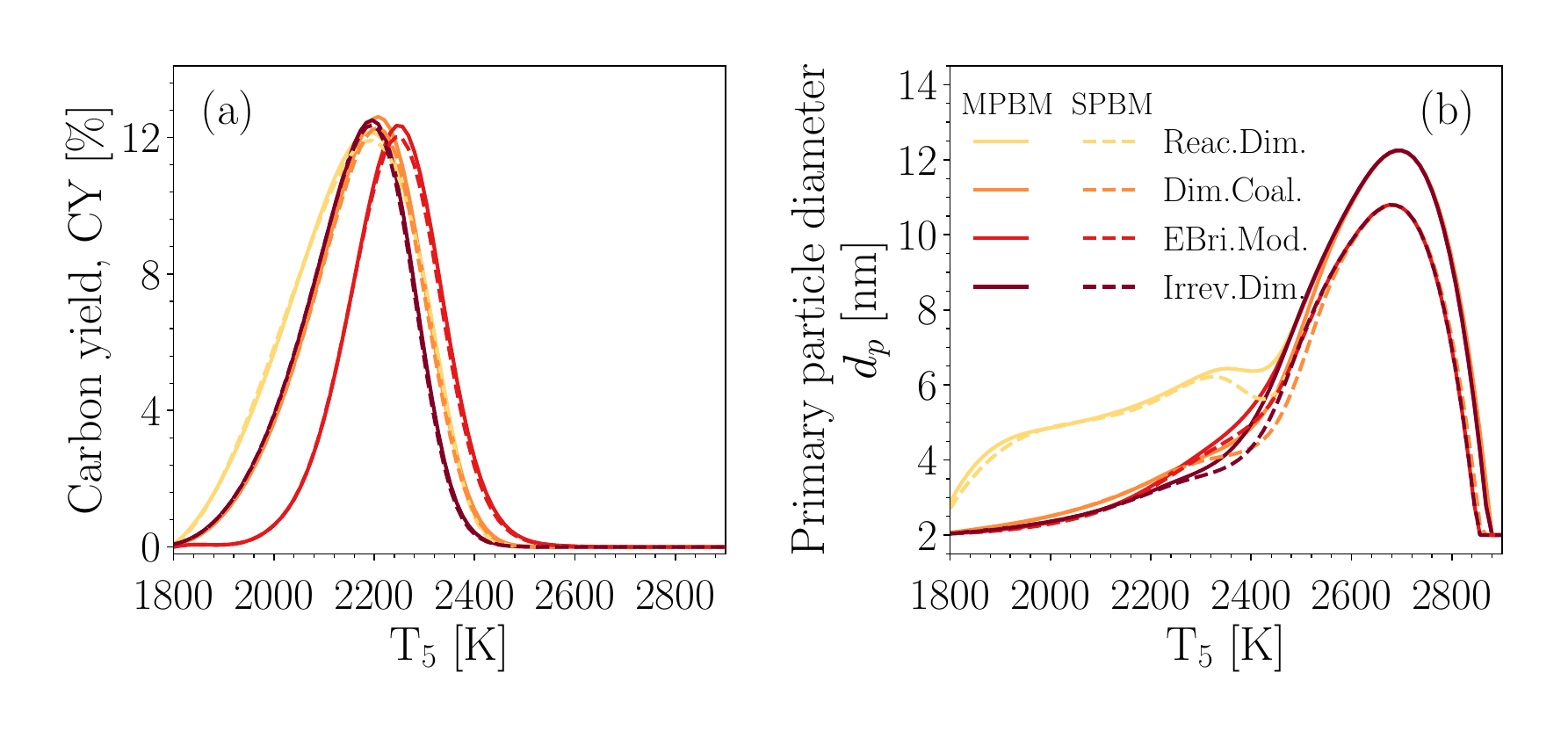};
	\caption{The comparison of CY (a) and primary particle diameter, $d_p$, at $t=$1.5 ms obtained using MPBM and SPBM models for the case optimized using equal adjustment factors to minimize the prediction error.}
	\label{fig:shockagof_yield_dp_cpr_pdynamics} 
\end{figure}

\begin{figure}[H]
	\centering
	\includegraphics[width=0.8\textwidth]{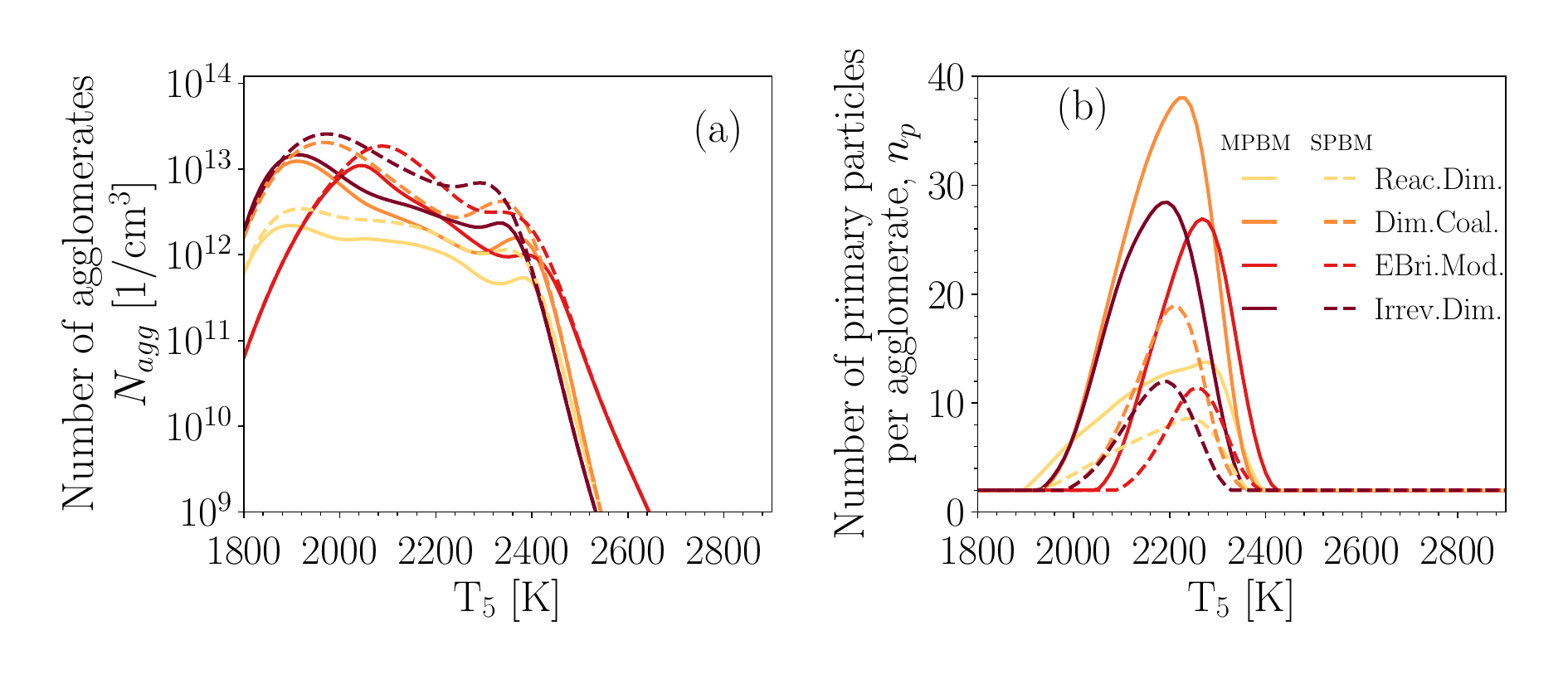};
	\caption{The temperature dependence of total number of agglomerates, $N_{agg}$ (a), number of primary particles per agglomerate, $n_p$ (b), at $t=$1.5 ms obtained using MPBM and SPBM models for the case optimized using equal adjustment factors to minimize the prediction error.}
	\label{fig:shockagof_N_agg_n_p_cpr_pdynamics} 
\end{figure}

\begin{figure}[H]
	\centering
	\includegraphics[width=0.8\textwidth]{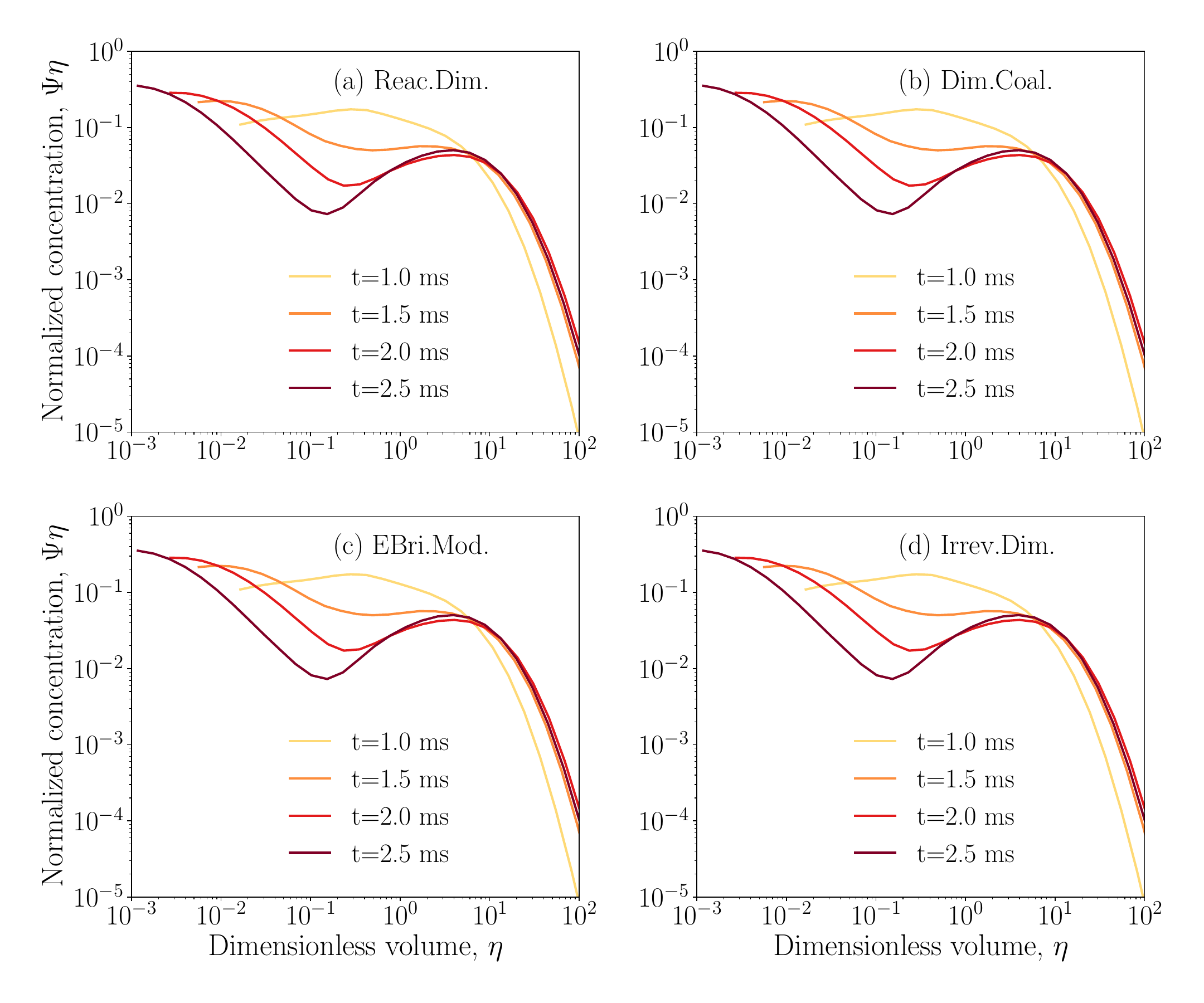};
	\caption{The non-dimensional particle size distribution during 5\%$\mathrm{CH_4}$-Ar at $\mathrm{T_5}=2200$ K that evolves from 1~ms to 2.5~ms indicating SPSD is not attained yet.}
	\label{fig:shockagof_psd} 
\end{figure}

\section{The Effect of Excluding Five-membered Rings}

\begin{figure}[H]
	\centering
	\begin{tikzpicture}
		\draw (0, 0) node[inner sep=0] 	{\includegraphics[width=0.8\textwidth]{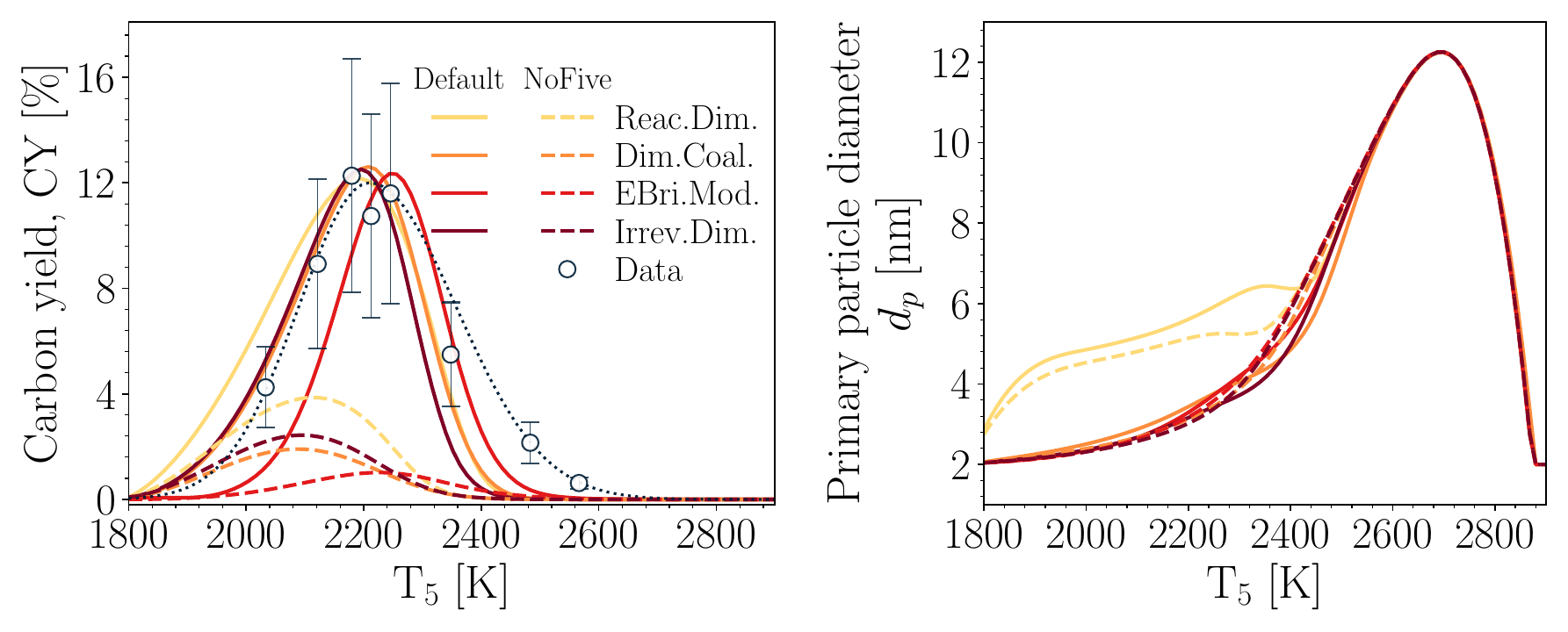}};
		\draw (-0.55, 0.29) node {\scriptsize{\cite{agafonov2016unified}}};
	\end{tikzpicture}
	\caption{The soot carbon yield, CY, at $t=$1.5 ms (a) and primary particle diameter, $d_p$ (b) obtained using the default soot precursors listed in Table~\ref{tab:precursors_list} (denoted by solid line and labeled as ``Default") compared with the same results when five-membered ring PAHs  excluded from soot precursors (denoted by dashed line and labeled as ``NoFive"). Both cases were obtained using Caltech mechanism and the same equal adjustment factors. The black dashed line was added to show the trend in the measurements~\citep{agafonov2016unified}.}
	\label{fig:shockagof_yieldspc_cpr} 
\end{figure}

\begin{figure}[H]
	\centering
	\includegraphics[width=0.8\textwidth]{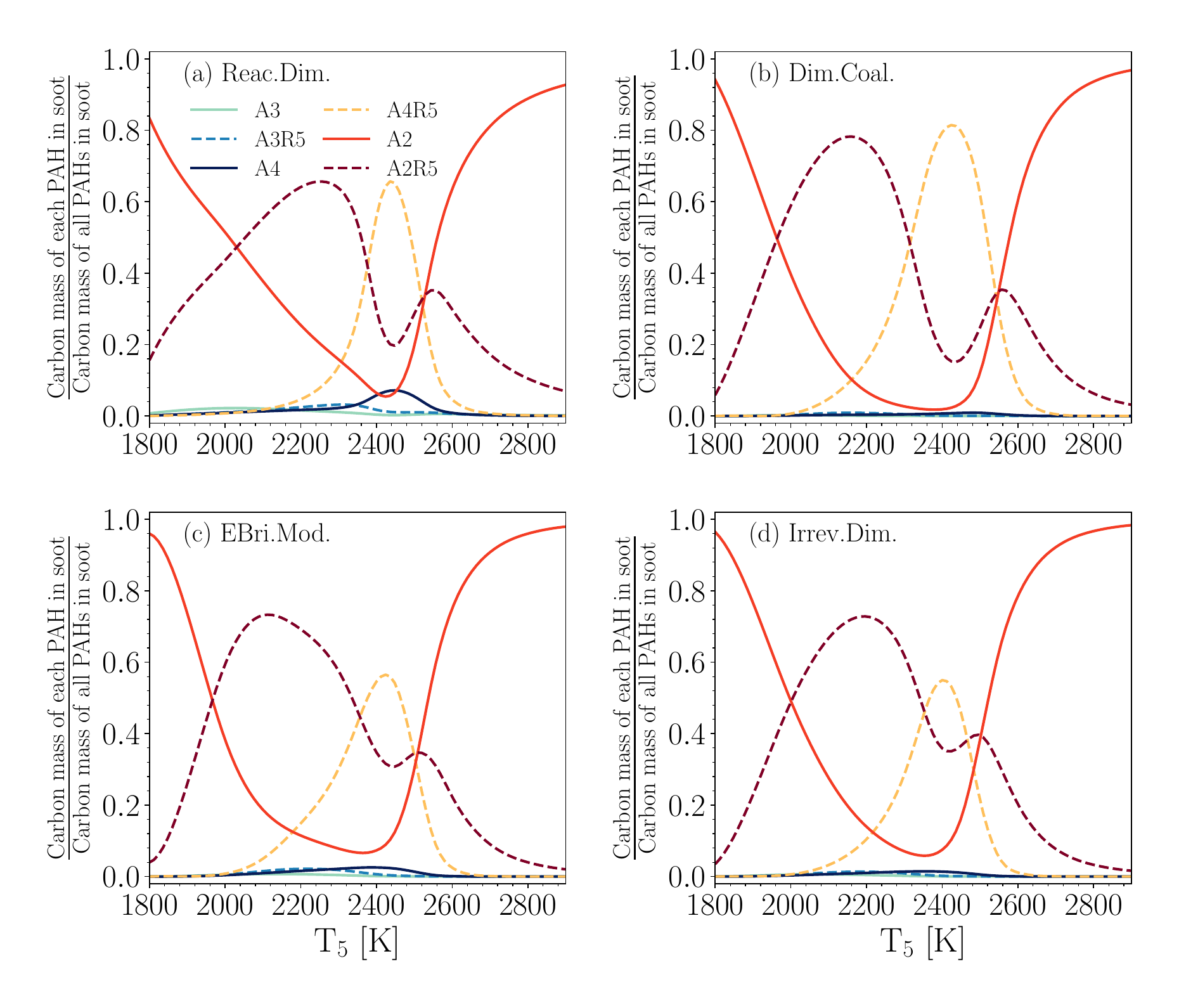}
	\caption{The contribution of each soot precursor to total carbon mass from precursors, at $t=$1.5 ms obtained using Caltech mechanisms, SPBM, and different inception models during 5\%$\mathrm{CH_4}$-Ar pyrolysis.}
	\label{fig:shockagof_spccont_cpr} 
\end{figure}

\section{The results of KAUST mechanisms for the shock-tube simulation}

\begin{figure}[H]
	\centering
	\begin{tikzpicture}
		\draw (0, 0) node[inner sep=0] 	{\includegraphics[width=0.8\textwidth]{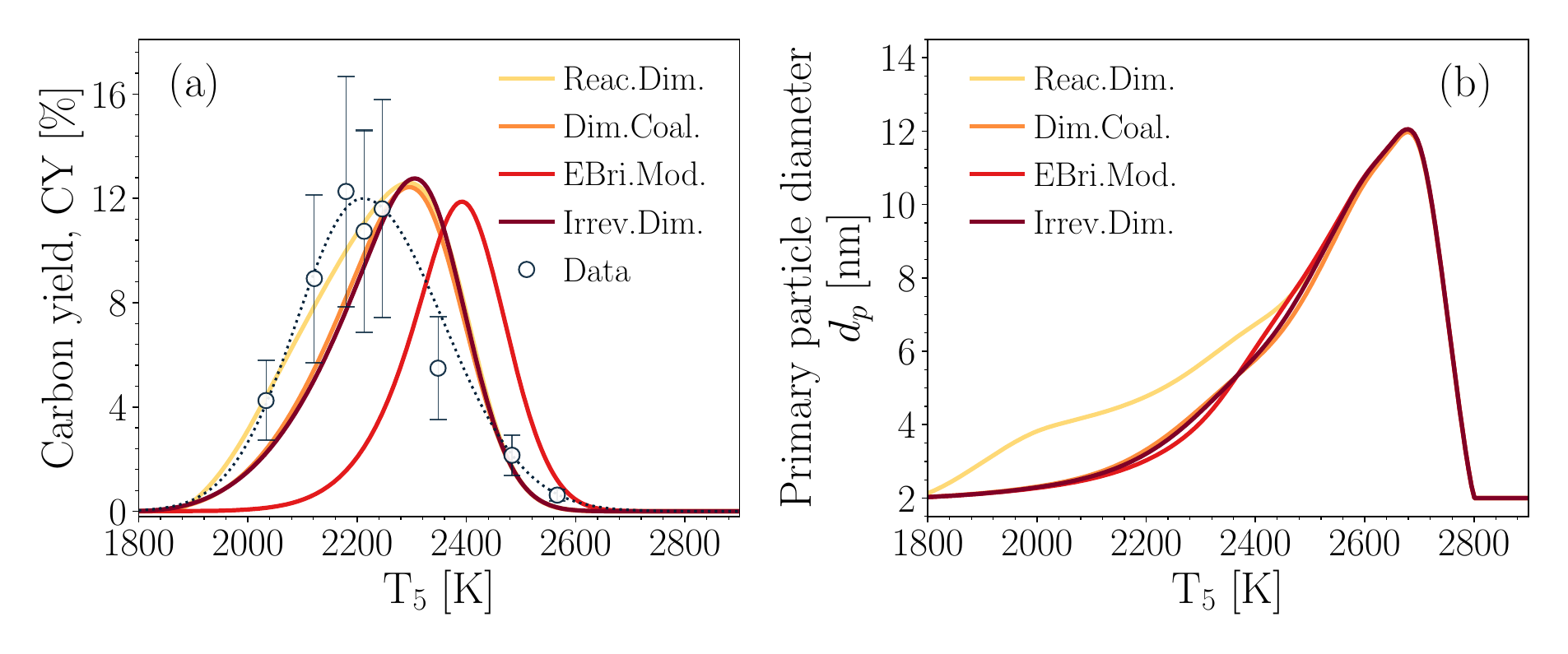}};
		\draw (-0.85, 0.42) node {\scriptsize{\cite{agafonov2016unified}}};
	\end{tikzpicture}
	\caption{The CY (a) and primary particle diameter, $d_p$, (b) at $t=1.5$~ms during the pyrolysis of 5\%~$\mathrm{CH_4}$-Ar obtained using KAUST mechanism, SPBM and different inception models optimized using equal adjustment factors to minimize the prediction error with extinction measurements~\citep{agafonov2016unified}. The dashed line was added to show the trend in the measurements.}
	\label{fig:shockagof_yield_dp_cpr_kaust} 
\end{figure}

\section{Ethylene pyrolysis in a flow reactor}

\begin{figure}[H]
	\centering
	\includegraphics[width=0.45\textwidth]{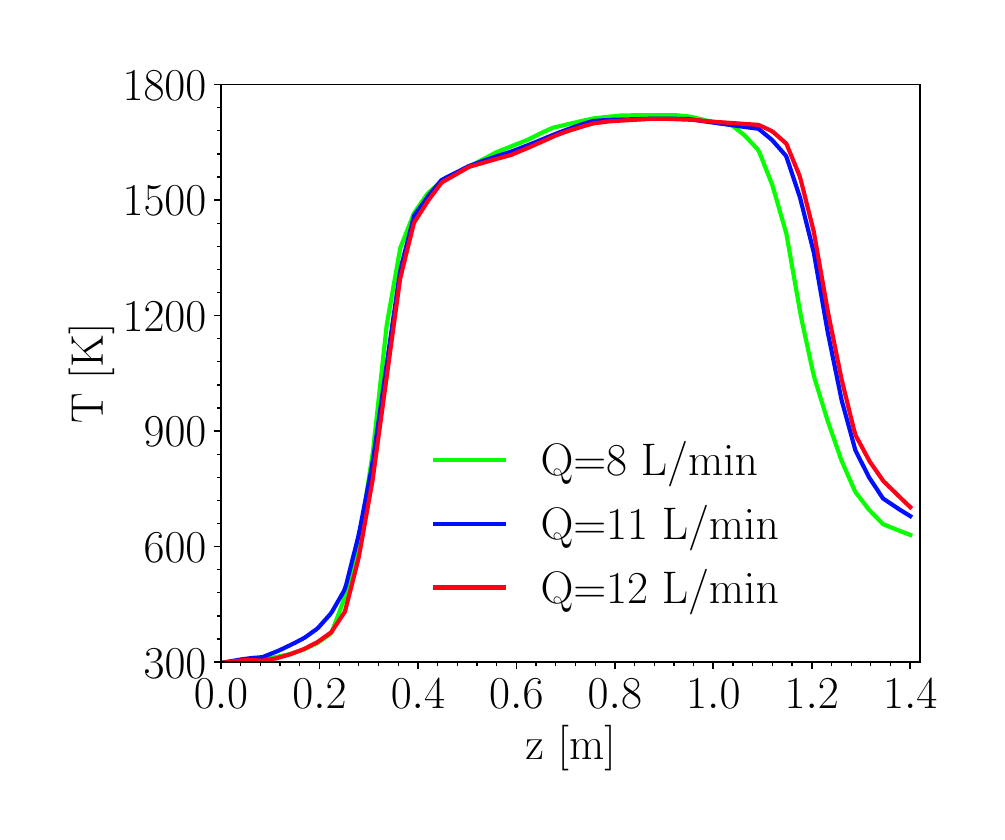}
	\caption{The centerline temperature along the reactor for $\mathrm{Q}=8$, 11, and 12 L/min interpolated from the thermocouple measurements~\citep{mei2019quantitative}.}
	\label{fig:pfr_temp} 
\end{figure}

\begin{figure}[H]
	\centering
	\begin{tikzpicture}
		\draw (0, 0) node[inner sep=0] 	{\includegraphics[width=1\textwidth]{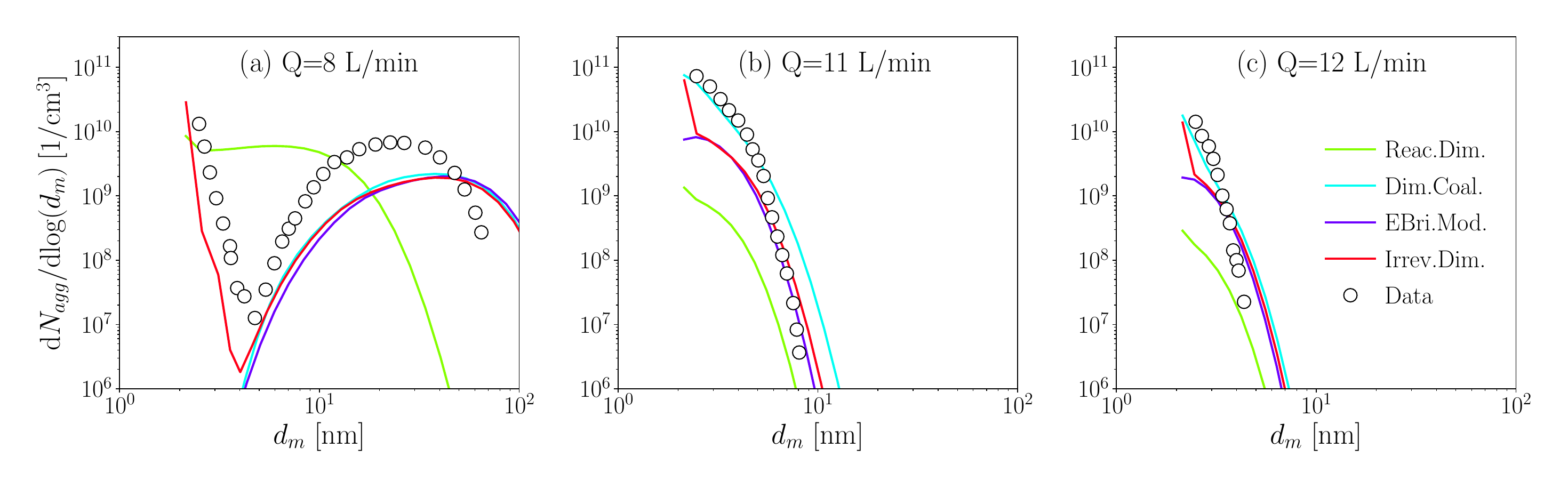}};
		\draw (6.63, -0.51) node {\scriptsize{\cite{mei2019quantitative}}};
	\end{tikzpicture}
	\caption{The particle size distribution at the end of PFR for $\mathrm{Q}=8$ (a), 11 (b), and 12 L/min (c) obtained using Caltech mechanism, SPBM and different inception models calibrated to match the predictions with measurement~\citep{mei2019quantitative}.}
	\label{fig:pfr_psd_caltech} 
\end{figure}

\begin{figure}[H]
	\centering
	\includegraphics[width=1\textwidth]{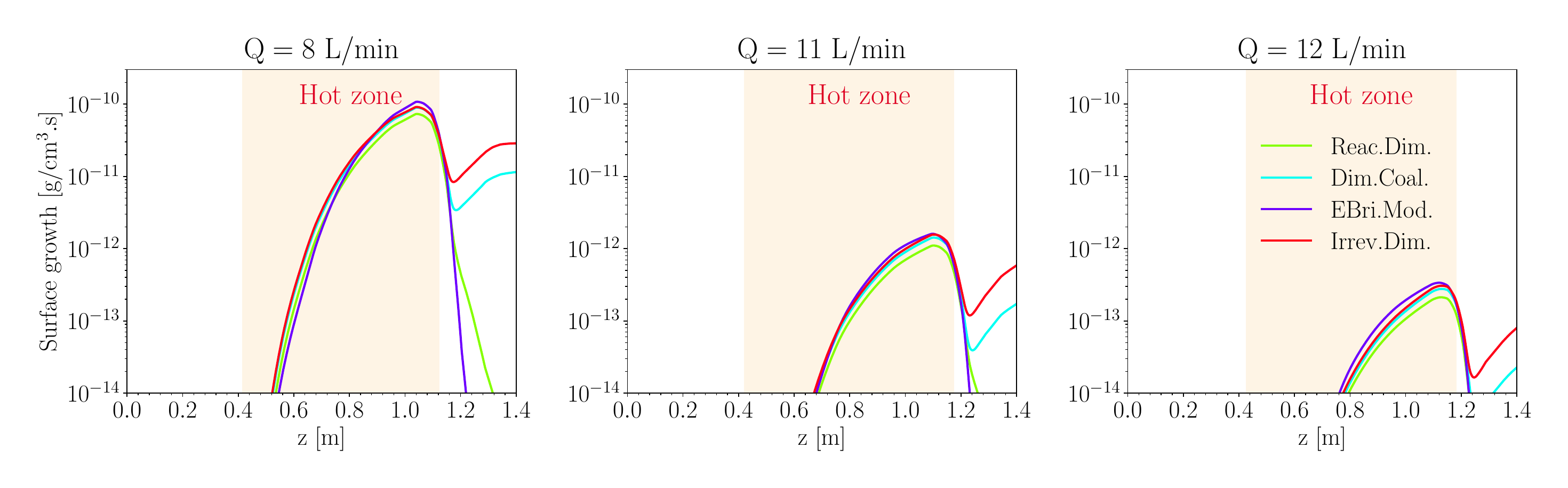}
	\caption{The surface growth by HACA and PAH adsorption combined along the PFR for $\mathrm{Q}=8$ (a), 11 (b), and 12 L/min (c) obtained using KAUST mechanism, SPBM and different inception models. The yellow area represents the hot zone ($\mathrm{T}>0.9\mathrm{T_{max}}$).}
	\label{fig:pfr_surfacegrowth} 
\end{figure}

\section{Ethylene oxidation in a perfectly stirred reactor}

\begin{table}[H]
	\centering
	\caption{The geometric mean mobility diameter, $d_{m,g}$, and the geometric mobility standard deviation, $\sigma_{m,g}$, obtained using different inception models compared with the value calculated from the measured PSD~\citep{manzello2007soot}.}
	\label{tab:psrpfr_morpcomp}
	\begin{tabular}{lllllll}
		\cline{2-7}
		& \multicolumn{2}{c}{$\phi=1.9$}                   & \multicolumn{2}{c}{$\phi=2.0$} & \multicolumn{2}{c}{$\phi=2.1$} \\ \cline{2-7} 
		& \multicolumn{1}{l} {$d_{m,g}$  [nm]} & $\sigma_{m,g}$ & \multicolumn{1}{l} {$d_{m,g}$  [nm]} &  $\sigma_{m,g}$ & \multicolumn{1}{l}{$d_{m,g}$  [nm]} & $\sigma_{m,g}$ \\ \hline
		\multicolumn{1}{l}{\textbf{Data}~\citep{manzello2007soot}}                      & \multicolumn{1}{l}{\textbf{6.04}}          &     \textbf{1.25}      & \multicolumn{1}{l}{\textbf{12.40}} &  \textbf{1.49} & \multicolumn{1}{l}{\textbf{17.66}} & \textbf{1.64} \\ 
		\multicolumn{1}{l}{Reactive Dimerization}     & \multicolumn{1}{l}{8.27}          &    1.14       & \multicolumn{1}{l}{11.10} & 1.21  & \multicolumn{1}{l}{16.88} & 1.58  \\ 
		\multicolumn{1}{l}{Dimer Coalescence}         & \multicolumn{1}{l}{8.18}          &      1.14     & \multicolumn{1}{l}{10.76} & 1.24 & \multicolumn{1}{l}{16.99} & 1.68 \\ 
		\multicolumn{1}{l}{E-Bridge Modified}          & \multicolumn{1}{l}{8.27}          &    1.15       & \multicolumn{1}{l}{11.02} & 1.27 & \multicolumn{1}{l}{14.56} & 1.58 \\ 
		\multicolumn{1}{l}{Irreversible Dimerization} & \multicolumn{1}{l}{8.19}          &      1.14     & \multicolumn{1}{l}{10.96} & 1.25 & \multicolumn{1}{l}{18.44} & 1.78 \\ \hline
	\end{tabular}
\end{table}

\end{document}